\documentclass[11pt,a4paper]{article}

\usepackage{jheppub}
\usepackage{amsmath}
\usepackage{multicol,bbm}
\usepackage{enumerate}
\usepackage{bibentry}

\usepackage{amssymb}
\usepackage{bm}
\usepackage{multirow}
\usepackage{caption}
\usepackage{tabularx}
\usepackage{enumitem}
\usepackage{extarrows}
\usepackage{tikz}
\usepackage{verbatim} % 多行注释
\usepackage{float} % 圖表固定位置

\graphicspath{{figures/}}  % 定义所有的 .eps 文件在 figures 子目录下

\def\be{\begin{equation}}
\def\ee{\end{equation}}
\def\ba{\begin{eqnarray}}
\def\ea{\end{eqnarray}}

\def\CP1{\mathbb{CP}^1}
\def\SL2C{\mathrm{SL}(2,\mathbb{C})}

\def\Z2{\mathbb{Z}_2}

\def\su2{{SU(2)}}

\def\[{\left[}
\def\]{\right]}

\def\({\left(}
\def\){\right)}
\def\[{\left[}
\def\]{\right]}

\def\<{\langle}
\def\>{\rangle}

\def\i2{\frac{i}{2}}

\def\2F1{\,_2{\rm F}_1}

\usepackage{tikz}
\usepackage{comment}
\begin{document}

% Use the \preprint command to place your local institutional report
% number in the upper righthand corner of the title page in preprint mode.
% Multiple \preprint commands are allowed.
% Use the 'preprintnumbers' class option to override journal defaults
% to display numbers if necessary
%\preprint{}

%Title of paper
\title{The ABJM Amplituhedron}

% repeat the \author .. \affiliation  etc. as needed
% \email, \thanks, \homepage, \altaffiliation all apply to the current
% author. Explanatory text should go in the []'s, actual e{-}mail
% address or url should go in the {}'s for \email and \homepage.
% Please use the appropriate macro foreach each type of information

% \affiliation command applies to all authors since the last
% \affiliation command. The \affiliation command should follow the
% other information
% \affiliation can be followed by \email, \homepage, \thanks as well.
%\author{}
%\email[]{Your e{-}mail address}
%\homepage[]{Your web page}
%\thanks{}
%\altaffiliation{}
%\affiliation{}
\date{\today}

\author[a,b,c]{Song He}
\author[d,e]{Yu-tin Huang}
\author[d,e]{Chia-Kai Kuo}

% The "\note" macro will give a warning: "Ignoring empty anchor..."
% you can safely ignore it.

\affiliation[a]{CAS Key Laboratory of Theoretical Physics, Institute of Theoretical Physics, Chinese Academy of Sciences, Beijing 100190, China}
\affiliation[b]{
School of Fundamental Physics and Mathematical Sciences,
Hangzhou Institute for Advanced Study; ICTP-AP International Centre for Theoretical Physics Asia-Pacific, UCAS, Hangzhou 310024, China}
\affiliation[c]%{School of Physical Sciences, University of Chinese Academy of Sciences, No.19A Yuquan Road, Beijing 100049, China}\affiliation[e]
{Peng Huanwu Center for Fundamental Theory, Hefei, Anhui 230026, P. R. China}
\affiliation[d]{Department of Physics and Center for Theoretical Physics, National Taiwan University, Taipei 10617, Taiwan}
\affiliation[e]{Physics Division, National Center for Theoretical Sciences, Taipei 10617, Taiwan}

% e-mail addresses: one for each author, in the same order as the authors

\emailAdd{songhe@itp.ac.cn}
\emailAdd{yutinyt@gmail.com}
\emailAdd{chiakaikuo@gmail.com}

\date{\today}

\abstract{In this paper, we take a major step towards the construction and applications of an all-loop, all-multiplicity amplituhedron for three-dimensional planar $\mathcal{N}=6$ Chern-Simons matter theory, or the $\textit{ABJM amplituhedron}$. We show that by simply changing the overall sign of the positive region of the original amplituhedron for four-dimensional planar $\mathcal{N}=4$ super-Yang-Mills (sYM) and performing a symplectic reduction, only three-dimensional kinematics in the middle sector of even-multiplicity survive. The resulting form of the geometry, combined with its parity images, gives the full loop integrand. This simple modification geometrically enforces the vanishing of odd-multiplicity cuts, and manifests the correct soft cuts as well as two-particle unitarity cuts. Furthermore, the so-called ``bipartite structures" of four-point all-loop negative geometries also directly generalize to all multiplicities. We introduce a novel approach for triangulating loop amplituhedra based on the kinematics of the tree region, resulting in local integrands tailored to ``prescriptive unitarity". This construction sheds fascinating new light on the interplay between loop and tree amplituhedra for both ABJM and $\mathcal{N}=4$ sYM: the loop geometry demands that the tree region must be dissected into $\textit{chambers}$, defined by the simultaneous positivity of maximal cuts. The loop geometry is then the ``fibration" of the tree region. Using the new construction, we give explicit results of one-loop integrands up to ten points and two-loop integrands up to eight points by computing the canonical form of ABJM loop amplituhedron.}

% insert suggested PACS numbers in braces on next line
%\pacs{}
% insert suggested keywords {-} APS authors don't need to do this
%\keywords{}

\maketitle
%must follow title, authors, abstract, \pacs, and \keywords
%\tableofcontents

\section{Introduction and summary of results}\label{sec:introduction}
The amplituhedron of planar ${\cal N}=4$ super Yang-Mills theory (sYM)~\cite{Arkani-Hamed:2013jha,Arkani-Hamed:2013kca, Arkani-Hamed:2017vfh} is a collection of {\it positive geometries} whose {\it canonical forms}~\cite{Arkani-Hamed:2017tmz} encode all-loop, all-multiplicity scattering amplitudes of the theory, where unitarity and causality {\it etc.} emerge from the geometries. In the past decade there has been a lot of progress in the mathematics and physics of the amplituhedron ({\it c.f.} \cite{Franco:2014csa, Ferro:2015grk, Bai:2015qoa, Dixon:2016apl,Karp:2016uax, Karp:2017ouj, Ferro:2018vpf, Galashin:2018fri, Arkani-Hamed:2018rsk,Salvatori:2018fjp, Kojima:2018qzz,Rao:2018uta, YelleshpurSrikant:2019meu, Langer:2019iuo,Lukowski:2019kqi, Herrmann:2020qlt, Kojima:2020gxs, Lukowski:2020dpn, Parisi:2021oql}), as well as in search for such positive geometries in other theories and contexts ({\it c.f.} \cite{Arkani-Hamed:2017fdk, Arkani-Hamed:2017mur, Eden:2017fow, Arkani-Hamed:2019mrd, Arkani-Hamed:2019vag, Arkani-Hamed:2019plo, He:2018okq, Damgaard:2019ztj, Ferro:2022abq, Huang:2021jlh, He:2021llb,  Huang:2018nqf,Arkani-Hamed:2018ign}). 

It has been long known that the scattering amplitude of three-dimensional ${\cal N}=6$ Chern-Simons-matter theory, or ABJM theory~\cite{Hosomichi:2008jb, Aharony:2008ug}, share many properties with ${\cal N}=4$ sYM. This includes the presence of Yangian (dual superconformal) symmetry for the leading singularities~\cite{Bargheer:2010hn, Huang:2010qy}, as well as its correspondence with cells of the positive (orthogonal) Grassmannian~\cite{Lee:2010du, Huang:2013owa, Huang:2014xza}. This leads to the natural question: \textit{is there a positive geometry associated with ABJM amplitudes, and if so, is the new geometry related to that of ${\cal N}=4$ sYM in a natural way?}  

Evidence for the existence of a geometry has been realized in~\cite{Huang:2021jlh, He:2021llb} where for particular component amplitudes, in the context of reduced supersymmetry, the answer can be identified as a momentum amplituhedron in a construction that is similar to the $\mathcal{N}=4$ sYM~\cite{Damgaard:2019ztj}. However, the component nature of the construction becomes an obstruction for generalizing to the loop level, since all components eventually appear in the cut of the amplitude. More recently, a new positive geometry was found in~\cite{He:2022cup}: remarkably the canonical form of the geometry gives the four-point planar integrand of ABJM theory to all loops. Not only has the geometry manifests various all-loop cuts such as soft cuts, unitarity cuts, and vanishing cuts due to odd-particle amplitudes, but one can also exploit it and compute four-point $L$-loop integrands up to $L=5$~\cite{He:2022cup} without much work. More recently, new integrated results, which contain {\it e.g.} $\Gamma_{\rm cusp}$ of the theory, have been obtained by integrating these four-point loop integrands up to $L=3$~\cite{He:2023exb,Henn:2023pkc} (see~\cite{Arkani-Hamed:2021iya, Chicherin:2022bov} for results in sYM).

The new-found geometry has a very simple relation with that of $\mathcal{N}=4$ sYM: one imposes a symplectic condition on all adjacent momentum twistors~\cite{Hodges:2009hk}, and an additional overall sign for all momentum-twistor four-brackets. The motivation for the former is straightforward: the symplectic condition reduces the original four-dimensional conformal group SU(4) to its three-dimensional counterpart Sp(4). The reason for the additional sign, however, is less clear. Furthermore, as the four-point tree geometry is trivial, it is unclear if the construction can be generalized to all multiplicity, where the subtle issue of branches, inherent in the definition of orthogonal Grassmannians, arises.

In this paper, we demonstrate that the answer is affirmative by introducing a novel interpretation of the amplituhedron. We begin by defining the ``tree-region" in the space of {\it momentum twistors}. This is the region for which the $n$ set of momentum twistors satisfying: 
\begin{equation}
   \boxed{\mathbb{T}_{n}:\quad\;\;  \begin{gathered}
         \langle i i{+}1 j j{+}1\rangle<0, \quad \left \{\langle 123 i\rangle\right \} \, \text{has $k$ sign flips},\\
         \langle\!\langle i i{+}1 \rangle\!\rangle=0 \ \text{for $i,j=1,2,\ldots,n$. }
     \end{gathered}\quad\,,  }
\end{equation}
where the four-bracket represents the $4\times 4$ determinant of $Z_i$s and the two-bracket represents contraction with Sp(4) metric. If we neglect the symplectic condition and reverse the overall sign of the four-bracket, we obtain the image of the tree $\mathcal{N}=4$ sYM amplituhedron in momentum twistor space. The change of sign leads to an immediate consequence: the region is non-empty {\it only} when $n=even$ and $k=\frac{n}{2}{-}2$. This is exactly the configuration where one has non-trivial amplitudes in ABJM. Thus the symplectic reduction and inversion of sign reduce the four-dimensional kinematics to a subspace of three-dimensional kinematics where ABJM has support! 

Next, for each point within $\mathbb{T}_n$ we can define its associated loop-region via:
\begin{equation} 
	\boxed{\mathbb{L}_n:\quad\begin{gathered}
	\quad 	\langle (AB)_a i i{+}1\rangle<0, \ \{\langle (AB)_a 1 i\rangle \} \ \text{has $k+2=\frac{n}{2}$ sign flip},\\
		\langle (AB)_a (AB)_b\rangle <0,\ \text{for any two loop $a,b\in 1,2,\ldots,L$},\\
		\langle\!\langle A_a B_a \rangle\!\rangle=0 \ \text{for $a\in 1,\ldots, L.$} 
	\end{gathered}}
\end{equation}
We claim that the planar loop-integrand of ABJM theory is given by the canonical form of $\mathbb{L}_n$. Note such geometric definition immediately manifests properties associated with the integrand of ABJM theory, such as reduction under soft-cuts or vanishing cuts with odd-multiplicity on one side. 

\begin{figure}[H]
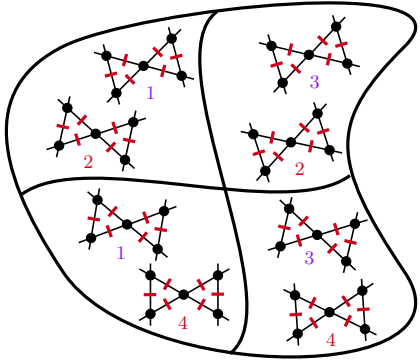
 
\begin{center}
$$
\vcenter{\hbox{\scalebox{0.8}{


}}}$$
\end{center}
\caption{An illustration of the chambers for the tree-region at eight points $\mathbb{T}_8$. Each chambers is characterized by the set of maximal cuts that are ``positive", i.e. that lives in the amplituhedron $\mathbb{L}_n$. }
\label{fig: Chambers} 
\end{figure}

Since we are considering canonical forms of a space ($\mathbb{L}_n$) that is dependent on where we are in $\mathbb{T}_n$, the form maybe distinct from point to point. This leads us to define a new structure within $\mathbb{T}_n$, which we termed ``\textbf{chambers}". As we traverse $\mathbb{T}_n$, a given chamber refers to regions whose loop-form are invariant. Thus $\mathbb{T}_n$ is dissected into several chambers. For example $\mathbb{T}_6$ has 1 chamber, $\mathbb{T}_8$ has 4 and $\mathbb{T}_{10}$ has 50. It turns out, this remarkable new feature of $\mathbb{T}_n$ has an invariant definition independent of loops. For each point in $\mathbb{T}_n$, consider the solution to
\begin{equation}
\sum_{j=1}^n \;(\mathcal{D}_{\Gamma_i})_{\alpha,j}Z_j=0,\quad \alpha=1,\cdots,k\,,
\end{equation}
where $(\mathcal{D}_{\Gamma_i})_{\alpha,j}$ are $4k$-dimensional cells in the momentum twistor orthogonal Grassmannian~\cite{Elvang:2014fja}. The collection of cells for which the solution is positive then defines the chambers. Thus the chambers can be considered as the ``overlap" of the cells. For example at eight-points, each of the four chambers are the overlap of two cells. As the $4k$-dimensional cells can also be identified as the kinematic solution for a $k$-loop maximal cut, the chambers can be said to be defined through the collection of maximal cut that resides in $\mathbb{L}_n$, see fig.~\ref{fig: Chambers}. Thus as one traverses $\mathbb{T}_n$, one sees that the loop-geometry is \textit{demanding} that the tree-region is further dissected into chambers. Note that sometimes distinct chambers might be degenerate at low loops, and are only fully resolved at higher loops: indeed the chambers for $\mathbb{T}_{10}$ are only fully resolved at two-loops. As we will see, this picture leads to a fibration of loop amplituhedron over the tree one:
\begin{equation}
\sum_{i} \mathcal{B}_i \times \Omega_i
\end{equation}
where $i$ labels the chambers of $\mathbb{T}_n$, $\mathbb{T}_n=\cup_{i} B_i$ and $\mathcal{B}_i, \Omega_i $ are the tree and loop-form associated with the chamber. We provide an illustration of this structure for eight-points:
$$\includegraphics[scale=0.6]{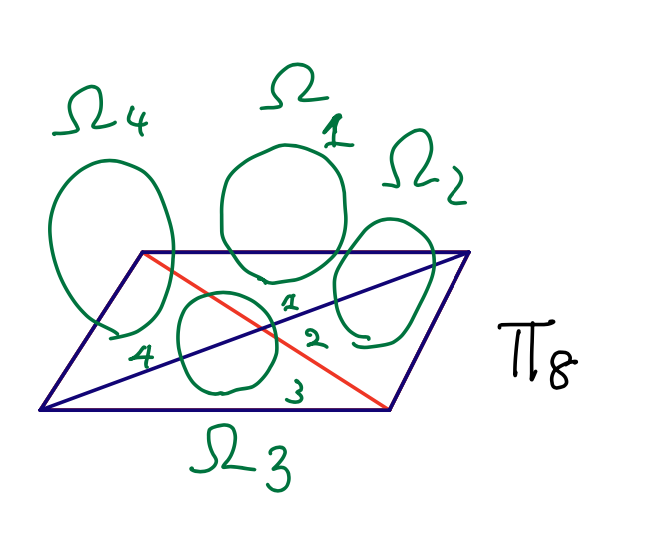}$$

Note that we are essentially triangulating the loop-region through $\mathbb{T}_n$, we do not introduce new boundaries in loop space. Thus the resulting forms are naturally local! As we will see that already for one loop ABJM, there is a unique loop form associated with each chamber, which turns out to be the sum of local integrals (boxes and triangles), matching the correct one-loop integrands up to $n=10$ as we have checked. For two loops, similarly we have obtained a unique two-loop form for each chamber as a linear combination of two-loop local integrals, which indeed produces correct results up to $n=8$ integrands~\cite{Caron-Huot:2012sos, He:2022lfz}. There is a single chamber/cell for $n=6$ (so any two points in the positive region give you the same loop form), and we have $4$ and $50$ such chambers by intersecting all possible BCFW chambers for $n=8,10$ respectively. As the distinct forms are associated with distinct maximal cut solutions, the representation is precisely that of ``prescriptive unitarity"~\cite{Bourjaily:2017wjl}. That is, the integrand consists of terms that evaluate to 1 on particular maximal cut solutions, and 0 on others. 

Note that our new way of triangulation is applicable to both ABJM and $\mathcal{N}=4$ sYM. Indeed it is an interesting mathematical problem of classifying such chambers both in ABJM and sYM, and we will also apply this new way of triangulating loop amplituhedron for sYM in~\cite{sYMchamber}. However, there is a subtlety that is special to ABJM. Just as the orthogonal group is a direct product of SO$(n)$ and parity reflection $\{\mathbb{I}, {-}\mathbb{I}\}$, the amplituhedron for ABJM also consists of the positive sector, and its parity conjugate image. For $k=\frac{n}{2}{-}2$, we have $2^{k}{-}1$ such images. This counting can be naturally understood from the number of maximal cut solutions for $k$-loops, which as discussed previously, are what characterizes the chambers in $\mathbb{T}_n$. Thus the full amplitude can be represented as:
\begin{equation}
\boxed{\quad  A_n^{L{-}loop}=\sum_{i,\alpha} \; \mathbf{\Pi}_{n,\alpha}\left[\mathcal{B}_i \times \mathbf{\Omega}_i \right]\quad }
\end{equation}
where $\mathbf{\Pi}_{n,\alpha}$ with $\alpha=0,1,\cdots, 2^{k}{-}1$ are the $n$-point parity operators. We have identified the resulting operators for $n=6,8$ (with $2^k=2,4$).

Independent of the interpretation as ABJM amplituhedron, already for $n=4$ the reduced geometry provides a nice simplified model for the original (all-loop) amplituhedron of ${\cal N}=4$ sYM. This becomes particularly clear when the amplituhedron is decomposed into the so-called {\it negative geometries}~\cite{Arkani-Hamed:2021iya}, which can be viewed as natural building blocks for multi-loop amplitudes in ${\cal N}=4$ sYM. As shown in~\cite{He:2022cup}, the reduction to $D=3$ has simplified negative geometries enormously: only those negative geometries corresponding to bipartite graphs survive the reduction, which not only drastically reduces the number of possible topologies but also put very strong constraints on their pole structure. We will see that essentially all these multi-loop structures are untouched when generalized to $n>4$, and we find that negative geometries again simplify the computation of loop forms for higher points. We will  explicitly construct the $L=2$ forms of negative geometries of each chamber of $n=6, 8$ cases, and explore consequences for some all-loop cuts. 

The rest of the paper is organized as follows. In section~\ref{sec:tree_geo} we motivate the tree geometry for ABJM theory in momentum twistor space by reviewing the momentum-space geometry implied by the Grassmannian integral and momentum-space amplituhedron. Equipped with the tree region, we proceed to define the loop amplituhedron for ABJM theory in section~\ref{sec:loop_geo}, where we demonstrate how the geometry manifests various consistency conditions unique to ABJM as well as the most important unitarity cuts; we then proceed to derive the loop-form for the positive sector for $n=6,8,10$ at one loop and $n=6,8$ at two loops. The full amplitude is then derived by introducing parity action on the positive sector in section~\ref{sec:parity}.

\section{Tree geometries}\label{sec:tree_geo}
It is instructive to begin with the tree-level amplitude, where its building blocks are related to $(n{-}3)$-dimensional cells of the orthogonal Grassmannian OG$^{>0}$($k',2k'$) where $n=2k'$. 
More precisely, the BCFW building blocks in the tree-level recursion can be written as residues of the integral formula~\cite{Lee:2010du}:
\begin{equation}\label{eq:Integral}
\int \frac{d^{k'\times 2k'}C}{ {\rm GL}(k')}\frac{\delta(C\cdot C^T)}{\prod_{i=1}^{k'}M_i} \delta^{2k'|3k'}(C\cdot\Lambda)\,,
\end{equation}
where $C$ denotes a $k'\times 2k'$ matrix that parameterizes $k'$-null planes in $2k'$ dimensional space; all inner products ``$\;\cdot\;$" are with respect to diagonal $n\times n$ metric with alternating signs and $M_i$ is the $i$-th $k'\times k'$ minor of $C$. The integral is localized to the $n{-}3$-dimensional subspace by evaluating the residues on the loci of vanishing minors $M_i$. Thus the subspace corresponds to the co-dimension $(k'{-}2)(k'{-}3)/2$ cells of the positive Grassmannian OG$^{>0}$($k',2k'$), {\it e.g.} for $n=4,6$ the amplitude is given by the top-cell while for $n=8$ by co-dimension one cells. The remaining $n{-}3$ degrees of freedom are then fixed by the following constraints ($2k'-3$ of them are independent) 
\begin{equation}\label{eq: Schubert}
\sum_{i=1}^n\; C_{\alpha, i}\lambda^a_i =0,\quad \alpha=1,\cdots,k'\,,
\end{equation} 
which imply momentum-conservation conditions $\sum_{i=1}^n (-1)^{i-1} \lambda_i^a \lambda_i^b=0$. Note importantly, the orthogonal condition $C\cdot C^T=0$ implies that the minors satisfy 
\begin{equation}
\frac{M_{i{+}k'}}{M_{i}}=\pm1\,.
\end{equation}
We will refer to the ratio being $+1$ ($-1$) as the positive (negative) branch.

It will be useful to utilize a graphical representation of these cells using on-shell diagrams, comprised of connected planar graphs of quartic vertices with $n$ external legs. For example for $n=6$, where the cell is top-dimensional, the representative graph is 
$$
\vcenter{\hbox{\scalebox{1}{


}}}$$
The $n{-}3=3$ degrees of freedom of this cell are associated with the three vertices. Each graph gives rise to a permutation pattern that encodes the linear dependency of the columns of $C_{\alpha, i}$, i.e. the pattern of vanishing minors. For a more detailed discussion, see~\cite{Huang:2013owa, Huang:2014xza}. For our purpose, it is sufficient to keep in mind that for a given on-shell diagram, solutions to eq.~\eqref{eq: Schubert} represents a configuration where all internal lines are on-shell for a given set of external kinematics. Said in another way, each vertex in the graph represents a four-point on-shell kinematics.

\paragraph{The tree region} The combination of cells in OG$(k',n)$ that ``tile" the tree amplitude can be mapped to a positive geometry through the moment map introduced in~\cite{He:2021llb, Huang:2021jlh} (see also~\cite{Lukowski:2021fkf}),  following that of ${\cal N}=4$ sYM ~\cite{Damgaard:2019ztj, Ferro:2020lgp}. In particular, a collection of ordered $2k'$ points on an $k{+}4$-dimensional moment can be represented as $\Lambda^A_i$ with $A=1,\cdots,2{+}k'$ and $i=1,\cdots,2k'$; then the momentum amplituhedron $Y_\alpha^A$ is the image of the top-cell in OG$(k',2k')$ through the following map: 
\begin{equation}
Y_\alpha^A=\sum_{i=1}^n \;C_{\alpha, i}\Lambda_i^A\,. 
\end{equation}
This space is equivalent to requiring the Gr($k',k'{+}2$) Grassmannian $Y_\alpha^A$ satisfying 
\begin{eqnarray}\label{MomAmp0}
\langle Y,i,i{+}1\rangle >0,&&\quad \nonumber\\
\sum_{i=1}^n\;(-1)^i\langle Y,i,a\rangle \langle Y,i,b\rangle=0,&&\;\forall a,b=1,\cdots,n\,\nonumber\\
\{\langle Y, 1, i\rangle\},&& \; {\rm has}\,k'\,{\rm sign\,flips} \,,
\end{eqnarray}
where $\langle Y,i,j\rangle\equiv \epsilon_{A_1A_2\cdots A_{k'{+}2}}Y_{\alpha_1}^{A_1}\cdots Y_{\alpha_{k'}}^{A_{k'}}\Lambda_{i}^{A_{k'{+}1}}\Lambda_{j}^{A_{k'{+}2}} $. 

The constraints in eq.~\eqref{MomAmp0} that defines the momentum amplituhedron can be directly framed in kinematic space $\lambda_i$. To see this, note that constraints on $\langle Y,i,j\rangle$ are really statements on the  components of $\Lambda_{i,j}$ that are perpendicular to $Y$. Thus it is natural to identify  
\begin{equation}\label{KinLam}
\lambda_i=Y^\perp (\Lambda_i^A)^T\,.
\end{equation}
Then we can simply identify $\langle Y,i,j\rangle=\langle i,j\rangle$, where $\langle i,j\rangle=\epsilon_{\alpha \beta}\lambda_i^{\alpha} \lambda_j^{\beta}$, with $\lambda_i$ defined through eq.~\eqref{KinLam}. Thus the tree amplituhedron is now defined through
\begin{equation}\label{MomAmp}
\boxed{\mathbb{K}_{n}:\quad\;\;\begin{gathered}
    \quad\langle i,i{+}1\rangle >0,\quad \{\langle  1, i\rangle\}, \; {\rm has}\,k'\,{\rm sign\,flips} \\
\sum_{i=1}^n\;(-1)^{i-1}\langle i,a\rangle \langle i,b\rangle=0,\;\forall a,b=1,\cdots,n\,.
\end{gathered}\;\;}
 \end{equation}
We will use $\mathbb{K}_n$ to denote any set of $n$ real spinors that satisfy the above sign constraints. Now, given a set of $\{\lambda_i\}$s, one can straightforwardly construct an associate set of \textit{momentum twistors} defined as 
\begin{equation}\label{Zmap}
Z_i=(-1)^{i-1}\left(\begin{array}{c} \lambda_i \\ \left(\sum_{j=1}^{i{-}1} p_j\right) \lambda_i\end{array}\right)\,.
\end{equation}
Then the sign patterns of eq.~\eqref{MomAmp} translates to:\footnote{Here $\langle i i{+}1 j j{+}1\rangle$ is defined with an extra $k$-dependent minus sign when $j j{+}1$ wraps around the last entry, i.e. we have $\langle i i{+}1  n1\rangle(-1)^{k-1}<0$.}
\begin{equation}\label{eq:tree_geo}
   \boxed{\mathbb{T}_{n}:\quad\;\;  \begin{gathered}
         \langle i i{+}1 j j{+}1\rangle<0, \quad \left \{\langle 123 i\rangle\right \} \, \text{has $k$ sign flips},\\
         \langle\!\langle i i{+}1 \rangle\!\rangle=0 \ \text{for $i,j=1,2,\ldots,n$. }
     \end{gathered}\quad\,.  }
\end{equation}
Here, $k=k'{-}2$ and the brackets are defined as $\langle ijkl\rangle := \epsilon_{IJKL} Z_i^I Z_j^J Z_k^K Z_l^L$ and $ \  \langle\!\langle ij\rangle\!\rangle := \Omega_{IJ}Z_i^I Z_j^J$. The Sp(4) symplectic metric $\Omega_{IJ}$ is given as 
\begin{equation}
    \begin{gathered} 
	\bold{\Omega}_{IJ}=\begin{pmatrix}
	           0 & \epsilon_{2\times 2} \\
	           \epsilon_{2\times 2} & 0 
	        \end{pmatrix}, \ 
	    \text{where} \    \epsilon_{2\times 2}=\begin{pmatrix}
	    0 & 1 \\
	    -1 & 0 
\end{pmatrix}.
    \end{gathered}
\end{equation}
From now on we will refer to this as the \textbf{tree (positive) region} $\mathbb{T}_{n}$. Note that under the constraint $\langle ii{+}1jj{+}1\rangle<0$, one sequence $\langle a a{+}1 a{+}2 i \rangle$ having the correct number of sign flips is sufficient to guarantee the same for others, and we have chosen $\langle 1 2 3 i \rangle$ as a representative. There is a similar T-dual relation for the cells of the orthogonal Grassmannian. Starting with a cell $C_{\Gamma}$ in ${\rm OG}(k',2k')$, on the support of eq.~\eqref{eq: Schubert}, one can map to a cell $\mathcal{D}_{\Gamma}$ in  $\widetilde{{\rm OG}}(k,2k{+}4)$, where the $\widetilde{\;\;}$ indicates the orthogonal condition is defined with respect to a kinematic dependent $n\times n$ metric. In particular, the co-dimension $(k'{-}2)(k'{-}3)/2$ cell $C_{\Gamma}$ will be mapped to a co-dimension $k(k{-}1)/2$ $\mathcal{D}_{\Gamma}$ and the $k'\times k'$ minor $M^{C}_i$ maps to $k\times k$ minor $M^{\mathcal{D}}_{i{+}1}$, where the superscript labels the Grassmannian. This relation can be derived on the level of the integral formula in eq.~\eqref{eq:Integral} which we review in appendix~\ref{sec:Tdual}.  

\paragraph{Absence of non-middle sectors and chambers} At this point, we have motivated the definition of the tree-region $\mathbb{T}_n$ as an image from the T-dualization of $\mathbb{K}_n$. As it turns out, $\mathbb{T}_n$ in itself is a natural starting point for the study of ABJM amplitudes in momentum twistor space, without ever referring to $\mathbb{K}_n$. To see this, first note that $\mathbb{T}_n$ is very similar to the tree-region for $\mathcal{N}=4$ sYM, except for two aspects: the symplectic constraint and the overall negative sign for $\langle ii{+}1jj{+}1\rangle$. The first is simply reflecting 3-dimensional kinematics, but remarkably, the second, {\it i.e.} whether $\langle ii{+}1jj{+}1\rangle$ is strictly positive or negative, determines which 3-dimensional theory we are talking about. Indeed with $\langle ii{+}1jj{+}1\rangle>0$, the sign flipping and symplectic constraint, we are simply considering the tree-level amplitude of $\mathcal{N}=4$ sYM, with the kinematics lying on a three-dimensional subspace. If we instead choose $\langle ii{+}1jj{+}1\rangle<0$, $\mathbb{T}_n$ is actually empty whenever $n$ is odd! Indeed using the identity 
\begin{equation}\label{eq:two-brackets formula}
\langle abcd\rangle= {-}\langle\!\langle ab \rangle\!\rangle \langle\!\langle cd \rangle\!\rangle{+}\langle\!\langle bc \rangle\!\rangle \langle\!\langle da \rangle\!\rangle{-} \langle\!\langle ac \rangle\!\rangle \langle\!\langle db \rangle\!\rangle\,,
\end{equation}
we have $\langle i i{+}1 i{+}2 i{+}3\rangle=\langle\!\langle i i{+}2 \rangle\!\rangle \langle\!\langle i{+}1 i{+}3 \rangle\!\rangle $ and thus 
\begin{equation}
\langle n{-}2 n{-}1 n \hat{1}\rangle\langle  n \hat{1} 23\rangle=\langle n{-}1 n 12\rangle\left(\frac{\prod_{i=1}^{n{-}3}\langle i i{+}1 i{+}2 i{+}3\rangle}{\prod_{i=2}^{n{-}3}\langle\!\langle i i{+}2 \rangle\!\rangle^2}\right)\,.
\end{equation}
For odd $n$ and $Z\in \mathbb{R}$, the parenthesis on the RHS is a definite positive sign. Thus if $\langle i i{+}1 j j{+}1\rangle<0$, the LHS is strictly positive while the RHS is negative, leading to a contradiction. For even $n$, one can similarly show that contradiction will arise unless $k=\frac{n}{2}-2$. We demonstrate this for $n=6,8$ in appendix~\ref{sec:AppendixTn}. Thus, all non-middle sectors are absent and $\mathbb{T}_n$ is non-trivial exactly when ABJM on-shell kinematics is defined! From now on, we will only require that $Z_i$ lives in $\mathbb{T}_n$, not necessarily derived from $\mathbb{K}_n$ using eq.~\eqref{Zmap}. So for example, the first two components of $Z_i$ do not necessarily correspond to the two-component spinors that satisfy momentum conservation.

\begin{figure}[H]
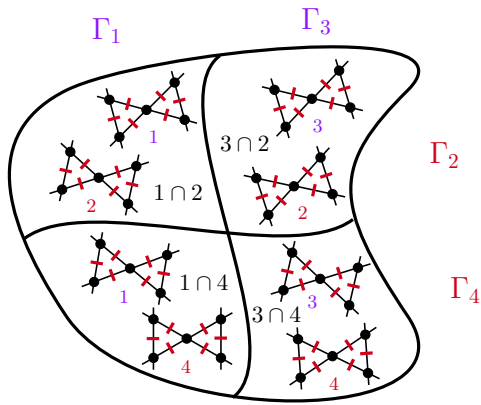
 
\begin{center}
$$
\vcenter{\hbox{\scalebox{0.8}{


}}}$$
\end{center}
\caption{An illustration of the chambers for the tree-region at eight points $\mathbb{T}_8$. Each chamber is characterized by the set of BCFW cells ($\Gamma_i$) that are positive under eq.~\eqref{eq:MomTwisGras} (or eq.~\eqref{eq: Schubert}). Equivalent, it is identified by the set of maximal cuts that are ``positive", i.e. there is one maximal cut solution that lives in the amplituhedron. }
\label{fig: Chambers2} 
\end{figure}

Now given a point in $\mathbb{T}_n$, we say that it is associated with a cell $\Gamma_i$ if solving 
\begin{equation}\label{eq:MomTwisGras}
\sum_{j=1}^n \;(\mathcal{D}_{\Gamma_i})_{\alpha,j}Z_j=0,\quad \alpha=1,\cdots,k\,,
\end{equation}
yields a positive $\mathcal{D}_{\Gamma_i}$, i.e. all the minors $M_i^{\mathcal{D}_{\Gamma_i}}$ are non-negative. Note that there are in total $2^k$ number of solutions for eq.~\eqref{eq:MomTwisGras}, so here we are referring to one of the solutions being positive. We will refer to this as the positive solution. For a given BCFW triangulation, the point can be associated with only one cell since such triangulations yield non-overlap tilling.\footnote{Since we do not have the tree amplituhedron in momentum twistor space, at this point, the non-overlapping feature can only be phrased in momentum amplituhedron.} However, it is perfectly acceptable for a point to simultaneously lie in multiple cells, as long as the set of ``overlapping" cells do not participate in a triangulation. Thus the maximal number of BCFW cells that can be associated with a given point is  the total number of distinct BCFW triangulations. Since for $n=4,6$ the tree amplitude corresponds to the top cell, we will use $n=8$ to illustrate this feature. The BCFW cells of $\widetilde{\rm OG}(2,8)$ can be categorized by the vanishing of its $2\times 2$ ordered minors:
\begin{equation}
\Gamma_1:\;\; M^{\mathcal{D}}_2=0,\quad \Gamma_2:\;\; M^{\mathcal{D}}_3=0,\quad \Gamma_3:\;\; M^{\mathcal{D}}_4=0,\quad \Gamma_4:\;\; M^{\mathcal{D}}_5=0\,.
\end{equation}
The tree amplitude is then given by the sum of the Grassmannian integral localized on cells $\Gamma_1$ and $\Gamma_3$ or $\Gamma_2$ and $\Gamma_4$. The two are equivalent via Cauchy's theorem. Now $\mathbb{T}_{8}$ can be schematically viewed as in fig.~\ref{fig: Chambers2}.

\noindent Points that lie in cell $\Gamma_1$ cannot lie in $\Gamma_3$, since the two combine to give the tree amplitude. On the other hand, it can simultaneously lie in $\Gamma_2$ or $\Gamma_4$, although not both due to the same reason. This suggests that we can use the intersection of BCFW cells to triangulate $\mathbb{T}_8$ into four \textit{chambers}:
\begin{equation}\label{8ptoverlap}
\mathbf{1}\cap\mathbf{2},\;\;\mathbf{1}\cap\mathbf{4},\;\;\mathbf{3}\cap\mathbf{2},\;\;\mathbf{3}\cap\mathbf{4}\,.
\end{equation}
These four chambers are related via cyclic rotation by one site. Importantly, this is not merely a choice of being more fine-grained in partitioning the tree-region, or being democratic to all BCFW triangulations. As we will see in the next section, such partitioning is demanded by the loop geometry. For 10-points, as we demonstrate in sec.~\ref{sec:tree_triangulate_loop}, we will have 50 chambers. To summarize, we determine the chamber associated with the external kinematic point by solving eq.~\eqref{eq:MomTwisGras} for all possible $4k$-cells $\Gamma_i$. The set of cells that are positive, i.e. the $\Gamma_i$s such that the solution for $\mathcal{D}_{\Gamma_i}$ have positive semi-definite minors, then defines the chamber. Thus the tree-region can be schematically represented as
\begin{equation}
\mathbb{T}_n=\bigcup_i B_i
\end{equation}
where $B_i$ represents the chambers. Note that here, we are considering BCFW cells. At $n=6,8,10$, it appears that all $4k$-dimensional cells are BCFW cells. It will be interesting to see if this is true for generic $k$. We will come back to this in the outlook.

Before moving on to loops, we briefly discuss other solutions to eq.~\eqref{eq:MomTwisGras}. Recall that when a point in $\mathbb{T}_n$ lies in $\Gamma_i$, only one of the $2^k$ solutions will lead to a positive $\mathcal{D}_{\Gamma_i}$. The other solutions, as it turns out, can be obtained by parity transformation ($\mathbf{\Pi}$) acting on subsets of the twistor variables, $Z_i\rightarrow -Z_i$ on a subset of twistors. The corresponding operation for $6,8$ points are:
\begin{eqnarray}
n=6\;\; && \mathbf{\Pi}_6: \{2,4,6\}\nonumber\\
n=8\;\; &&\mathbf{\Pi}_{8,1}:\{2,4,6\},\; \mathbf{\Pi}_{8,2}:\{2,6,8\},\; \mathbf{\Pi}_{8,3}\equiv\mathbf{\Pi}_{8,2}\mathbf{\Pi}_{8,1}:\{4,8\}\,,
\end{eqnarray}
where the curly bracket denotes the twistors on which the parity operator acts, and the last entry in the second line represents the requisite parity flip can be viewed as consecutive action of $\mathbf{\Pi}_{8,2}\mathbf{\Pi}_{8,1}$. With these parity operations, the remaining solutions can be converted into the positive solution. The availability of such an operation will be crucial for us to obtain the full amplitude.

\section{ABJM integrands from loop geometries}\label{sec:loop_geo}
We have established a natural region $\mathbb{T}_n$ in momentum twistor space which characterizes the kinematics of tree-level scattering in ABJM. On the one hand it is the direct image of momentum amplituhedron via T-duality, and on the other it is non-trivial precisely for the kinematic configuration that ABJM amplitudes have support. 
Given this region, we would now like to ``attach" some loop geometries whose canonical forms give the loop integrand. At four points, we only have four twistors and a single four-bracket, $\langle 1234\rangle$. Thus up to an Sp(4) rotation all points in $\mathbb{T}_4$ is projectively equivalent to 
\begin{equation}\label{eq:4ptZ}
\{Z\}=\begin{pmatrix}
	    0 & 0 & 0 & -1 \\
	    0 & 0 & -1 & 0 \\
	    -1 & 0 & 0 & 0 \\
	    0 & -1 & 0 & 0 
\end{pmatrix}.
\end{equation}
The geometry for four-point all loop amplitude was found in~\cite{He:2022cup} where one considers the canonical form for the loop-region
\begin{equation} \begin{gathered}
	\quad 	\langle (AB)_a i i{+}1\rangle<0, \\
		\langle (AB)_a (AB)_b\rangle <0,\ \text{for any two loop $a,b\in 1,2,\ldots,L$},\\
		\langle\!\langle A_a B_a \rangle\!\rangle=0 \ \text{for $a\in 1,\ldots, L.$} 
	\end{gathered}
\end{equation}
where $A, B$ denotes the loop twistors parameterized with respect to the columns of eq.~\eqref{eq:4ptZ}. Once again the four-point ABJM loop amplituhedron is defined in a fashion similar to $\mathcal{N}=4$ super Yang-Mills, just with  a definite \textit{negative} sign for the four-bracket and a symplectic constraint for the loop twistors. As we now show, by considering twistors living in $\mathbb{T}_n$, extending the definition to $n$-points with an appropriate assignment of sign flips, we conjecture to have (positive branch of) the $n$-point ABJM amplituhedron which gives the correct $n$-point loop integrand!  
\subsection{Definition and some all-loop cuts of the loop amplituhedron}
Let us begin with the definition of the amplituhedron for $n$-points ABJM loop amplitude. Starting with $Z_i\in \mathbb{T}_n$, and we define the loop-region $\mathbb{L}_n$ by requiring
\begin{equation} \label{eq:loop_space}
	\boxed{\mathbb{L}_n:\quad\begin{gathered}
	\quad 	\langle (AB)_a i i{+}1\rangle<0, \ \{\langle (AB)_a 1 i\rangle \} \ \text{has $k+2=\frac{n}{2}$ sign flip},\\
		\langle (AB)_a (AB)_b\rangle <0,\ \text{for any two loop $a,b\in 1,2,\ldots,L$},\\
		\langle\!\langle A_a B_a \rangle\!\rangle=0 \ \text{for $a\in 1,\ldots, L.$} 
	\end{gathered}}
\end{equation}
Here once again, we can equivalently choose any other reference point for the sign flip conditions, i.e. for the sequence $\{\langle (AB)_a j i\rangle | i\neq j\}$ with distinct $j$s. First of all, the degree of this space matches the degree of freedom for the loop integrand. Indeed for each loop one has $4\times 2{-}4{-}1=3$ degrees of freedom, where $-4$ and $-1$ corresponds to the GL(2) redundancy and symplectic constraint. Parameterizing the loop twistors by 
\begin{equation}\label{eq:LoopParam}
    \begin{split}
        (Z_A)_a&=Z_{i-1}+x_a Z_{i}-w_a Z_{j}\\
        (Z_B)_a&=y_a Z_{i}+Z_{j-1}+z_a Z_{j}\,.
    \end{split}
\end{equation}
The symplectic condition $\langle\!\langle AB\rangle\!\rangle=0$ can be used to solve $x_a$. Thus for each ``point" in $\mathbb{T}_n$, eq.~\eqref{eq:loop_space} defines a $3L$-dimensional geometry in $(w_a, y_a, z_a)$ whose canonical form we will identify as the $L$-loop integrand.

\noindent \textbf{Soft-Cuts}:

Before studying loop forms in detail, one can already see that the region defined in eq.~\eqref{eq:loop_space} geometrically manifests many properties of the loop integrand of ABJM theory. For example it is known that due to the Chern-Simons interaction, the triple cut of the $L$-loop integrand has milder behaviour and simply reduces to the $(L{-}1)$-loop integrand, 
\begin{equation}\label{SoftCond}
	\left.A_{n}^{\ell-\text{loop}}\right|_{\substack{\textrm{cut}\\i{-}1,\,i,\,i{+}1}}=(-1)^{i}A_{n}^{(\ell-1)-\text{loop}}
\end{equation}
where we have used the dual regions to define the cut
$$
\vcenter{\hbox{\scalebox{0.75}{


}}}$$
Choosing the parameterization $(i{-}2,i{-}1,i,i{+}1)$ for the loop twistors, the constraints 
\begin{equation}
\langle (AB)_a\,i{-}2\, i{-}1\rangle=\langle (AB)_a\, i{-}1\, i\rangle=\langle (AB)_a\, i\, i{+}1\rangle=0
\end{equation}
lead to $(AB)_a$ being identical to the line $(i{-}1, i)$. On this solution, the mutual negativity between the cut and any remaining loop variable in eq.~\eqref{eq:loop_space} becomes:
%\begin{equation}
%    \begin{gathered}
 %       \langle (AB)_a j j{+}1\rangle=\langle i{-}1\, i\, j\, j{+}1\rangle<0,\\
 %       \{\langle (AB)_a i{+}1 j\rangle \}= \{\langle i{-}1\, i\, i{+}1\, j\rangle \} \ \text{has $k+2$ sign flip}.
%    \end{gathered}
%\end{equation}
%The first line is satisfied simply due to us choosing a point in $\mathbb{T}_n$, while the second line is satisfied due to the positive projection, which is discussed in section 5.1 of~\cite{Arkani-Hamed:2017vfh}. 

\begin{equation}
    \langle(AB)_a (AB)_b\rangle=\langle   i{-}1 i  (AB)_b\rangle<0\,,
\end{equation}
which is nothing but the inequalities that define the $(L{-}1)$-loop amplituhedron. 

\noindent \textbf{Vanishing Cuts}:

Slightly more non-trivial, are cuts which isolate odd-point kinematics, which should vanish due to the absence of odd-point amplitudes. Note that since we've already shown that $\mathbb{T}_n$ is empty for $n=odd$, all we need to show is that on any cut which isolates odd-kinematics, eq.~\eqref{eq:loop_space} completely factorizes into two parts which consist of odd regions. Consider for example a two-particle cut $\langle AB i{-}1 i\rangle=\langle AB j{-}1 j\rangle=0$
$$
\vcenter{\hbox{\scalebox{0.75}{


}}}$$
As shown in~\cite{YelleshpurSrikant:2019meu}, on such cut the inequalities for the amplituhedron of $\mathcal{N}=4$ sYM can be recast  into a union of ``left"  and ``right"  amplituhedron inequalities. Recall the difference between the definition of ABJM amplituhedron in eq.~\eqref{eq:loop_space} and the $\mathcal{N}=4$ counterpart is the overall sign of the four-brackets and the symplectic constraint. The former does not affect the analysis of~\cite{YelleshpurSrikant:2019meu}, since it only changes the overall sign, while the latter immediately sets odd-factorizations to zero.   

For multi-loop cuts, we need to take into account the mutual negativity constraint for the loop twistors. Using the familiar parametrization of $n=4$, we expand each loop variable using $Z_1, Z_2, Z_3, Z_4$:
\begin{equation}
(A B)_a=(Z_1+ x_a Z_2- w_a Z_4, y_a Z_2+ Z_3+z_a Z_4) ,
\end{equation}
The condition $\langle\!\langle AB \rangle\!\rangle=0$ becomes 
\begin{equation}
x_a z_a + y_a w_a{+}z_a \frac{\langle\!\langle 14\rangle\!\rangle}{\langle\!\langle 24\rangle\!\rangle}=-\frac{\langle\!\langle 13\rangle\!\rangle}{\langle\!\langle 24\rangle\!\rangle}\,.
\end{equation}
Since $\langle 1234\rangle=\langle\!\langle13 \rangle\!\rangle \langle\!\langle 24 \rangle\!\rangle<0$, we can rescale the RHS to 1. Note that here for $n>4$ we no longer have $\langle\!\langle 14 \rangle\!\rangle=0$, however the mutual negativity condition 
\begin{equation}\label{eq: MutualNeg}
\frac{\langle (AB)_{a}(AB)_{b}\rangle}{\langle 2341\rangle}=(w_a-w_b)(y_a-y_b)+(z_a-z_b)(x_a-x_b)<0
\end{equation}
is unaffected by this extra term if we solve for $x$. Thus the implications of the mutual negativity explored in~\cite{He:2022cup} for $n=4$ carries over to general multiplicity. In particular, solving for $x_a$ and using the fact that $z_a>0$ as implied from $\langle AB23\rangle<0$, we instead have 
\begin{equation}\label{eq: MutualNeg1}
d_{a,b}\equiv(w_az_b{-}w_bz_a)(y_az_b{-}y_bz_a)-(z_{a,b})^2<0.
\end{equation}
Now consider the following cut, which incises a three-point kinematics:
$$
\vcenter{\hbox{\scalebox{0.75}{


}}}$$
With $i=2$, we have $\langle (AB)_a12\rangle=\langle (AB)_b12\rangle=0$ which implies $w_a=w_b=0$. For such case, $d_{a,b}=-(z_{a,b})^2<0$ which is always satisfied, and hence $d_{a,b}=0$ is not a boundary. Thus this cut does not appear as a boundary. 

Finally, consider a three-particle cut with one of the mutual propagators put on-shell:
\begin{equation}
\langle (AB)_a j{-}1 j\rangle=\langle (AB)_b i{-}1 i\rangle=\langle (AB)_a (AB)_b\rangle=0.
\end{equation}
Schematically we have:
$$
\vcenter{\hbox{\scalebox{0.75}{
%

}}}$$
We would like to show that this boundary does not exist if one has an odd-multiplicity on both sides of the cut. Using the parameterization in eq.~\eqref{eq:LoopParam} it is straight forward to see that on the cut $\langle (AB)_a j{-}1 j\rangle=\langle (AB)_b i{-}1 i\rangle=0$, the mutual negativity condition becomes 
\begin{equation}\label{eq:MutualNegOddCut}
    \begin{split}
        \langle (AB)_a (AB)_b\rangle&=\langle i{-}1\,i\,j{-}1\,j\rangle\left(w_ay_b-z_{a,b}^2\frac{\langle i{-}1\,i\,j{-}1\,j\rangle}{\langle\!\langle i{+}1B_{a}\rangle\!\rangle\langle\!\langle i{+}1B_{b}\rangle\!\rangle}\right)\\
&=\langle i{-}1\,i\,j{-}1\,j\rangle\left(w_ay_b-\left(1-\frac{z_b}{z_a}\right)^2\frac{\langle (AB)_aj{-}2j{-}1\rangle}{\langle (AB)_bii{+}1\rangle}\frac{\langle\!\langle i{-}1 i{+}1\rangle\!\rangle}{\langle\!\langle j{-}2\,j\rangle\!\rangle}\right)<0\,.
    \end{split}
\end{equation}
The first term in the parenthesis is strictly positive since $w,y>0$. The sign of the second term will be determined by the ratio of $\frac{\langle\!\langle i{-}1i{+}1\rangle\!\rangle}{\langle\!\langle j{-}2\,j\rangle\!\rangle}$. 

To understand the sign of this ratio let's digress and consider further implications of the definite-sign condition for $\langle r\, r{+}1\, s\, s{+}1\rangle$:
\begin{equation}\label{eq:consecutive twistors}
    \langle r\, r{+}1\, r{+}2\, r{+}3 \rangle,\  (-1)^{k-1}\langle n{-}2\,n{-}1\,n\,1\rangle,\  (-1)^{k-1}\langle  n123\rangle<0. 
\end{equation}
Note that when the entries are consecutive, the symplectic condition and   \eqref{eq:two-brackets formula} will lead to complete factorization of the four-brackets: $\langle r\, r{+}1\, r{+}2\, r{+}3 \rangle=\langle \!\langle  r\, r{+}2 \rangle \! \rangle \langle \!\langle  r{+}1\, r{+}3 \rangle \! \rangle$. The definite-sign condition will then lead to the following two sets of solutions for the two-brackets: 
\begin{equation}
    \begin{split}
        &\langle\!\langle 13 \rangle\!\rangle, \ \langle\!\langle 35 \rangle\!\rangle, \ \cdots,\  (-)^{k-1}\langle\!\langle n{-}1 1\rangle\!\rangle\gtrless 0,\\
        &\langle\!\langle 24 \rangle\!\rangle, \ \langle\!\langle 46 \rangle\!\rangle, \ \cdots,\  (-)^{k-1}\langle\!\langle n 2\rangle\!\rangle\lessgtr0.
    \end{split}
\end{equation}
Thus we see that $\langle\!\langle rr{+}2\rangle\!\rangle$ will have the same sign for all odd $r$ and the opposites sign for all even $r$. If one has an odd-multiplicity on both sides of the cut then $\frac{\langle\!\langle i{-}1i{+}1\rangle\!\rangle}{\langle\!\langle j{-}2\,j\rangle\!\rangle}$ will be strictly negative and for even-multiplicity it will be strictly positive. Thus we see that for an odd-multiplicity, all terms in the parenthesis of eq.~\eqref{eq:MutualNegOddCut} are strictly positive and hence when combined with $\langle r\,r{+}1\,s\,s{+}1\rangle<0$ trivializes the sign, and hence it is no longer a boundary. It is amusing that properties of $\mathbb{T}_n$ are essential to ensure the absence of all odd-multiplicity cuts. This phenomenon of $\mathbb{T}_n$ telling loops how to ``behave", will be the central theme of this paper and we will see how it manifests when determining the loop-form.

\noindent\textbf{Two-particle cuts}:

Finally, let us consider two-particle cuts, where the amplitude factorizes into ``left" and ``right" counterparts. We will show that this simply follows from the geometrical factorization of the amplituhedron. Before going into the detail, it is crucial to note that the complete amplitude is the sum of various sectors obtained from parity operators $\Pi$s (to be discussed in sec.~\ref{sec:parity}). As the parity operators are defined from the interchanging of maximal cut solutions, their presence under unitarity cuts factorizes. Thus for simplicity, here we will simply focus on the positive sector.

In the context of $\mathcal{N}=4$ sYM, the unitarity of the amplitude can be translated into geometrical factorization of the amplituhedron for all $n$ and degrees $k$. More precisely, under the cut $\langle AB i{-}1 i\rangle=\langle AB j{-}1 j\rangle=0$, geometry factorization refers to the inequalities of the amplituhedron being equivalent to summing over unions of inequalities associated with the left and right amplituhedron, with the external legs suitably arranged. In the sum, one includes left/right amplituhedron will run over all possible values of $L_{1(2)}$, $n_{1(2)}$ and $k_{1(2)}$ that satisfy
\begin{equation}
    \begin{split}
        L=L_1+L_2+1,\quad n_{1}+n_2=n+2, \quad k=k_1+k_2, 
    \end{split}
\end{equation}
and the external legs of the left and right sets could be chosen as 
\begin{equation}
    \begin{gathered}
            \mathcal{L}=\{Z_1,\cdots, Z_i, A,B, Z_{j-1},\cdots,Z_n\},\ \mathcal{R}=\{A,Z_{i+1},\cdots,Z_j, B\}.
    \end{gathered}
\end{equation}
Note that since any rescaling of $Z_i$s does not change the lines defined by pairs of momentum twistors when considering the factorized geometry, one should allow for rescaling $Z_i\in \mathcal{L}, \mathcal{R}$ by factors $Z_i\rightarrow \sigma_{\mathcal{L}(\mathcal{R})}(i)Z_i$. The central point of geometric factorization is that \textit{there exists} a set of rescaling such that on the cut, the inequalities of the amplituhedron can be factorized into a union of inequalities which can be identified as the rescaled twistors living inside left and right amplituhedron. This was beautifully established in~\cite{YelleshpurSrikant:2019meu}.

As the ABJM amplituhedron can be viewed as a reduction of the $\mathcal{N}=4$ sYM amplituhedron, with an additional change of sign, geometrical factorization simply follows. Indeed such a discussion for $n=4$ was already presented in~\cite{He:2023exb}. For $n>4$, note that due to the negative sign, the available geometry automatically satisfies $k_i=\frac{n_i}{2}-2$ with even $n_i$ for $i=1,2$. For example, for $n=6$, on the cut $\langle AB 23\rangle=\langle AB45\rangle=0$ for 6-4 channel, $(AB)$ can be parameterized by
\begin{equation}\label{eq: cutsol}
    A=Z_2+x Z_3, \quad B=Z_4+z Z_5. 
\end{equation}
Here,  $x=-(\langle\!\langle25\rangle\!\rangle z+\langle\!\langle24\rangle\!\rangle)/\langle\!\langle35\rangle\!\rangle z$ from solving $\langle\!\langle AB \rangle\!\rangle=0$. The inequalities associated with $\{Z_i,A,B\}$ in  $\mathbb{T}_6$ and $\mathbb{L}_6$ can be shown to be equivalent to the intersection of inequalities involving external left and right sets  $\mathcal{L}=\{1,2,A,B,5,6\}$, $\mathcal{R}=\{A,3,4,B\}$: 
\begin{equation}
    \begin{gathered}
        \text{Left:}\ \langle i\, i{+}1\, j\, j{+}1 \rangle_{\mathcal{L}}<0,\ \{\langle 123i \rangle_{\mathcal{L}}\}\ \text{has $1$ sign flip}, \ \langle\!\langle i i{+}1\rangle\!\rangle_{\mathcal{L}}=0.\\
        \text{Right:}\ \langle 1234\rangle_{\mathcal{R}}<0,\ \langle\!\langle i i{+}1\rangle\!\rangle_{\mathcal{R}}=0.
    \end{gathered}
\end{equation}
Here the labelling $i$ in $\langle\cdots \rangle_{\mathcal{L}(\mathcal{R})}$ denotes the $i$-th twistor in the set $\mathcal{L}$($\mathcal{R}$). Furthermore, the inequalities involving the remaining uncut loops $(AB)_a$ can be replaced by the union of those defining the loop space with external sets $\mathcal{L}$ and $\mathcal{R}$, according to the distribution of $L-1$ loop on the left and right sets:
\begin{equation}\label{eq:loop_2_cut}
\begin{gathered}
    \left(\begin{gathered}
		\langle (AB)_a i i{+}1\rangle_\mathcal{L}<0, \ \{\langle (AB)_a 1 i\rangle \}_\mathcal{L} \ \text{has $k_1+2$ sign flip},\\
		\langle (AB)_a (AB)_b\rangle_\mathcal{L} <0,\ \text{for any two loop $a,b\in \{L_1\}$}. 
	\end{gathered}\right)\land \Bigg( \mathcal{L},\,k_1,\,L_1 \rightarrow \mathcal{R},\,k_2,\,L_2\Bigg),
% \text{with}\, \langle\!\langle A_a B_a \rangle\!\rangle=0 \ \text{for}\  a\in 1,\ldots, L.
\end{gathered}
\end{equation}
satisfying $\langle\!\langle A_a B_a \rangle\!\rangle=0$ for $a\in 1,\ldots, L-1$ and with $k_1=1$, $k_2=0$. Here, $\{L_1\}\subset \{1,2,\cdots, L-1\}$ has $L_1$ elements and $\{L_2\}$ is the complement with $L_2=L{-}1{-}L_1$ elements.

%For $n=8$, on the double cut $\langle AB34\rangle=\langle AB56\rangle=0$, the amplituhedron geometry can be replaced by the inequalities involving external left and right sets  $\mathcal{L}=\{1,2,3,A,B,6,7,8\}$, $\mathcal{R}=\{A,4,5,B\}$,  living in the region of: 
%\begin{equation}
%    \begin{gathered}
%        \text{Left:}\ \langle i\, i{+}1\, j\, j{+}1 \rangle_{\mathcal{L}}<0,\ \{\langle 123i \rangle_{\mathcal{L}}\}\ \text{has $2$ sign flip}, \ \langle\!\langle i i{+}1\rangle\!\rangle_{\mathcal{L}}=0.\\
%        \text{Right:}\ \langle 1234\rangle_{\mathcal{R}}<0,\ \langle\!\langle i i{+}1\rangle\!\rangle_{\mathcal{R}}=0\,.
%    \end{gathered}
%\end{equation}

For $n=8$, the double cut for the 6-6 channel is $\langle A B 23\rangle=\langle A B 67\rangle=0$, and the double-cut geometry can be replaced by the inequalities involving external left and right sets  $\mathcal{L}=\{1,2,A,-B,-7,-8\}$, $\mathcal{R}=\{A,3,4,5,6,B\}$,  living in the region: 
\begin{equation}
    \begin{gathered}
        \text{Left:}\ \langle i\, i{+}1\, j\, j{+}1 \rangle_{\mathcal{L}}<0,\ \{\langle 123i \rangle_{\mathcal{L}}\}\ \text{has $1$ sign flip}, \ \langle\!\langle i i{+}1\rangle\!\rangle_{\mathcal{L}}=0.\\
        \text{Right:}\ \langle i\, i{+}1\, j\, j{+}1 \rangle_{\mathcal{R}}<0,\ \{\langle 123i \rangle_{\mathcal{R}}\}\ \text{has $1$ sign flip}, \ \langle\!\langle i i{+}1\rangle\!\rangle_{\mathcal{R}}=0.
    \end{gathered}
\end{equation}
The remaining inequalities for the uncut loop variables can be replaced as~\eqref{eq:loop_2_cut} by adjusting $k_1=1,\, k_2=1$.

Not only does the region factorize, but the resulting form can also be matched with the explicit computation in momentum space, which we present in detail in appendix~\ref{sec:TwoParticle}.
\subsection{Triangulating loop amplituhedron and one-loop integrands}\label{sec:tree_triangulate_loop}
We now proceed to construct the loop integrand by deriving the canonical form for the positive geometry defined in eq.~\eqref{eq:loop_space}, starting at one-loop. We will use a point in $\mathbb{T}_n$ as input, {\it i.e.} a set of momentum twistors $\{Z\}$ satisfying eq.~\eqref{eq:tree_geo}. Then using this set of $\{Z\}$ for eq.~\eqref{eq:loop_space} we can derive a canonical form. 

Note that given two points in $\mathbb{T}_n$, the canonical form derived from eq.~\eqref{eq:loop_space} need not be the same. Remarkably, \textit{points from the same chamber in $\mathbb{T}_n$ yields the same canonical form!} This is extremely non-trivial since while two points in the same chamber give identical forms, their sign-flipping triangulation yields different blocks. Writing 
\begin{equation}
\mathbb{T}_n=\cup_i \;B_i
\end{equation}
where $B_i$ represents the chambers. For example at eight points, $i=1,2,3,4$ representing the four distinct overlaps in eq.~\eqref{8ptoverlap}. Then the full loop amplitude should be written as 
\begin{equation}
\sum_i \; \mathcal{B}_i \times \mathbf{\Omega}_i 
\end{equation}
where $\mathbf{\Omega}_i$ is the canonical form from eq.~\eqref{eq:loop_space} triangulated by the point in $B_i$, and $\mathcal{B}_i$ is the associated tree-form for $B_i$. In other words, we have a product geometry, schematically taking the form:
$$\includegraphics[scale=0.6]{Product.png}$$
Note that we have not properly defined what the tree-form $\mathcal{B}_i$ is. As the chambers are defined as the overlap of various $(n{-}3)$-dimensional cells, one expects that linear combinations of $\mathcal{B}_i$ simply match to various leading singularities. For example at eight-point, where we have four chambers $\mathbf{1}\cap\mathbf{2},\;\;\mathbf{1}\cap\mathbf{4},\;\;\mathbf{3}\cap\mathbf{2},\;\;\mathbf{3}\cap\mathbf{4}\,$, one expects the sum of tree-forms $\mathcal{B}_{\mathbf{1}\cap\mathbf{2}},\;\;\mathcal{B}_{\mathbf{1}\cap\mathbf{4}}$ gives the leading singularity associated with the cell $\Gamma_1$. This is almost correct except for one caveat: there are four leading singularities associated with cell $\Gamma_1$, which one should be assigned to $\mathcal{B}_{\mathbf{1}\cap\mathbf{2}}\oplus\mathcal{B}_{\mathbf{1}\cap\mathbf{4}}\,$? As mentioned previously, for a given point in $B_i$ there are $2^k$ solutions to eq.~\eqref{eq: Schubert} of which only one is positive for cells that overlap in the chosen chamber. The remaining becomes positive after parity transformation $\Pi$. Thus we conclude that the combination of tree-forms $\mathcal{B}_i$ gives the positive leading singularity, while further parity action yields the remaining $2^k-1$ counterparts. The full integrand of ABJM amplitude is then given as
\begin{equation}\label{eq: loopdef}
\boxed{\quad  A_n^{L{-}loop}=\sum_{i,\alpha} \; \mathbf{\Pi}_{n,\alpha}\left[\mathcal{B}_i \times \mathbf{\Omega}_i \right]\quad }
\end{equation}
where $\alpha=0,1,\cdots 2^{k}{-}1$ labels the $2^{k}{-}1$ parity operators with $\mathbf{\Pi}_{n,0}$ being the identity. Notice at loop level the parity operation acts on the tree-form \textit{and} the loop-form: its action on loop forms is simply equivalent to the canonical form of eq.~\eqref{eq:loop_space} where $\mathbf{\Pi}$ acts on the external momentum twistors. We will often refer to the ``bare" geometry, i.e. acted by $\mathbf{\Pi}_{n,0}$, as \textbf{positive sector}. Note that the amplitude in eq.~\eqref{eq: loopdef} is the momentum twistor space integrand, which is equivalent to the momentum space integrand with an overall   $(\prod_{i=1}^n \langle i i{+}1\rangle)^{-\frac{k{+}3}{2}}$ stripped off. This prefactor is identified through the ``T-duality" transformation that relates the momentum space Grassmannian to the momentum twistor space Grassmannian in appendix~\ref{sec:Tdual}.

To better illustrate the proposal, in this section we will study ABJM loop amplituhedron and focus on the positive part first, {\it i.e.} $\Pi_{n,0}$ part of eq.~\eqref{eq: loopdef}. 
We start with one-loop case, and at this point we collect all possible sign patterns for $\langle AB 1 i\rangle$ that are compatible with the sign flipping criteria in eq.~\eqref{eq:loop_space}. This gives a collection of regions defined by inequalities of the remaining parameters $(y,z,w)$. The sum of canonical forms (3-form) for these regions gives the loop integrand in the parameterization of eq.~\eqref{eq:LoopParam}. To reconstruct the integrand, we introduce the following  local integral basis:  
\begin{equation}\label{eq: 1loopLocal}
    \begin{split}
        &I_{box}(i,j,k):=\int_a \frac{\epsilon(a,i,j,k,X)}{(a\cdot i)(a\cdot j)(a\cdot k)(a\cdot X)}\\
        &I_{tri}(i,j,k):=\int_a \frac{\sqrt{(i\cdot j)(j\cdot k)(k\cdot i)}}{(a\cdot i)(a\cdot j)(a\cdot k)}\, .
    \end{split}
\end{equation}
The integrals are presented in their dual form, using the notation of~\cite{Caron-Huot:2012sos, He:2022lfz}. Although we write the integrals in dual variables, as they take simpler form, their translation to twistor variables are straight forward. For example, the epsilon numerators can be rewritten as:
%\begin{equation}
%    \begin{gathered}
%        \epsilon(i,j,k,l,m)=\Big(\big(\langle\!\langle i{-}1 j\rangle\!\rangle\langle\!\langle j{-}1 k\rangle\!\rangle \langle\!\langle k{-}1 l\rangle\!\rangle \langle\!\langle l{-}1 m\rangle\!\rangle \langle\!\langle m{-}1 i\rangle\!\rangle\\
%        - (i{-}1 \leftrightarrow i)\big)-(k{-}1 \leftrightarrow k) \Big)\cdots-(m{-}1\leftrightarrow m)
%    \end{gathered}
%\end{equation}
\begin{equation}
    \begin{gathered}
        \epsilon(i,j,k,l,m)=\frac{1}{\prod_{s=i,j,k,l,m} \langle s{-}1 s\rangle}\Big(\big(\big(\langle\!\langle i{-}1 j\rangle\!\rangle\langle\!\langle j{-}1 k\rangle\!\rangle \langle\!\langle k{-}1 l\rangle\!\rangle \langle\!\langle l{-}1 m\rangle\!\rangle \langle\!\langle m{-}1 i\rangle\!\rangle\\
        - (i{-}1 \leftrightarrow i)\big)-(k{-}1 \leftrightarrow k) \big)\cdots-(m{-}1\leftrightarrow m)\Big).
    \end{gathered}
\end{equation}
The coefficients of these integrals can now be fixed by using the explicit parameterization for the loop variables in eq.~\eqref{eq:LoopParam} and matching to the 3-form derived from the geometry. As it turns out, these coefficients are completely fixed already by matching to the co-dimension one boundary of the geometry, {\it i.e.} single cuts. Such local representation allows us to directly compare the result determined by the geometry with the known results in the literature~\cite{Brandhuber:2012un, Caron-Huot:2012sos}.

The fact that the result obtained from the geometry can be  directly matched with a local integrand basis is not surprising. After all, we are ``triangulating" the loop-form by breaking $\mathbb{T}_n$ into chambers and don't introduce new boundaries in the loop space $(AB)$. Thus our procedure gives rise to a representation satisfying prescriptive unitarity~\cite{Bourjaily:2017wjl}, $\textit{i.e.}$ local integrands that contain only local singularities and match to only one maximal cut solution while vanishing on others. Note that from this point of view, it is natural that $\mathbb{T}_n$ is dissected into chambers that are identified from overlapping cells. This is the result of the tension between 1. prescriptive unitarity, where all leading singularities (cells) are realized once, and 2. the fact that $\mathbb{T}_n$ can be tiled by a distinct subset of cells. Thus to realize prescriptive unitarity, $\mathbb{T}_n$ must be dissected into units that represent the overlap of cells, $\textit{i.e.}$ chambers.

\paragraph{Six-point case} 
Here we present the construction in detail for the six-point one-loop amplitude. The eight- and ten-point amplitude can be derived following similar steps, and we will only spell out the differences originating from the more intricate triangulation of $\mathbb{T}_n$. 

We will use on-shell diagrams to represent BCFW cells. At six-point, the two available on-shell diagrams are equivalent, reflecting the fact that the tree amplitude corresponds to the top cell:
%$$\includegraphics[scale=0.8]{Equiv}$$
$$
\vcenter{\hbox{\scalebox{1.0}{


}}}$$
Let us begin with an arbitrary point in $\mathbb{T}_6$. Since the tree amplitude is given by the top-cell, there is only one chamber. This implies that all points in $\mathbb{T}_6$ should yield the same loop-form through eq.~\eqref{eq:loop_space}. As we will see, while distinct points lead to the same form, \textit{their sign-flipping triangulation yields distinct representations!} We will now demonstrate these properties using two different points:

\noindent
\textbf{Point 1.}
\begin{equation}\label{eq:kim1}
\{Z\}=\left(
\begin{array}{cccccc}
 -\frac{193133}{3848} & -\frac{501011}{11544} & \frac{14121}{481} & \frac{15545}{481} & -\frac{665245}{11544} & -\frac{236387}{3848} \\
 -\frac{30635}{7696} & -\frac{27015}{7696} & \frac{1975}{962} & \frac{2263}{962} & -\frac{28521}{7696} & -\frac{30725}{7696} \\
 0 & \frac{1907188375}{11105328} & \frac{183507575}{2082249} & -\frac{85910836}{2082249} & -\frac{1640762167}{11105328} & 0 \\
 0 & \frac{302520625}{22210656} & \frac{2794125}{462722} & -\frac{763400}{231361} & -\frac{213262225}{22210656} & 0 \\
\end{array}
\right)\,.
\end{equation}
One can check that these twistors indeed satisfy eq.~\eqref{eq:tree_geo} and are in $\mathbb{T}_6$. For simplicity, let's focus on the two-form living on the co-dimension-one boundary $\langle AB 34\rangle=0$. Choosing $i=2$ and $j=4$ in eq.~\eqref{eq:LoopParam}, this boundary translate to $y=0$ and parameterized by $w,z$. We can triangulate this space by considering all sign patterns consistent with three sign flips. There are two consistent sign-flipping patterns: 
\begin{equation}
\{\langle (AB) 1 i\rangle\}=\{-,+,-,-,+\},\quad \{-,+,-,+,+\}\,.
\end{equation}
Other sign flipping such as $\{-,-,+,-,+\}$ and $\{-,+,+,-,+\}$ are not consistent with eq.~\eqref{eq:kim1}. For the first pattern, we have the following triangulation:  
\begin{equation}
    \begin{split}
    \{-,+,-,-,+&\}:     \quad\bigg( -\frac{\langle 1235\rangle}{\langle1245\rangle}<z<-\frac{\langle 1235\rangle}{\langle1245\rangle}+\frac{\langle \!\langle 13\rangle\!\rangle \langle 1256\rangle \sqrt{-\langle 1234\rangle\langle3456\rangle\langle 12 56\rangle}}{\langle 1236\rangle \langle 1245\rangle \langle 1456\rangle}\\
        &\quad \quad  \land 0<w< \frac{2\langle 1235\rangle}{\langle2345\rangle}-\frac{\langle1235\rangle\langle1456\rangle}{\langle1345\rangle\langle2456\rangle}+\frac{\langle1245\rangle}{\langle2345\rangle}z+\frac{\langle1235\rangle\langle1236\rangle}{\langle1246\rangle\langle2345\rangle }\frac{1}{z} \bigg)\\
        \lor &\bigg( {-}\frac{\langle 1235\rangle}{\langle1245\rangle}{+}\frac{\langle \!\langle 13\rangle\!\rangle \langle 1256\rangle \sqrt{-\langle 1234\rangle\langle3456\rangle\langle 12 56\rangle}}{\langle 1236\rangle \langle 1245\rangle \langle 1456\rangle}<z  
          \land \frac{2\langle 1235\rangle}{\langle 2345\rangle}{+}\frac{\langle1245\rangle}{\langle2345\rangle}z\\
        \quad {+}&\frac{\langle 1235\rangle\langle 2356\rangle}{\langle 2345\rangle\langle 2456\rangle}\frac{1}{z}<w<\frac{2\langle 1235\rangle}{\langle2345\rangle}{-}\frac{\langle1235\rangle\langle1456\rangle}{\langle1345\rangle\langle2456\rangle}{+}\frac{\langle1245\rangle}{\langle2345\rangle}z{+}\frac{\langle1235\rangle\langle1236\rangle}{\langle1246\rangle\langle2345\rangle }\frac{1}{z}   \bigg).
    \end{split}
\end{equation}
\noindent
It is straightforward to compute the associated 2-form:
\begin{equation}\label{eq:kim_form1}
    \begin{split}
       -&\frac{dz}{z{+}\frac{\langle 1235\rangle}{\langle1245\rangle}{-}\frac{\langle \!\langle 13\rangle\!\rangle \langle 1256\rangle \sqrt{-\langle 1234\rangle\langle3456\rangle\langle 12 56\rangle}}{\langle 1236\rangle \langle 1245\rangle \langle 1456\rangle}} \wedge \Bigg(\frac{dw}{w}-\frac{dw}{w{-}\frac{2\langle 1235\rangle}{\langle 2345\rangle}{-}\frac{\langle1245\rangle}{\langle2345\rangle}z{-}\frac{\langle 1235\rangle\langle 2356\rangle}{\langle 2345\rangle\langle 2456\rangle}\frac{1}{z}}\Bigg)\\
       &\quad +\frac{dz}{z+\frac{\langle 1235\rangle}{\langle1245\rangle}} \wedge \Bigg(\frac{dw}{w}-\frac{dw}{w-\frac{2\langle 1235\rangle}{\langle2345\rangle}+\frac{\langle1235\rangle\langle1456\rangle}{\langle1345\rangle\langle2456\rangle}-\frac{\langle1245\rangle}{\langle2345\rangle}z-\frac{\langle1235\rangle\langle1236\rangle}{\langle1246\rangle\langle2345\rangle }\frac{1}{z}}\Bigg)\,. \\
    \end{split}
\end{equation}

\noindent
For the other sign flip pattern, we have
\begin{equation}
    \begin{split}
    \{-,+,-,+,+\}:     & \qquad   \Big( 0<z<-\frac{\langle 1235\rangle}{\langle 1245\rangle}\land w>0\Big)
        \lor \Big( z>-\frac{\langle1235\rangle}{\langle1245\rangle}\\
         \land &\frac{2\langle 1235\rangle}{\langle2345\rangle}{-}\frac{\langle1235\rangle\langle1456\rangle}{\langle1345\rangle\langle2456\rangle}{+}\frac{\langle1245\rangle}{\langle2345\rangle}z{+}\frac{\langle1235\rangle\langle1236\rangle}{\langle1246\rangle\langle2345\rangle }\frac{1}{z} <w\Big)\,,\\
    \end{split}
\end{equation}
\noindent
with the associated 2-form 
\begin{equation}\label{eq:kim_form2}
    \begin{split}
       \frac{dz \wedge dw }{z w}-\frac{dz}{z+\frac{\langle 1235\rangle}{\langle1245\rangle}} \wedge \Bigg(\frac{dw}{w}-\frac{dw}{w-\frac{2\langle 1235\rangle}{\langle2345\rangle}+\frac{\langle1235\rangle\langle1456\rangle}{\langle1345\rangle\langle2456\rangle}-\frac{\langle1245\rangle}{\langle2345\rangle}z-\frac{\langle1235\rangle\langle1236\rangle}{\langle1246\rangle\langle2345\rangle }\frac{1}{z}}\Bigg)\,.
    \end{split}
\end{equation}
By taking the sum of eqs.~\eqref{eq:kim_form1} and \eqref{eq:kim_form2}, the spurious poles in these two subregions are cancelled, resulting in a complete 2-form on the boundary $\langle AB34\rangle=0$. 

We can now find the combination of local integral basis that matches the above form. Since we are considering the geometry for the boundary $\langle AB 34\rangle=0$, not surprisingly only integrals involving dual region $4$ will participate. Substituting eq.~\eqref{eq:LoopParam} into the integrals and taking into account the appropriate Jacobian factors we find that the result:
\begin{equation}\label{eq:Answ1}
-I_{box}(2,3,4)+I_{box}(3,4,5)-I_{box}(4,5,6)-I_{box}(2,4,6)+I_{tri}(2,4,6)\,. 
\end{equation}

\noindent
\textbf{Point 2.}
Let us now consider another point,
\begin{equation}\label{eq:kim2}
\{Z\}=\left(
\begin{array}{cccccc}
 -\frac{17457}{305} & -\frac{92931}{2135} & \frac{73236}{2135} & \frac{53184}{2135} & -\frac{106989}{2135} & -\frac{20343}{305} \\
 -\frac{2703}{610} & -\frac{14949}{4270} & \frac{5022}{2135} & \frac{4008}{2135} & -\frac{13611}{4270} & -\frac{2697}{610} \\
 0 & \frac{55932228}{130235} & \frac{29095632}{130235} & -\frac{1232138304}{4558225} & -\frac{388876788}{651175} & 0 \\
 0 & \frac{4330206}{130235} & \frac{1950264}{130235} & -\frac{90652608}{4558225} & -\frac{25777926}{651175} & 0 \\
\end{array}
\right)\,
\end{equation}
Now the allowed sign flipping patterns are $\{-,-,+,-,+\}, \{-,+,-,-,+\}$. For the first, we have:
\begin{equation}
    \begin{gathered}
 \{-,-,+,-,+\}:\quad      z>-\frac{\langle 1235\rangle\langle 3456\rangle}{\langle1345\rangle\langle2456\rangle} \land w>\frac{2\langle 1235\rangle}{\langle 2345\rangle}+\frac{\langle1245\rangle}{\langle2345\rangle}z+\frac{\langle 1235\rangle\langle 2356\rangle}{\langle 2345\rangle\langle 2456\rangle}\frac{1}{z} 
    \end{gathered}
\end{equation}
where the 2-form is
\begin{equation}\label{eq:kim2_form1}
    \begin{split}
       \frac{dz}{z+\frac{\langle 1235\rangle\langle3456\rangle}{\langle1345\rangle\langle2456\rangle}} \wedge \frac{dw}{w-\frac{2\langle 1235\rangle}{\langle2345\rangle}-\frac{\langle1245\rangle}{\langle2345\rangle}z-\frac{\langle1235\rangle\langle2356\rangle}{\langle2345\rangle\langle2456\rangle }\frac{1}{z}}.
    \end{split}
\end{equation}
For the other pattern:

\begin{equation}
    \begin{split}
        &\bigg( 0<z<{-}\frac{\langle 1235\rangle}{\langle1245\rangle}{+}\frac{\langle \!\langle 13\rangle\!\rangle \langle 1256\rangle \sqrt{{-}\langle 1234\rangle\langle3456\rangle\langle 12 56\rangle}}{\langle 1236\rangle \langle 1245\rangle \langle 1456\rangle} \land w>0 \bigg)\\
         \{-,+,-,-,+\}:\  \lor &\bigg( {-}\frac{\langle 1235\rangle}{\langle1245\rangle}{+}\frac{\langle \!\langle 13\rangle\!\rangle \langle 1256\rangle \sqrt{-\langle 1234\rangle\langle3456\rangle\langle 12 56\rangle}}{\langle 1236\rangle \langle 1245\rangle \langle 1456\rangle}<z< {-}\frac{\langle 1235\rangle \langle 3456\rangle}{\langle 1345\rangle \langle 2456\rangle} \\ 
        & \qquad \qquad \qquad \land w> \frac{2\langle 1235\rangle}{\langle 2345\rangle}+\frac{\langle1245\rangle}{\langle2345\rangle}z+\frac{\langle 1235\rangle\langle 2356\rangle}{\langle 2345\rangle\langle 2456\rangle}\frac{1}{z} \bigg).
    \end{split}
\end{equation}
Its 2-form is
\begin{equation}\label{eq:kim2_form2}
    \begin{split}
       -&\frac{dz}{z{+}\frac{\langle 1235\rangle}{\langle1245\rangle}{-}\frac{\langle \!\langle 13\rangle\!\rangle \langle 1256\rangle \sqrt{-\langle 1234\rangle\langle3456\rangle\langle 12 56\rangle}}{\langle 1236\rangle \langle 1245\rangle \langle 1456\rangle}} \wedge \Bigg(\frac{dw}{w}-\frac{dw}{w{-}\frac{2\langle 1235\rangle}{\langle 2345\rangle}{-}\frac{\langle1245\rangle}{\langle2345\rangle}z{-}\frac{\langle 1235\rangle\langle 2356\rangle}{\langle 2345\rangle\langle 2456\rangle}\frac{1}{z}}\Bigg)\\
       &\quad +\frac{dz\wedge dw}{ z w}-\frac{dz}{z+\frac{\langle 1235\rangle\langle3456\rangle}{\langle1345\rangle\langle2456\rangle}} \wedge \frac{dw}{w-\frac{2\langle 1235\rangle}{\langle2345\rangle}-\frac{\langle1245\rangle}{\langle2345\rangle}z-\frac{\langle1235\rangle\langle2356\rangle}{\langle2345\rangle\langle2456\rangle }\frac{1}{z}}.
    \end{split}
\end{equation}
We see that the triangulation arising from kinematic point eq.~\eqref{eq:kim2} is distinct from \eqref{eq:kim1}. However, the sum of eqs.~\eqref{eq:kim2_form1} and \eqref{eq:kim2_form2} gives the same as the sum of eqs.~\eqref{eq:kim_form1} and \eqref{eq:kim_form2}.

The local form that contributes to the single cut $\langle AB34\rangle=0$ in eq.~\eqref{eq:Answ1}, can be compared to the known complete result:
\begin{equation}\label{eq:one-6pt}
    \begin{split}
        &  \text{LS}_{6,+} \left(I_{box}(1,3,5){+}I_{tri}(1,3,5){-}I_{box}(2,4,6){+}I_{tri}(2,4,6){+}\sum_{i=1}^6 ({-})^iI_{box}(i{-}1,i,i{+}1)\right)\\
    +& \text{LS}_{6,-} \left(I_{box}(1,3,5){-}I_{tri}(1,3,5){-}I_{box}(2,4,6){-}I_{tri}(2,4,6){+}\sum_{i=1}^6 ({-})^iI_{box}(i{-}1,i,i{+}1)\right)\,,\\
    \end{split}
\end{equation}
where $ \text{LS}_{6,\pm}$ correspond to the leading singularity associated with the positive and negative branch of the orthogonal Grassmannian. For convenience, we give their explicit form in momentum (spinor) space:
\begin{eqnarray}\label{eq:LS6}
 \text{LS}_{6,\pm}=\frac{\delta^{6}(Q)\left(\sum_{i,j,k=I} \epsilon_{ijk}\langle ij\rangle\eta_k\pm \{1,3,5\rightarrow 2,4,6\} \right)}{A^\pm_{52}A^\pm_{36}A^\pm_{14}},\quad I=\{1,3,5\} 
\end{eqnarray}
where $A^\pm_{a,b}\equiv \sum_{i\in I}\langle ia\rangle\langle ib\rangle\pm\langle a{+}2a{-}2\rangle\langle b{-}2b{+}2\rangle$. 
It is easy to see that the canonical form in eq.~\eqref{eq:Answ1} for the boundary $\langle AB34\rangle=0$ indeed matches the terms proportional to $ \text{LS}_{6,+}$ that contains such a single cut. To obtain the remaining part proportional to $\text{LS}_{6,-}$ requires the action of parity operators, which we leave to section~\ref{sec:parity} for detail.

\paragraph{Eight-point case.}
We now move on to the eight-point one-loop amplitude. At eight-points the BCFW representation for the tree-level amplitude consists of two terms, each corresponding to an on-shell diagram that can be interpreted as a two-loop maximal cut. There are in fact four in-equivalent  diagrams:
%$$\includegraphics[scale=0.4]{OnShell1}\,,$$
$$
\vcenter{\hbox{\scalebox{0.8}{


}}}$$
where $i=1,\cdots,4$. Here the subscript for the cells $\Gamma_i$, also represents the vanishing $2\times 2$ minor for $\widetilde{OG}(2,8)$. The tree amplitude is given by the sum of the residues from the Grassmannian integral in eq.~\eqref{eq:Integral} on these cells:
\begin{equation}
    \begin{gathered}
        \sum_{\text{sol.}}\Gamma_1+\Gamma_{3}=\sum_{\text{sol.}}\Gamma_2+\Gamma_{4}=A_{8}^{\text{tree}}\,,
    \end{gathered}
\end{equation}
where $\sum_{\text{sol.}}$ sums over the $2^2=4$ solutions of eq.~\eqref{eq: Schubert}. As discussed previously, there are four chambers for $\mathbb{T}_8$:
$$
\vcenter{\hbox{\scalebox{0.8}{


}}}$$
Unlike the six-point case, where any point in $\mathbb{T}_6$ yields the same canonical form, here the four chambers in $\mathbb{T}_8$ yield distinct forms! For example for points in the chamber $1\cap 2$, the corresponding canonical form can be matched to:
\begin{equation}
\vcenter{\hbox{\scalebox{1}{
 %
}}}
\end{equation}

\noindent
Note that the last four integrands on the first line contain the one-loop maximal cut $(1,3,5)$ and $(5,7,1)$, whose cut is precisely the leading singularity associated with cell $\Gamma_1$
 %$$\includegraphics[scale=0.3]{OnShell2}\,.$$
$$
\vcenter{\hbox{\scalebox{0.9}{
%

}}}$$
Similarly, the integrands on the second line are associated with that of cell $\Gamma_2$. This is a general phenomenon: when a point is associated with a cell $\Gamma_p$ the resulting loop-form will include integrands that participate in the cuts contained by its on-shell diagram. Said in another way, when we declare a point in $\mathbb{T}_n$ lies in $\Gamma_p$, this implies that cuts contained in the on-shell diagram of $\Gamma_p$ can be realized as the boundary of some positive geometry, i.e. the loop amplituhedron. Thus the canonical form for eq.~\eqref{eq:loop_space}, from $\mathbf{p}\cap\mathbf{q}$ includes integrands that participate in maximal cuts of $\Gamma_{p}$ and $\Gamma_{q}$.

For convenience, let us consider the following combination of integrands as building blocks:
\begin{equation}\label{eq:L=1 n=8 block function}
    \begin{split}
        &I^{L{=}1}_{8}(0)=\sum _{i=1}^{8}(-1)^{i} \, I_{box}( i{-}1,i,i{+}1),\\
        &I^{L{=}1}_{8}(1)=I_{box}( 1,3,5) +I_{tri}( 1,3,5)  +I_{box}( 5,7,1)+I_{tri}( 5,7,1),\\
        &I^{L{=}1}_{8}(2)=-I_{box}( 2,4,6)+I_{tri}( 2,4,6) -I_{box}( 6,8,2) +I_{tri}( 6,8,2) ,\\
        &I^{L{=}1}_{8}(3)= I_{box}( 3,5,7)+I_{tri}( 3,5,7)  +I_{box}( 7,1,3)+I_{tri}( 7,1,3),\\
        &I^{L{=}1}_{8}(4)= -I_{box}( 4,6,8)+I_{tri}( 4,6,8)  -I_{box}( 8,2,4)+I_{tri}( 8,2,4).
    \end{split}
\end{equation}
With these shorthand notations, the total canonical form is
\begin{equation}\label{eq:3d_8_loop_form}
\sum_{p,q}\;\left[p\cap q\right]_+ \times \Big( I^{L{=}1}_{8}(0){+}I^{L{=}1}_{8}(p){+}I^{L{=}1}_{8}(q)\Big),\quad\quad \{p,q\}\in \{1,2\}, \{1,4\}, \{3,2\}, \{3,4\}\,.
\end{equation}
Once again, the subscript $\left[\;\right]_+$ indicates we are considering the positive sector. Note that as the chambers are related by cyclic rotation, while the loop amplituhedron defined in eq.~\eqref{eq:loop_space} is cyclic invariant. Thus the combination $I^{L{=}1}_{8}(p){+}I^{L{=}1}_{8}(q)$ for each chamber are related by cyclic rotation, as one can explicitly check. Since $I^{L{=}1}_{8}(p)$ appears in all intersections involving $p$, collecting the terms proportional to $I^{L{=}1}_{8}(p)$ simply gives $\Gamma_p$. Thus in total, we obtain
\begin{equation}
    (\Gamma_{2,+}+\Gamma_{4,+})\times I^{L{=}1}_{8}(0)+ \sum_{i=1}^4\, \Gamma_{i,+} \times I^{L{=}1}_{8}(i)\,,
\end{equation}
where $\Gamma_{i,+}$ represents the positive solution for the $i$-th cell. This yields the correct positive sector of the one-loop integrand.

\paragraph{Ten-point case}  The anatomy of the tree region at $n=6,8$ is relatively simple: the $\mathbb{T}_6$ consists of one chamber and $\mathbb{T}_8$ four. The four chambers are shown to be inherently distinct by their extended one-loop form. As we will see, at ten-points we will need to go to two-loops to fully distinguish all chambers.

At ten points, there exist three distinct topologies of on-shell diagrams, displayed below along with their equivalent representation under the triangle move:

\begin{equation}
\Gamma_{T_i}:\vcenter{\hbox{\scalebox{0.65}{


}}}
\end{equation}
Including its cyclic images, there are a total of $10+10+5=25$ in-equivalent BCFW cells. From the recursion formula, the ten-point tree amplitude can be expressed as the sum of three factorization channels: 8-4, 6-6, and 4-8. The residues on these cells from the Grassmannian integral in eq.~\eqref{eq:Integral} have the following relations:
\begin{equation}\label{eq:10pt}
    \begin{gathered}
        \sum_{\text{sol.}}T_i+T_{i+2}+T_{i+4}+T_{i+6}+T_{i+8}=A_{10}^{\text{tree}},\\
        \sum_{\text{sol.}} T_{i}+T_{i+4}=\sum_{\text{sol.}} B_{i}+P_{i},\ \sum_{\text{sol.}} T_{i+2}+T_{i+8}=\sum_{\text{sol.}} B_{i+3}+P_{i+8},\quad 
    \end{gathered}
\end{equation}
where the summation is taken over the $2^3=8$ solutions of eq.~\eqref{eq: Schubert}.  The second line of equality reflects the fact that in the 8-4 and 4-8 channels, the eight-point sub-amplitude can be expressed in two ways
%$$\includegraphics[scale=0.4]{OnShell3}$$
$$
\vcenter{\hbox{\scalebox{0.5}{


}}}$$
Thus we have in total 22 inequivalent BCFW triangulations. 

We will now use these cells to dissect $\mathbb{T}_{10}$. Note that, unlike eight points, now cells can appear in multiple triangulations.  As the cells within a single BCFW triangulation do not overlap, we can use this information to work out the overlapping regions throughout $\mathbb{T}_{10}$. For example consider a point residing in $\Gamma_{T_1}$, the first line in eq.~\eqref{eq:10pt} tells us that it cannot lie in $\Gamma_{T_{i}}$ for $i=3,5,7,9$ at the same time. Now consider the BCFW triangulation emerging from replacing $T_1, T_5$ using the second line of eq.~\eqref{eq:10pt}: $\{T_3,T_7,T_9,B_1,P_1\}$. Since the point has been excluded from the images of $T_3$, $T_7$, and $T_9$, it must be located within one of the images of either $B_1$ or $P_1$. This tells us that $T_1$ can be further separated into $T_1\cap B_1$ and $T_1\cap P_1$. Iteratively scanning through the 22 distinct BCFW triangulations, one results in $\mathbb{T}_{10}$ being decomposed into fifty chambers! For example, the ten of chambers involving $T_1$ are:
\begin{equation}\label{eq: 10Intersection}
    \begin{gathered}
        T_{1,2}\cap P_{1,2,7,8},\ T_{1,2}\cap P_{1,8}\cap B_2, \ T_{1,4}\cap P_{1,4,7,10}, \ T_{1,6}\cap B_{1,2}, \ T_{1,6}\cap P_{1,2,6,7},\\  
        T_{1,6}\cap P_{1,6}\cap B_2,\ T_{1,6}\cap P_{2,7}\cap B_1, \  T_{1,8}\cap P_{1,4,7,8}, \ T_{1,10}\cap P_{1,6,7,10}, \\ 
        \text{and}\quad  T_{1,10}\cap P_{7,10}\cap B_1\,,
    \end{gathered}
\end{equation}
where $T_{1,2}\cap P_{1,8}\cap B_2$ is a short hand notation for $T_1\cap T_2\cap P_1\cap P_8\cap B_2$. The remaining 40 corresponds to four cyclic shifts of the above by 2. The chambers in eq.~\eqref{eq: 10Intersection} can be represented diagrammatically as follows:
\begin{figure}[H]
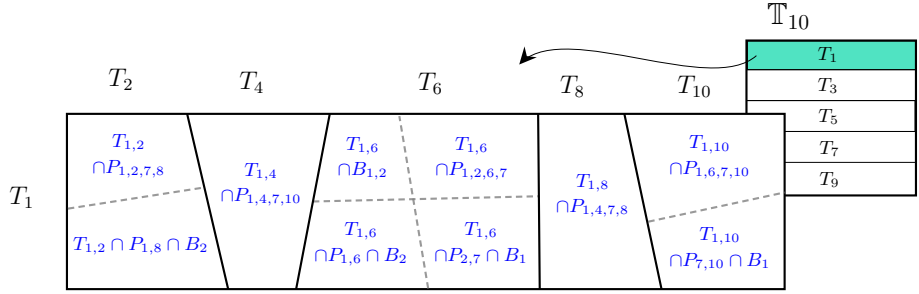
 
\begin{center}
$$
\vcenter{\hbox{\scalebox{0.8}{


}}}
$$
\end{center}
\caption{Here we present the chamber structure of $\mathbb{T}_{10}$. Beginning with the BCFW triangulation of $\mathbb{T}_{10}$  in terms of 5 pieces, as displayed in the upper right corner, each piece is further dissected into smaller pieces via its overlap with other BCFW cells. These smaller pieces constitute the chambers of $\mathbb{T}_{10}$.}
\label{fig:10tree_subregion_3d} 
\end{figure}

After listing the chambers of $\mathbb{T}_{10}$  we now can use it to geometrically obtain the one-loop ten-point integrand. Unlike $n=8$, we find that some chambers are degenerate at the one-loop, i.e. they yield the same form. For example chambers $T_{1,2}\cap P_{1,2,7,8}$ and $T_{1,2}\cap P_{1,8}\cap B_2$. In fact, this can be understood from the one-loop cuts visible form the cells $T_{1,2}$

$$
\vcenter{\hbox{\scalebox{0.65}{


}}}$$
We see that the cells $T_1$ and $T_2$ already contain the one-loop boundaries in $P_i$ with $i=1,2,7,8$, as well as $B_2$. Thus further dissection of the region $T_1\cap T_2$ with $P_i$s or $B_2$ is undetectable at one-loop. It turns out the intersection of two sets $\{T_i\}_{i=1,3,5,7,9}$ and $\{T_i\}_{i=2,4,6,8,10}$, are sufficient to characterize the one loop geometry of the ten-point. For example $T_3\cap T_6$ covers another set of boundaries distinct from $T_1\cap T_2$:

$$
\vcenter{\hbox{\scalebox{0.65}{
%

}}}$$
Thus from the one-loop geometry point of view, it is sufficient to write $\mathbb{T}_{10}$ as:
\begin{equation}
    \mathbb{T}_{10}=\bigcup_{i,j=1,2,\cdots,5} [2i{-}1\cap 2j]
\end{equation}
where $2i{-}1\cap 2j:=T_{2i-1}\cap T_{2j}$. These are the regions bounded by the black contours in fig.~\ref{fig:10tree_subregion_3d} and its cyclic image. There are total of 25 sub-regions in this choice, and using them we can obtain the one-loop ten-point integrand geometrically. In the regions $[1\cap 2]$, the corresponding loop canonical form is
\begin{equation}\label{eq: 8pt1loop}
\vcenter{\hbox{\scalebox{1}{

}}}
\end{equation}
Again, we can see that the loop-form is separated into three parts: the common soft part, integrands related to $T_1$, and integrands related to $T_2$. We point out that $I_{tri/box}(1,3,5)$, $I_{tri/box}(1,5,7)$, and $I_{tri/box}(1,7,9)$ produce the three one-loop maximal cuts in $T_1$ while  $I_{tri/box}(2,4,6)$, $I_{tri/box}(2,6,8)$, and $I_{tri/box}(2,8,10)$ produce the three one-loop maximal cuts in $T_2$:

Similarly, we can define the block functions to express the canonical form of other regions:
\begin{equation}
    \begin{split}
        &I^{L{=}1}_{10}(0):=\sum _{i=1}^{10}(-1)^{i} \, I_{box}( i{-}1,i,i{+}1),\\
        &I^{L{=}1}_{10}(2i{-}1):= I_{box}( 2i{-}1,2i{+}1,2i{+}3){+}I_{tri}( 2i{-}1,2i{+}1,2i{+}3){+}I_{box}( 2i{-}1,2i{+}3,2i{+}5) \\
        &\quad \quad \quad \quad  {+}I_{tri}( 2i{-}1,2i{+}3,2i{+}5) {+}I_{box}( 2i{-}1,2i{+}5,2i{+}7) {+}I_{tri}( 2i{-}1,2i{+}5,2i{+}7) ,\\
        &I^{L{=}1}_{10}(2i):= {-}I_{box}( 2i,2i{+}2,2i{+}4){+}I_{tri}( 2i,2i{+}2,2i{+}4){-}I_{box}( 2i,2i{+}4,2i{+}6)  \\
        &\quad \quad \quad \quad{+}I_{tri}( 2i,2i{+}4,2i{+}6)  {-}I_{box}( 2i,2i{+}6,2i{+}8) {+}I_{tri}( 2i,2i{+}6,2i{+}8).
    \end{split}
\end{equation}
and the canonical form of the region $2i{-}1\cap 2j$ is
\begin{equation}
    I^{L{=}1}_{10}(0)+I^{L{=}1}_{10}(2i{-}1)+I^{L{=}1}_{10}(2j).
\end{equation}
Summing over the 25 regions will give the one-loop ten-point amplitude in the positive sector.

We see that the poles in the integrand, associated with an intersection region, exactly match the local poles of the on-shell diagrams that define the intersection! As the number of cells involved in an intersection increases, we will need higher loops to expose all possible combinations of distinct boundaries. Indeed we have seen that for 10 points, some chambers are degenerate at one loop. Such degeneracy is lifted as the loop order increases. In the previous example of one-loop degeneracy for $T_{1,2}\cap P_{1,2,7,8}$ and $T_{1,2}\cap P_{1,8}\cap B_2$, will be lifted at two loops. Indeed $P_2$ contains the set of poles for the two-loop maximal cut $\langle AB 12\rangle=\langle AB 34\rangle=\langle AB 56\rangle=\langle AB CD\rangle=\langle CD 67\rangle=\langle CD 89\rangle=0$ that is not covered by   $T_{1,2}\cap P_{1,8}\cap B_2$, and thus we expect the degeneracy to be lifted. Indeed explicit computations show that the two-loop forms for individual chambers are distinct. It will be interesting to see if these chambers are sufficient to characterize all distinct two-loop forms. We leave this to future work.

In summary, we find that at $n=6,8,10$ there are one, four, and fifty chambers respectively. The chambers yield distinct loop-forms at one-loop for $n=8$ and at two-loops for $n=10$. This is illustrated below:

\begin{figure}[H]
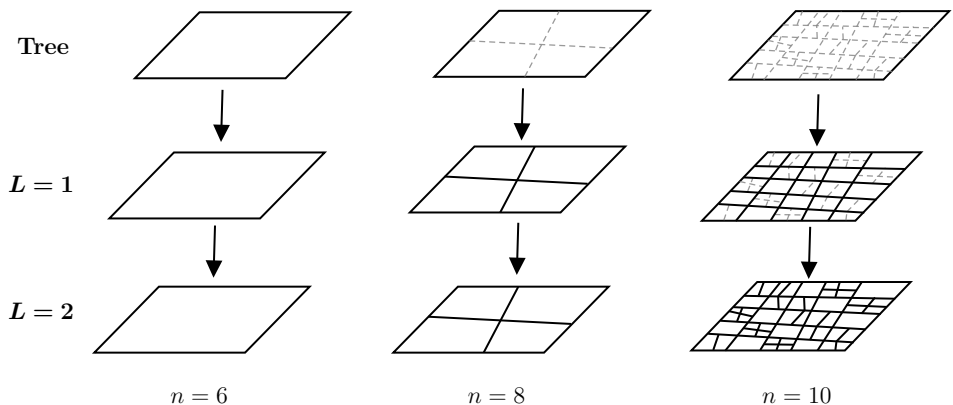
 
\begin{center}
$$
\vcenter{\hbox{\scalebox{0.75}{


}}}$$
\end{center}
\caption{The loop-form for each distinct chamber might be degenerate at low loops. For example at 10-points, the one-loop geometry can only distinguish between 25 subregions. The degeneracy is lifted at two-loops, revealing the 50 chambers.}
\label{fig:tree_anatomy} 
\end{figure}
\noindent This analysis invites us to introduce a new characterization of the chambers. Recall that each BCFW diagram also represents a set of maximal cuts. For $n$-points the diagram represents an $n/2{-}2=k$-loop maximal cut. This means that when solving for eq.~\eqref{eq:MomTwisGras} (or equivalently \eqref{eq: Schubert}), we are also obtaining a maximal cut solution associated with the given point in $\mathbb{T}_n$. Thus asking for the solution to be positive is equivalent to asking for the cut solution, which can be translated into a particular configuration of $(A,B)$, to reside on the boundary of the loop amplituhedron eq.~\eqref{eq:loop_space}. Said in another way, \textit{the chambers of $\mathbb{T}_n$ are defined by the set of maximal cut that can simultaneously reside in the loop amplituhedron!} This also explains why we need to go to $k$-loops to fully resolve the degeneracy of the chambers. Beyond $k$-loops, since the maximal cuts are associated with on-shell diagrams that are reducible, the chambers will no longer bifurcate. At this point, we have used the overlapping of BCFW cells to define the chambers. The question of whether BCFW cells are sufficient simply translate to whether all of the $n{-}3$-dimensional cells of OG$^{>0}$($k',2k'$) are BCFW cells. For $k'=3$, the only 3-dimension cells is the top cell. For $k'=4$ we have four distinct co-dimension 1 cells that are 5-dimensional which exactly correspond to our BCFW cells. At $k'=5$ we have. It will be interesting to clarify whether BCFW cells constitute all possible $n{-}3$-dimensional cells for general OG$^{>0}$($k',2k'$).

\subsection{Higher loops and bipartite negative geometries}

Now we move to higher-loops. We can proceed as before, simply triangulating the loop-region using the chambers in $\mathbb{T}_n$. As we will see for $n=6$ and $n=8$, there is no need for further dissection, the loop-form continues to be invariant in each chamber. However, at higher-loops, we can verify some important features of the multi-loop geometries defined in eq.~\eqref{eq:loop_space}, which in turn provide some evidence that they indeed give all-loop ABJM amplituhedron for all multiplicities. Recall that just as in the four-point case, the loop geometry admits a natural decomposition by writing each ``mutual positivity condition" as ``no condition" minus ``negative condition"; in this way, our loop geometry becomes a sum of ``negative geometries" in $D=3$: each negative geometry is represented by a labelled graph with $L$ nodes and $E$ edges (edge $(i j)$ for $\langle (AB)_i (AB)_j\rangle>0$ since we reversed all signs, and no condition otherwise), with an overall sign factor $(-)^E$. We sum over all graphs with $L$ nodes without $2$-cycles, 
\begin{equation}
{\cal A}_L=\sum_g (-)^{E(g)} {\cal A}(g)
\end{equation} 
where ${\cal A}(g)$ denotes the (oriented) geometry for graph $g$. It suffices to consider all {\it connected} graphs, whose (signed) sum gives the geometry for the logarithm of amplitudes~\cite{Arkani-Hamed:2021iya}. Such a decomposition is useful since each $A_g$ is simpler, whose canonical form is easier to compute. For example, the canonical forms for $L=2,3$ are
%It is trivial for two loops: ${\cal A}_2$ is given by the square of ${\cal A}_1$ (symmetrized over the two loops, with $E=0$), {\it minus} the only connected graph (with one negative condition,  $E=1$) which gives the $\log$ of amplitude $\tilde{\Omega}_2:=\Omega_2-\frac 1 2 \Omega_1^2$. Similarly the three-loop form $\Omega_3$ is given by the sum of $\Omega_1^3/3!$ ($E=0$), $-\tilde{\Omega}_2 \Omega_1/2$ ($E=1$), and the connected part $\tilde{\Omega}_3$ (with chain and triangle graphs):
\begin{equation}
\vcenter{\hbox{\scalebox{0.8}{


}}}
\end{equation}
where the connected part, or the integrand for logarithm of the amplitude, is denoted as $\tilde{\Omega}_L$, {\it e.g.} $\tilde{\Omega}_2:=\Omega_2-\frac 1 2 \Omega_1^2$. Similarly, the connected part of $L=4$ is given by the sum of graphs with $6$ topologies (and so on for higher $L$),
\begin{equation}
    \tilde{\Omega}_4=
    \vcenter{\hbox{\scalebox{0.8}{
%

}}}
\end{equation}

Now we show that exactly as in the $n=4$ case, only those negative geometries with {\it bipartite} graphs survive in $D=3$, thus most negative geometries do not contribute. For example, for $\tilde{\Omega}_3$, the chain graph contributes but the triangle does not, {\it i.e.}
\begin{equation}\label{eq:3loopg}
    \underline{\tilde{\Omega}}_3=
\vcenter{\hbox{\scalebox{0.9}{
%

}}}
\end{equation}
For $\tilde{\Omega}_4$, only the two kinds of tree graphs and the box contribute. Note that the fraction of bipartite graphs in all graphs tends to zero quickly as $L$ increases: for $L=2,\dots, 7$, the number of topologies for connected graphs are $1,2,6,21,112, 853$, while that of bipartite topologies is $1,1,3, 5, 17, 44$.

The reason that only bipartite negative geometries survive in $D=3$ is exactly the same as that for the $n=4$ case, since only mutual conditions are needed in this reduction, which we've proven in eq.~\eqref{eq: MutualNeg1} is still of the form $w_{a,b} y_{a,b}-z_{a,b}^2<0$~\footnote{Here, we change the coordinate to the Poincare patch $\textit{i.e.}$ $y^\prime=y/z$, $w^\prime=w/z$, $z^\prime=1/z$. Then eq.~\eqref{eq: MutualNeg1} is proportional to $w_{a,b} y_{a,b}-z_{a,b}^2<0$.}. Thus once again all mutual negativity conditions $\langle (AB)_a (AB)_b\rangle>0$ translates to the inequality
$w_{a,b} y_{a,b}-z_{a,b}^2>0$ and one concludes that $(w_a-w_b)$ and $(y_a-y_b)$ must have the same sign.

The remaining part of the proof goes exactly as for $n=4$ (see appendix A of \cite{He:2022lfz} for details): for each edge of the graph we have an ``arrow" which leads to ``time ordering" with no closed loop, and by the beautiful {\it transitive reduction} we find that except for bipartite graphs where a black/white node denotes source/sink of the ``ordering”, all other graphs cancel in the sum~\cite{He:2022lfz}! Moreover, just as in $n=4$ and as we will see for $L=2$ $n=6,8$,  the canonical form for these bipartite negative geometries has remarkably simple pole structures. Since a sink (white node) can never have $y, w\to 0$, it cannot have poles $\langle AB 12\rangle=0$ or $\langle AB34\rangle=0$; similarly, a source (black node) cannot have poles $\langle AB 23\rangle=0$ (note that $\langle AB14\rangle=0$ is not a physical pole for $n>4$). One can immediately generalize this by using other parametrizations, {\it e.g.} $Z_3, Z_4, Z_5, Z_6$, which means that each sink cannot have poles in $\langle AB 56\rangle=0$, and each source cannot have poles in $\langle AB34\rangle=0$. In this way, we reach the conclusion, that each source/sink (black/white node) can only have poles $\langle AB i{-}1 i\rangle=0$ for $i$ odd/even, respectively (the latter includes the special case $i=n$, which can only be a pole for white node).

Let us verify the above description by computing $\tilde{\Omega}^\alpha_2:=\Omega^\alpha_2- \frac 1 2 (\Omega^\alpha_1)^2$ for each chamber labelled by $\alpha$. The bipartite nature is reflected in that each black/white node for $\ell$ can only have inverse propagators $(\ell \cdot I)$ for $I$ odd/even, which we will show matches nicely with local integrands for $\Omega_2^\alpha$~\cite{He:2022cup}. 
%\textbf{Tree regions of $n=6,8$ are not further dissected. Definition of bipartite graphs for $n>4$, not log amplitude anymore. Computation of the negative geometry at $n=6$? Compatibility.}

\paragraph{Two-loop six-point integrand} 
We start with $n=6$ and first record the loop form $\Omega^\alpha_2$ for the top cell of the positive branch (LS${}_{6,+}$):
\begin{equation}\label{eq: 2loop6pt}
    \begin{split}
        \sum_{i=1}^6 -\frac{1}{2} I^{\text{critter}}(i)+I^{\text{crab}}(i)+I^{\text{2mh}}_+ (i)-\sum_{i=1,3,5}I^{bt}_1(i)+\sum_{i=2,4,6}I^{bt}_2(i)
    \end{split}
\end{equation}
where the local (planar) integrands are given in the appendix~\ref{app:local_integral}. 
Next, we reproduce this result by computing canonical forms for the {\it negative geometry}, $\tilde{\Omega}^\alpha_2=\Omega^\alpha_2-1/2 (\Omega^\alpha_1)^2$, where the one-loop form is for the same $\alpha$, or LS${}_+$. The result is naturally given in terms of some integrands that look like "bipartite graphs".

The two-loop six-point form for negative geometry of the positive region is given by a sum of three types of topologies, which are (non-planar) double-box, box-triangle, and double-triangle integrals, and we denote them by $I_a (B; W), I_b (B; W), I_c (B; W)$, respectively:
\begin{equation}
    \begin{gathered}
        I_a(135, 246)+ \left[ I_b(135, 24) + (24 \to 46, 26) + I_b(13, 246) + (13\to 35, 15)\right]\\ 
+[I_c (13, 24) +(24 \to 26)+ (13 \to 35; 24 \to 24, 46)+ (13 \to 15; 24 \to 26, 46)\\
+I_c (13, 46)+(13 \to 35, 15; 46 \to 26, 24)]+ (\ell_1 \leftrightarrow \ell_2)\,,
    \end{gathered}
\end{equation}
where $B \subset \{1,3,5\}$ and $W\subset \{2,4,6\}$ denote external points of the propagators involving $\ell_1$ and $\ell_2$, {\it e.g.} 
\begin{equation}
I_a(135, 246)=\frac{N_a (135,246)}{(\ell_1 \cdot 1)(\ell_1 \cdot 3)(\ell_1 \cdot 5)(\ell_1 \cdot \ell_2)(\ell_2 \cdot 2)(\ell_2 \cdot 4)(\ell_2 \cdot 6)},  
\end{equation}
and similarly for $I_b, I_c$; note that we have $1$ $I_a$ integral, $2\times 3=6$ $I_b$ integrals, and $3+6=9$ $I_c$ integrals ($6$  cyclic images of $I_c(13,24)$ and $3$ of $I_c(13,46)$, and in the end we symmetrize over $\ell_1, \ell_2$. We give explicitly the numerator of these integrals
\begin{equation}\label{eq:negative_2Loop6}
    \vcenter{\hbox{\scalebox{1}{

}}}$};
% Text Node
\draw (31,137) node [anchor=north west][inner sep=0.75pt]  [font=\footnotesize] [align=left] {$\displaystyle 1,3,5$};
% Text Node
\draw (96,137) node [anchor=north west][inner sep=0.75pt]  [font=\footnotesize] [align=left] {$\displaystyle 2,4$};
% Text Node
\draw (120.8,110) node [anchor=north west][inner sep=0.75pt]  [font=\footnotesize] [align=left] {$\displaystyle :=N_b(135,24)= ( \ell_1\cdot y_{4,( 23) \bigcap ( 145)})(1\cdot 3) -\frac{\epsilon ( \ell_1,2,3,4,5)\sqrt{( 1\cdot 3\cdot 5)} }{( 3\cdot 5)},$};

% bipartite graph (13,246)
\draw (37,164) node [anchor=north west][inner sep=0.75pt]  [font=\footnotesize] [align=left] {$\vcenter{\hbox{\scalebox{1}{
\begin{tikzpicture}[x=0.75pt,y=0.75pt,yscale=-1,xscale=1]
%uncomment if require: \path (0,460); %set diagram left start at 0, and has height of 460
%Straight Lines [id:da028659140963990914] 
\draw [line width=0.75]    (39.9,60.4) -- (99.9,60.4) ;
%Shape: Circle [id:dp3880150352528413] 
\draw  [fill={rgb, 255:red, 255; green, 255; blue, 255 }  ,fill opacity=1 ] (95.82,60.76) .. controls (95.62,58.5) and (97.29,56.51) .. (99.54,56.32) .. controls (101.8,56.12) and (103.79,57.79) .. (103.98,60.04) .. controls (104.18,62.3) and (102.51,64.29) .. (100.26,64.48) .. controls (98,64.68) and (96.01,63.01) .. (95.82,60.76) -- cycle ;
%Shape: Circle [id:dp3802493835591214] 
\draw  [fill={rgb, 255:red, 0; green, 0; blue, 0 }  ,fill opacity=1 ] (35.82,60.76) .. controls (35.62,58.5) and (37.29,56.51) .. (39.54,56.32) .. controls (41.8,56.12) and (43.79,57.79) .. (43.98,60.04) .. controls (44.18,62.3) and (42.51,64.29) .. (40.26,64.48) .. controls (38,64.68) and (36.01,63.01) .. (35.82,60.76) -- cycle ;
\end{tikzpicture}
}}}$};
% Text Node
\draw (35,179) node [anchor=north west][inner sep=0.75pt]  [font=\footnotesize] [align=left] {$\displaystyle 1,3$};
% Text Node
\draw (90,179) node [anchor=north west][inner sep=0.75pt]  [font=\footnotesize] [align=left] {$\displaystyle 2,4,6$};
% Text Node
\draw (120.8,152) node [anchor=north west][inner sep=0.75pt]  [font=\footnotesize] [align=left] {$\displaystyle :=N_b(13,246)=( \ell_2\cdot y_{3,( 12) \bigcap ( 634)})(2\cdot 6) -\frac{\epsilon ( \ell_2,6,1,2,3)\sqrt{( 2\cdot 4\cdot 6)}} {( 2\cdot 6)},$};

% bipartite graph (13,24)
\draw (37,204) node [anchor=north west][inner sep=0.75pt]  [font=\footnotesize] [align=left] {$\vcenter{\hbox{\scalebox{1}{
\begin{tikzpicture}[x=0.75pt,y=0.75pt,yscale=-1,xscale=1]
%uncomment if require: \path (0,460); %set diagram left start at 0, and has height of 460
%Straight Lines [id:da028659140963990914] 
\draw [line width=0.75]    (39.9,60.4) -- (99.9,60.4) ;
%Shape: Circle [id:dp3880150352528413] 
\draw  [fill={rgb, 255:red, 255; green, 255; blue, 255 }  ,fill opacity=1 ] (95.82,60.76) .. controls (95.62,58.5) and (97.29,56.51) .. (99.54,56.32) .. controls (101.8,56.12) and (103.79,57.79) .. (103.98,60.04) .. controls (104.18,62.3) and (102.51,64.29) .. (100.26,64.48) .. controls (98,64.68) and (96.01,63.01) .. (95.82,60.76) -- cycle ;
%Shape: Circle [id:dp3802493835591214] 
\draw  [fill={rgb, 255:red, 0; green, 0; blue, 0 }  ,fill opacity=1 ] (35.82,60.76) .. controls (35.62,58.5) and (37.29,56.51) .. (39.54,56.32) .. controls (41.8,56.12) and (43.79,57.79) .. (43.98,60.04) .. controls (44.18,62.3) and (42.51,64.29) .. (40.26,64.48) .. controls (38,64.68) and (36.01,63.01) .. (35.82,60.76) -- cycle ;
\end{tikzpicture}
}}}$};
% Text Node
\draw (35,219) node [anchor=north west][inner sep=0.75pt]  [font=\footnotesize] [align=left] {$\displaystyle 1,3$};
% Text Node
\draw (96,219) node [anchor=north west][inner sep=0.75pt]  [font=\footnotesize] [align=left] {$\displaystyle 2,4$};
% Text Node
\draw (120.8,195) node [anchor=north west][inner sep=0.75pt]  [font=\footnotesize] [align=left] {$\displaystyle :=N_c(13,24)=-\frac{( 1\cdot 3)( 2\cdot 4)}{2},$};

% bipartite graph (13,46)
\draw (37,247) node [anchor=north west][inner sep=0.75pt]  [font=\footnotesize] [align=left] {$\vcenter{\hbox{\scalebox{1}{
\begin{tikzpicture}[x=0.75pt,y=0.75pt,yscale=-1,xscale=1]
%uncomment if require: \path (0,460); %set diagram left start at 0, and has height of 460
%Straight Lines [id:da028659140963990914] 
\draw [line width=0.75]    (39.9,60.4) -- (99.9,60.4) ;
%Shape: Circle [id:dp3880150352528413] 
\draw  [fill={rgb, 255:red, 255; green, 255; blue, 255 }  ,fill opacity=1 ] (95.82,60.76) .. controls (95.62,58.5) and (97.29,56.51) .. (99.54,56.32) .. controls (101.8,56.12) and (103.79,57.79) .. (103.98,60.04) .. controls (104.18,62.3) and (102.51,64.29) .. (100.26,64.48) .. controls (98,64.68) and (96.01,63.01) .. (95.82,60.76) -- cycle ;
%Shape: Circle [id:dp3802493835591214] 
\draw  [fill={rgb, 255:red, 0; green, 0; blue, 0 }  ,fill opacity=1 ] (35.82,60.76) .. controls (35.62,58.5) and (37.29,56.51) .. (39.54,56.32) .. controls (41.8,56.12) and (43.79,57.79) .. (43.98,60.04) .. controls (44.18,62.3) and (42.51,64.29) .. (40.26,64.48) .. controls (38,64.68) and (36.01,63.01) .. (35.82,60.76) -- cycle ;
\end{tikzpicture}
}}}$};
% Text Node
\draw (35,262) node [anchor=north west][inner sep=0.75pt]  [font=\footnotesize] [align=left] {$\displaystyle 1,3$};
% Text Node
\draw (96,262) node [anchor=north west][inner sep=0.75pt]  [font=\footnotesize] [align=left] {$\displaystyle 4,6$};
% Text Node
\draw (120.8,238) node [anchor=north west][inner sep=0.75pt]  [font=\footnotesize] [align=left] {$\displaystyle :=N_c(13,46) =\frac{( 1\cdot 3)( 4\cdot 6)}{2}.$};

\end{tikzpicture}
}}}
\end{equation}
which are also denoted in terms of bipartite graphs. Here  $(i, j)\bigcap (l,m,n)$ indicates the intersection of a line $(i,j)\equiv(Z_i,Z_j)$ and a plane $(l,m,n)\equiv(Z_l,Z_m,Z_n)$ is given by 
\begin{equation}
    (i,j)\bigcap(l,m,n)=Z_i \langle j, l,m,n\rangle-Z_j \langle i, l,m,n\rangle,
\end{equation}
and $y_{p,q}$ is a dual vector corresponding to the  line $(Z_p,Z_q)$. In twistor space, the numerator has very compact forms. For example, $(\ell_1\cdot y_{4,(2,3)\bigcap(1,4,5)})$ in twistor space (removing angle brackets) is  
\begin{equation}
    (\ell_1\cdot y_{4, (2,3)\bigcap (1,4,5)}) \rightarrow
    \begin{gathered}
        \langle AB 4 \big((2,3)\cap (1,4,5)\big) \rangle
       = \langle AB 24 \rangle\langle1345\rangle{-}\langle AB 34\rangle\langle1245\rangle.
    \end{gathered}
\end{equation}

\paragraph{Two-loop eight-point integrand}

Next, we consider the two-loop eight-point case, where the tree positive region has four chambers, labelled as $(p \cap q)$ for $p=1,3$ and $q=2,4$. We focus on the first one, $1\cap 2$, and we first give the expected loop form associated with it:
\begin{equation}\label{eq: 8pt2loop}
\vcenter{\hbox{\scalebox{1}{


}}}
\end{equation}
and for convenience let us consider the following combination of integrands as building blocks:
\begin{equation}\label{eq:L=2 n=8 block function}
    \begin{split}
        &I_8^{L{=}2}(0)=\sum _{p=1}^{8} {-}I_{A}^{db}( p) {+}I_{B}^{db}( p) {+}\frac{1}{2} I_{C}^{db}( p),\\
        &I_8^{L{=}2}(i)=\sum _{p=1,5} I_{\overline{D} ,+}^{db}( i{+}p) {+}I_{\overline{E} ,+}^{db}( i{+}p) +\sum _{p=3,7} I_{\underline{D} ,+}^{db}( i{+}p) {+}I_{\underline{E} ,+}^{db}( i{+}p)+\sum _{p=2,6} {-}I_{F,+,+}^{db}( i{+}p) \\ 
        &\quad \quad {+}\frac{1}{2} I_{G,+,+}^{db}( i{+}p) +({-})^i\sum _{p=4,8} I_{\overline{A}}^{bt}( i{+}p) {+} I_{\underline{A}}^{bt}( i{+}p) +({-})^i\sum _{i=2,6} I_{B}^{bt}( i{+}p).
    \end{split}
\end{equation}
These two-loop integrands are defined in appendix~\ref{app:local_integral}. In complete analogy with the six-point case above, we will reproduce these loop forms by 
computing  canonical forms for the negative geometry, $\tilde{\Omega}_2=\Omega_2-\frac 1 2 \Omega_1^2$ for the chamber $1 \cap 2$. These results are again naturally given in terms of loop forms for bipartite graphs, and we divide them into four groups according to their pole structures, each of which resembles the six-point result. The first group consists of those integrands with pole structures denoted by  $(135,246)$:

\begin{equation}\label{eq:negative_2Loop8_(135,246)}
    \vcenter{\hbox{\scalebox{1}{

}}}$};
% Text Node
\draw (31,137.5) node [anchor=north west][inner sep=0.75pt]  [font=\footnotesize] [align=left] {$\displaystyle 1,3,5$};
% Text Node
\draw (96,137.5) node [anchor=north west][inner sep=0.75pt]  [font=\footnotesize] [align=left] {$\displaystyle 2,4$};
% Text Node
%\draw (120.8,110) node [anchor=north west][inner sep=0.75pt]  [font=\footnotesize] [align=left] {$\displaystyle :=N_b(135,24)= -(\ell_1\cdot 2)(1\cdot 4)(3\cdot 5)/2+(\ell_1\cdot 3\big((1\cdot 4)(2\cdot 5)-(1\cdot 5)(2\cdot 4)\big)/2 +\frac{\epsilon ( \ell_1,2,3,4,5)\sqrt{( 1\cdot 3\cdot 5)} }{( 3\cdot 5)},$};
% Text Node
\draw (120.8,119) node [anchor=north west][inner sep=0.75pt]  [font=\footnotesize] [align=left] {$\displaystyle
\begin{array}{{>{\displaystyle}l}}
:=N_b(135,24)= -(\ell_1\cdot 2)(1\cdot 4)(3\cdot 5)/2+(\ell_1\cdot 3)\big((1\cdot 4)(2\cdot 5){-}(1\cdot 5)(2\cdot 4)\big)/2 \\
\qquad \qquad \qquad \quad \ \ -(\ell_1\cdot 4)(1\cdot 3)(2\cdot 5)-\epsilon ( \ell_1,2,3,4,5)\sqrt{( 1\cdot 3\cdot 5)}/ ( 3\cdot 5),
\end{array}%
$};

\end{tikzpicture}
}}}
\end{equation}

We denote the sum of these integrands for bipartite graphs in the first group as $I(135,246)$, which is essentially the same as the six-point case:
\begin{equation}
    \begin{split}
        I(135,246)=&I_a(135,246)+[I_b(135,24)+I_b[135,46]+I_b(13,246)+I_b(35,246)]\\
        &+[I_c(13,24)+(13\rightarrow 35)+(24\rightarrow46)+(13\rightarrow35;24\rightarrow46)].
    \end{split}
\end{equation}
where {\it e.g.} $I_a(135,246)=N_a(135,246)/((\ell_1 \cdot 1)(\ell_1 \cdot 3)(\ell_1 \cdot 5)(\ell_2 \cdot 2)(\ell_2 \cdot 4)(\ell_2 \cdot 6)$. Similarly, there is a second group of bipartite graphs, related to the first ones by cyclic shifts, and we denote the sum of these integrands as
\begin{equation}
    I(157,268)=I(135,246)\Big|_{(135,246)\rightarrow(571,682)}
\end{equation}

The third group of integrands consists of the following bipartite graphs with pole structures denoted by $(135,268)$:
\begin{equation}\label{eq:negative_2Loop8_(135,268)}
    \vcenter{\hbox{\scalebox{1}{
$};

\end{tikzpicture}
}}}
\end{equation}
The full integrand is given by the sum of these building blocks:
\begin{equation}
    \begin{split}
        I(135,268)=&I_a(135,268)+[I_b(135,28)+I_b(135,68)+I_b(13,268)+I_b(35,268)]\\
        &+[I_c(13,28)+I_c(35,68)]
    \end{split}
\end{equation}
and finally the integrand for the last group is related by cyclic shift:
\begin{equation}
    I(157,246)=I(135,268)\Big|_{(135,268)\rightarrow(571,624)}
\end{equation}

The complete loop form associated with the chamber $1 \cap 2$ is given by the sum of these four groups: $I(135,246)+I(157, 268)+I(135, 268)+I(157, 246) + (\ell_1 \leftrightarrow \ell_2)$. 

%\textbf{Definition of bipartite graphs of the eight-point, compatible}

\paragraph{Comments on all loops}

The bipartite structures of negative geometries, together with the even/odd pole structures, have important implications for all loops. As we have seen for $L=2$ they greatly simplify canonical forms of these loop geometries, and they impose stronger and stronger constraints for higher $L$. Exactly as in the $n=4$ case, if we take the log of the loop form for each chamber $\tilde{\Omega}_\alpha$ as the sum of (connected) negative geometries, our prediction is that for $L=3$ only chain topologies are needed for all multiplicities since triangle (or any odd-gon) cannot be made bipartite. Similarly, we have chain, star, and box topologies for $L=4$ and $5$ out of $21$ topologies for $L=5$~\cite{He:2022lfz}. Moreover, in each case, the black/white node only has odd/even poles as we have seen for $L=2$, which reduces the space of possible integrands enormously. For example, we expect that the $L=3$, $n=6$ integrand can be determined very similar to what we have done for $L=2$. Since only chain graphs are needed for the negative geometries, we can write down an ansatz with (a subset of) $1,3,5$ assigned for any black node, and (a subset of) $2,4,6$ for any white node; such constraining mutual conditions and pole structures make it very plausible that the full $L=3$ integrand can be computed from geometries without the need of other input. We leave this computation (and possible ones for higher $n$ and $L$) to future works.

Last but not least, as we have discussed for $n=4$ case in~\cite{He:2023exb}, these bipartite structures are clearly related to the special feature that no odd-point amplitudes exist in ABJM theory. As we have seen earlier, the fact that three-particle and five-particle cuts for all-loop, all-multiplicity amplitudes vanish clearly follow from the geometry. More generally, we find that arbitrarily complicated all-loop cuts of $n$-point geometry must vanish as long as it isolates any (purely internal or partly external) odd-point amplitude. For purely internal case with only $D_{ij}$ cut, this is simply because any bipartite graphs (which applies to negative geometries for all $n$) cannot contain an odd cycle; for general cases involving certain $x,y,z,w$ cut, one can also generalize the argument given in~\cite{He:2023exb} to all $n$.

\section{Beyond positive solutions and parity action}\label{sec:parity}
One of the central themes of this paper is the dissection of $\mathbb{T}_n$ into chambers by analyzing the positivity of the solution to, 
\begin{equation}
\sum_i (\mathcal{D}_{\Gamma})_{\alpha,i}Z_i=0\,.
\end{equation}
As mentioned previously, there are in general more than one solution and the remaining are crucial in obtaining the full amplitude. For example at six-points, where the BCFW cell is the top cell and there's no dissection, we have two solutions corresponding to the positive and negative branch of the Grassmannian. Denoting the number of solutions as $\Gamma[\mathcal{D}_{\Gamma}]$, for BCFW cells in ABJM theory we have $\Gamma[\mathcal{D}_{\Gamma}]=2^{k}$. Note that this is distinct from $\mathcal{N}=4$ sYM, as one finds $\Gamma[\mathcal{D}]=1$.\footnote{There are certainly non-BCFW cells for $\mathcal{N}=4$ that has $\Gamma[\mathcal{D}_{\Gamma}]>1$. We will discuss their relevance for the dissection of $\mathbb{T}_n$ in the outlook. } This can be attributed to the fact that BCFW shifts in four-dimensions are linear in the deformation variable, while quadratic for three-dimensions:
$$
\vcenter{\hbox{\scalebox{0.8}{


}}}$$
There is another way to understand the $2^{k}$ counting for ABJM. Each tree-level BCFW term can be represented as an on-shell diagram with $k$-loops. Recall that each on-shell diagram can be understood as representing multi-loop maximal cut for a set of fixed external kinematics. As there are two solutions for each loop we have $2^k$ solutions. 

Since for each loop the cut solutions are that to a quadratic equation, they are simply related by taking different branches of the square root of the discriminant. For example for the following maximal cut 
$$
\vcenter{\hbox{\scalebox{1}{
%

}}}$$
the square root is simply 
\begin{equation}
\sqrt{{-}(p_1{+}p_2)^2(p_3{+}p_4)^2(p_5{+}p_6)^2}=\langle12\rangle\langle 34\rangle\langle 56\rangle
\end{equation}
thus to change between the two solutions will require us to change a sign in front of the square root, which in this case can be achieved by simply applying a parity operation $\Pi_6$ $Z_i\rightarrow -Z_i$ on legs $\{2,4,6\}$ (equivalently $\{1,3,5\}$). In the following, we will use the six- and eight-point examples to demonstrate how combining eq.~\eqref{eq:loop_space} and parity operations generate the full amplitude.

\subsection{The complete six-point amplitude}
Previously we've motivated the parity operation from interchanging between solutions of the maximal cut. Here we will directly see how the action acts on the analytic solution of the BCFW cell, which is straightforward at six-point since the cell is a top cell. At six-point $\Gamma[C_6]=2$, and the positive branch is given as:
\begin{equation}
C_{6,+}=\left(
\begin{array}{cccccc}
\lambda_1^1 & -\lambda_2^1 & \lambda_3^1 &-\lambda_4^1 &\lambda_5^1 &-\lambda_6^1\\
    \lambda_1^2 & -\lambda_2^2 & \lambda_3^2 &-\lambda_4^2 &\lambda_5^2 &-\lambda_6^2\\
 \langle35\rangle & \langle46\rangle & \langle51\rangle & \langle62\rangle & \langle13\rangle & \langle24\rangle \\
\end{array}
\right)\,.
\end{equation}
Indeed one can check that with the two brackets living in $\mathbb{K}_n$, i.e. eq.~\eqref{MomAmp}, $C_{6,+}\in OG^{>0}$. The other solution is given as:
\begin{equation}
C_{6,-}=\left(
\begin{array}{cccccc}
\lambda_1^1 & -\lambda_2^1 & \lambda_3^1 &-\lambda_4^1 &\lambda_5^1 &-\lambda_6^1\\
    \lambda_1^2 & -\lambda_2^2 & \lambda_3^2 &-\lambda_4^2 &\lambda_5^2 &-\lambda_6^2\\
 \langle35\rangle & -\langle46\rangle & \langle51\rangle & -\langle62\rangle & \langle13\rangle & -\langle24\rangle \\
\end{array}
\right)\, .
\end{equation}
Indeed the two cells can be interchanged by the action of $\Pi_6$ (with overall scale a sign on the columns 2, 4, 6). Similarly the corresponding leading singularities in eq.~\eqref{eq:LS6} interchanges under $\Pi_6$ 
\begin{equation}
   \text{LS}_{6,+} \xleftrightarrow{\{2,4,6\}} \text{LS}_{6,-}.
\end{equation}
The fact that the two solutions are interchanged under $\Pi_6$, also tells us that the negative solution actually becomes positive once $\Pi_6$ acts on the external twistors. Since the solutions also correspond to the solution of maximal cuts, the parity operation changes the solution of the maximal cut. Said in another way, the negative cut solution lives on the boundary of loop amplituhedron, i.e. eq.~\eqref{eq:loop_space}, if the external twistors are acted upon by $\Pi_6$. 

Let us see this play out explicitly. The maximal cut requires:
\begin{equation}
    \langle AB 12\rangle=\langle AB 34\rangle=\langle AB 56\rangle=0.
\end{equation}
The solution is given by
\begin{equation}
    \begin{split}
        Z_A=Z_1+\alpha Z_2, \quad Z_B=\beta Z_3+Z_4
    \end{split}
\end{equation}
with
\begin{equation}
    \begin{split}
        \alpha=\frac{\langle3561\rangle}{\langle2356\rangle}\pm\frac{\sqrt{-\langle 1234\rangle\langle3456\rangle\langle5612\rangle}}{\langle\!\langle 24\rangle\!\rangle \langle2356\rangle},\ \beta=-\frac{\langle 4561\rangle+\alpha \langle2456\rangle}{\langle3561\rangle+\alpha\langle2356\rangle}.
    \end{split}
\end{equation}
%After acting the parity operator $\Pi_6$ on the two  cut solutions
%\begin{equation}
 %   \begin{split}
  %      Z_A=Z_1-\alpha Z_2, \quad Z_B=\beta Z_3-Z_4
   % \end{split}
%\end{equation}
%with
%\begin{equation}
 %   \begin{split}
  %      \alpha=-\frac{\langle3561\rangle}{\langle2356\rangle}\pm\frac{\sqrt{-\langle 1234\rangle\langle3456\rangle\langle5612\rangle}}{\langle\!\langle 24\rangle\!\rangle \langle2356\rangle},\ \beta=-\frac{\langle 4561\rangle-\alpha \langle2456\rangle}{-\langle3561\rangle+\alpha\langle2356\rangle}.
   % \end{split}
%\end{equation}
We can see that the parity operator $\Pi_6$ changes the relative sign in $\alpha$, resulting in two solutions interchanging. The positive cut solution will locate inside the amplituhedron~\eqref{eq:loop_space}, while the negative one sits in the counterpart where  the parity operator acts on the amplituhedron, explicitly the region defined by (with $\hat{Z}_i:=-Z_i$ for $i=2,4,6$) $\langle\!\langle A B \rangle\!\rangle=0$, $\langle (A B)_a (A B)_b\rangle<0$ and
\begin{equation}\label{eq:modified_region_n=6} 
	\begin{gathered}
	 \begin{gathered}
	    \langle (AB)_a 1 \hat{2}\rangle<0,\, \langle (AB)_a  \hat{2} 3\rangle<0,  \langle (AB)_a  3 \hat{4} \rangle<0, \\
     \langle (AB)_a  \hat{4} 5\rangle<0,  \langle (AB)_a  5 \hat{6} \rangle<0,  \langle (AB)_a  \hat{6} 1\rangle<0,
	\end{gathered}\\
 \left\{\begin{gathered}
     \langle (AB)_a 1 \hat{2}\rangle, \langle (AB)_a 1 3\rangle, \langle (AB)_a 1 \hat{4}\rangle,\\
     \langle (AB)_a 1 5\rangle, \langle (AB)_a 1 \hat{6}\rangle
 \end{gathered}\right\} \  \ \text{has $3$ sign flip}\,. 
	\end{gathered}
\end{equation}
Remarkably the associated canonical form from the above is the same as the unprojected one \textbf{except} for a change of sign for the triangle integrals at one-loop in eq.~\eqref{eq:one-6pt} and $I^{\text{2mh}}, I^{bt}_j$ at two-loops:
\begin{equation}\label{eq: 2loop6pt}
    \begin{split}
   A^{2{-}loop}_6=\text{LS}_{6,+}\left[\sum_{i=1}^6 -\frac{1}{2} I^{\text{critter}}(i)+I^{\text{crab}}(i)+I^{\text{2mh}}_+ (i)-\sum_{i=1,3,5}I^{bt}_1(i)+\sum_{i=2,4,6}I^{bt}_2(i)\right]\\
   {+}
   \text{LS}_{6,-}\left[\sum_{i=1}^6 -\frac{1}{2} I^{\text{critter}}(i)+I^{\text{crab}}(i)+I^{\text{2mh}}_{\textcolor{red}{-}} (i)\textcolor{red}{+}\sum_{i=1,3,5}I^{bt}_1(i)\textcolor{red}{-}\sum_{i=2,4,6}I^{bt}_2(i)\right]
    \end{split}
\end{equation}
Let's understand this change of sign from two aspects. First, the momentum space image of the $\Pi_6$ is  $\lambda_i\rightarrow (-)^{i{-}1}\lambda_i$. Thus only spinor brackets $\langle ij\rangle$ are affected and not vector inner products. As the only integrand in our local integrand basis that carries spinor brackets are precisely the one-loop triangles and the two-loop box-triangles and (parity odd part of) two-mass hard box (see eq.~\eqref{eq: 1loopLocal} and appendix~\ref{app:local_integral}).

Before moving on to the eight-points, we note that we could have defined the parity operation to act on legs $\{1,3,5\}$ instead. As one can straightforwardly check that $C_{6,\pm}$ interchanges amongst itself.

\subsection{The complete eight-point amplitude}
At eight points, $\Gamma[C_{8}]=4$ we have four solutions for any given cell. The relevant on-shell diagrams are of the form 
$$
\vcenter{\hbox{\scalebox{0.9}{


}}}$$
We would like to define a parity operation that interchanges the solutions of all four BCFW cells $\Gamma_i$. Since the square root for each (two-loop) maximal cuts
\begin{equation}
\langle ii{+}1\rangle\langle i{+}2 i{+}3\rangle\sqrt{-p^2_{i{+}4,i{+}5,i{-}2,i{-}1}},\quad \langle i{+}4i{+}5\rangle\langle i{-}2 i{-}1\rangle\sqrt{-p^2_{i,i{+}1,i{+}2,i{+}3}}\,
\end{equation}
($p_I:=\sum_{i\in I} p_i$), it is easy to see that parity flips $\{ i, i{+}2, i{+}4\}$, $\{ i, i{+}6, i{+}4\}$ introduces the requisite sign. We will choose $\Pi_{8,1}:\;\{2,4,6\}$ and $\Pi_{8,2}:\;\{ 2,6,8\}$. To keep track of how parity acts on the solutions for each cell, we denote the solutions as $\text{LS}_{8,\pm,\pm}[ i]$, where $\text{LS}_{8,+,+}[ i]$ is the positive solution for cell $\Gamma_i$. The remaining three are then defined through the following sequence of projections:
\begin{equation}\label{eq:8-pt LS parity}
\vcenter{\hbox{\scalebox{1}{


}}}
\end{equation}
where $\Pi_{8,1}\Pi_{8,2}$ is equivalent to just flipping $\{4,8\}$, which we will simply denote as $\Pi_{8,3}$. Note that we will use the same operation for all four-cells.

At the eight-point, we have four chambers. The loop-form associated with the positive sector is given for chamber $1\cap 2$ in eq.~\eqref{eq: 8pt1loop} and \eqref{eq: 8pt2loop}. The form for chambers $3\cap 2$, $3\cap 4$, and $1\cap 4$ are related by cyclic rotation. The full amplitude is then given by 
\begin{equation}
A^{2{-}loop}_8=\sum_{i}\mathcal{B}_{i}\times \Omega_i{+}\Pi_{8,1}\left[\mathcal{B}_{i}\times \Omega_i\right]{+}\Pi_{8,2}\left[\mathcal{B}_{i}\times \Omega_i\right]{+}\Pi_{8,3}\left[\mathcal{B}_{i}\times \Omega_i\right]
\end{equation}
where $i=1\cap 2,\, 1\cap 4,\, 3\cap 2,\, 3\cap 4$. In the positive sector, we identify 
\begin{eqnarray}\label{eq: BtoLS}
\mathcal{B}_{1\cap 2}{+}\mathcal{B}_{1\cap 4}=\text{LS}_{8,+,+}[1],\quad \mathcal{B}_{3\cap 2}{+}\mathcal{B}_{3\cap 4}=\text{LS}_{8,+,+}[3],\nonumber\\
\mathcal{B}_{1\cap 2}{+}\mathcal{B}_{3\cap 2}=\text{LS}_{8,+,+}[2],\quad \mathcal{B}_{1\cap 4}{+}\mathcal{B}_{3\cap 4}=\text{LS}_{8,+,+}[4],
\end{eqnarray}
Thus the action of parity operators on $\mathcal{B}_i$ can be read off from their action on $\text{LS}$s. The action of $\Pi_{8,i}$ on the loop-forms is again obtained by computing the canonical form for the region in eq.~\eqref{eq:loop_space} with a subset of twistors replaced by $Z_i\rightarrow \hat{Z}_i=-Z_i$. In the following we give the loop-form associated with chamber $1 \cap 2$, the remaining are obtained by cyclic shifts. 

\noindent\textbf{One-loop forms}:
The one-loop forms for chamber $1\cap 2$ under parity operation $\Pi_{8,1}$ are:
\begin{equation}
    \begin{split}
         \ & I^{L{=}1}_{8}(0)=\sum _{i=1}^{8}(-1)^{i} \, I_{box}( i{-}1,i,i{+}1)\, ,\\
        \ & I_{8}^{L{=}1}(1)=   I_{box}( 1,3,5) +I_{tri}( 1,3,5)  +I_{box}( 5,7,1)-I_{tri}( 5,7,1)\, ,\\
      \  & I_{8}^{L{=}1}(2)=   -I_{box}( 2,4,6)+I_{tri}( 2,4,6) -I_{box}( 6,8,2) -I_{tri}( 6,8,2) \, ,
    \end{split}
\end{equation}
and $\Pi_{8,2}$,
\begin{equation}
    \begin{split}
     \ &    I^{L{=}1}_{8}(0)=\sum _{i=1}^{8}(-1)^{i} \, I_{box}( i{-}1,i,i{+}1)\, ,\\
       \ & I^{L{=}1}_{8}(1):   I_{box}( 1,3,5) -I_{tri}( 1,3,5)  +I_{box}( 5,7,1)+I_{tri}( 5,7,1)\, ,\\
        \ & I^{L{=}1}_{8}(2)=   -I_{box}( 2,4,6)-I_{tri}( 2,4,6) -I_{box}( 6,8,2) +I_{tri}( 6,8,2) \, ,
    \end{split}
\end{equation}
and $\Pi_{8,3}$,
\begin{equation}
    \begin{split}
       \ & I^{L{=}1}_{8}(0)=\sum _{i=1}^{8}(-1)^{i} \, I_{box}( i{-}1,i,i{+}1)\, ,\\
        \ & I^{L{=}1}_{8}(1)=  I_{box}( 1,3,5) -I_{tri}( 1,3,5)  +I_{box}( 5,7,1)-I_{tri}( 5,7,1)\, ,\\
       \ & I^{L{=}1}_{8}(2)=   -I_{box}( 2,4,6)-I_{tri}( 2,4,6) -I_{box}( 6,8,2) -I_{tri}( 6,8,2) \, .
    \end{split}
\end{equation}

\noindent\textbf{Two-loop forms}:
The two-loop forms for chamber $1\cap 2$ under parity operation $\Pi_{8,1}$ is given as,
\begin{equation}
    \begin{split}
         I^{L{=}2}_{8}(0)=\ &\sum _{i=1}^{8} -I_{A}^{db}( i) +I_{B}^{db}( i) +\frac{1}{2} I_{C}^{db}( i)\, ,\\
          I^{L{=}2}_{8}(1)=\ &   I_{\overline{D} ,+}^{db}( 2)+I_{\overline{D} ,-}^{db}( 6)    + I_{\underline{D} ,+}^{db}( 4)+ I_{\underline{D} ,-}^{db}( 8)+ I_{\overline{E} ,-}^{db}( 2)+ I_{\overline{E} ,+}^{db}( 6)\\
        &\ {+}I_{\underline{E} ,-}^{db}( 4){+}I_{\underline{E} ,+}^{db}( 8)  {-}I_{F,-,+}^{db}( 3){-}I_{F,+,-}^{db}( 7) {+} \frac{1}{2} I_{G,-,+}^{db}( 3){+} \frac{1}{2} I_{G,+,-}^{db}( 7)\\
        &\   +I_{\overline{A}}^{bt}( 1)-I_{\overline{A}}^{bt}( 5) -I_{\underline{A}}^{bt}( 1)  +I_{\underline{A}}^{bt}( 5) - I_{B}^{bt}( 3) + I_{B}^{bt}( 7)\, ,\\
         I^{L{=}2}_{8}(2)=\ &  I_{\overline{D} ,+}^{db}( 3)+I_{\overline{D} ,-}^{db}( 7) +I_{\underline{D} ,-}^{db}( 1)+I_{\underline{D} ,+}^{db}( 5)   + I_{\overline{E} ,-}^{db}( 3)+ I_{\overline{E} ,+}^{db}( 7)  \\
        &\ {+}I_{\underline{E} ,+}^{db}( 1){+}I_{\underline{E} ,-}^{db}( 5) {-}I_{F,-,+}^{db}( 4) {-}I_{F,+,-}^{db}( 8) {+} \frac{1}{2} I_{G,-,+}^{db}( 4){+} \frac{1}{2} I_{G,+,-}^{db}( 8)\\
        &\ - I_{\overline{A}}^{bt}( 2)+ I_{\overline{A}}^{bt}( 6) +I_{\underline{A}}^{bt}( 2) -I_{\underline{A}}^{bt}( 6) + I_{B}^{bt}( 4)- I_{B}^{bt}( 8)\, ,\\
    \end{split}
\end{equation}
and $\Pi_{8,2}$,
\begin{equation}
    \begin{split}
         I^{L{=}2}_{8}(0)=\ &\sum _{i=1}^{8} -I_{A}^{db}( i) +I_{B}^{db}( i) +\frac{1}{2} I_{C}^{db}( i)\, ,\\
         I^{L{=}2}_{8}(1)=\ &   I_{\overline{D} ,-}^{db}( 2)+I_{\overline{D} ,+}^{db}( 6)    + I_{\underline{D} ,-}^{db}( 4)+ I_{\underline{D} ,+}^{db}( 8)+ I_{\overline{E} ,+}^{db}( 2)+ I_{\overline{E} ,-}^{db}( 6)\\
        &\ {+}I_{\underline{E} ,+}^{db}( 4){+}I_{\underline{E} ,-}^{db}( 8)  {-}I_{F,+,-}^{db}( 3){-}I_{F,-,+}^{db}( 7) {+} \frac{1}{2} I_{G,+,-}^{db}( 3){+} \frac{1}{2} I_{G,-,+}^{db}( 7)\\
        &\   -I_{\overline{A}}^{bt}( 1)+I_{\overline{A}}^{bt}( 5) +I_{\underline{A}}^{bt}( 1)  -I_{\underline{A}}^{bt}( 5) + I_{B}^{bt}( 3) - I_{B}^{bt}( 7)\, ,\\
         I^{L{=}2}_{8}(2)=\ &   I_{\overline{D} ,-}^{db}( 3)+I_{\overline{D} ,+}^{db}( 7) +I_{\underline{D} ,+}^{db}( 1)+I_{\underline{D} ,-}^{db}( 5)   + I_{\overline{E} ,+}^{db}( 3)+ I_{\overline{E} ,-}^{db}( 7)  \\
        &\ {+}I_{\underline{E} ,-}^{db}( 1){+}I_{\underline{E} ,+}^{db}( 5) {-}I_{F,+,-}^{db}( 4) {-}I_{F,-,+}^{db}( 8) {+} \frac{1}{2} I_{G,+,-}^{db}( 4){+} \frac{1}{2} I_{G,-,+}^{db}( 8)\\
        &\ + I_{\overline{A}}^{bt}( 2)- I_{\overline{A}}^{bt}( 6) -I_{\underline{A}}^{bt}( 2) +I_{\underline{A}}^{bt}( 6) - I_{B}^{bt}( 4)+ I_{B}^{bt}( 8)\, ,\\
    \end{split}
\end{equation}
and $\Pi_{8,3}$,
\begin{equation}
    \begin{split}
         I^{L{=}2}_{8}(0)=\ &\sum _{i=1}^{8} {-}I_{A}^{db}( i) {+}I_{B}^{db}( i) {+}\frac{1}{2} I_{C}^{db}( i)\, ,\\
         I^{L{=}2}_{8}(1)=\ & \sum _{i=2,6}  I_{\overline{D} ,-}^{db}( i)   {+} I_{\overline{E} ,-}^{db}( i){+}\sum _{i=4,8}  I_{\underline{D} ,-}^{db}( i) {+}I_{\underline{E} ,-}^{db}( i) {+}\sum _{i=3,7} {-}I_{F,-,-}^{db}( i) \\
        &\  {+}\sum _{i=3,7} \frac{1}{2} I_{G,-,-}^{db}( i){+}\sum _{i=1,5} I_{\overline{A}}^{bt}( i) {+}I_{\underline{A}}^{bt}( i) {+}\sum _{i=3,7} I_{B}^{bt}( i)\, ,\\
         I^{L{=}2}_{8}(2)=\ & \sum _{i=3,7}  I_{\overline{D} ,-}^{db}( i)   {+} I_{\overline{E} ,-}^{db}( i){+}\sum _{i=1,5}  I_{\underline{D} ,-}^{db}( i) {+}I_{\underline{E} ,-}^{db}( i) {+}\sum _{i=4,8} {-}I_{F,-,-}^{db}( i) \\
        &\ {+}\sum _{i=4,8} \frac{1}{2} I_{G,-,-}^{db}( i){-}\sum _{i=2,6} I_{\overline{A}}^{bt}( i) {+}I_{\underline{A}}^{bt}( i) {-}\sum _{i=4,8} I_{B}^{bt}( i)\, .\\
    \end{split}
\end{equation}

\noindent \textbf{The complete eight-point integrand}:
The complete one/two-loop integrand is 
\begin{equation}
    \begin{split}
   A^{1{-}loop}_8=&\sum_{a,b=\pm,\pm}\left(\text{LS}_{8,a,b}[1]+\text{LS}_{8,a,b}[3]\right)\times \left(\sum _{i=1}^{8}({-})^{i} \, I_{box}( i{-}1,i,i{+}1)\right)\\
   &+\sum_{a,b=\pm,\pm}\sum_{i=1}^4\text{LS}_{8,a,b}[i] \times \Big( I_{a}(i,i{+}2,i{+}4)+I_{b}(i{+}4,i{+}6,i)\Big)\, 
    \end{split}
\end{equation}
with $I_\pm(i,i{+}2,i{+}4):=({-})^{i-1}I_{box}(i,i{+}2,i{+}4)\pm I_{tri}(i,i{+}2,i{+}4)$, and 
\begin{equation}\label{eq: 2loop8pt}
    \begin{split}
   &A^{2{-}loop}_8=\sum_{a,b=\pm,\pm}\Bigg((\text{LS}_{8,a,b}[1]{+}\text{LS}_{8,a,b}[3])\times \sum _{i=1}^{8}\left( {-}I_{A}^{db}( i) {+}I_{B}^{db}( i) {+}\frac{1}{2} I_{C}^{db}( i)\right)\\
   &\  {+}\sum_{i=1}^4\text{LS}_{8,a,b}[i] \times \Big( I_{\overline{D} ,a}^{db}( i{+}1) {+} I_{\overline{D} ,b}^{db}( i{+}5)    {+} I_{\underline{D} ,a}^{db}( i{+}3) {+} I_{\underline{D} ,b}^{db}( i{+}7) {+} I_{\overline{E} ,b}^{db}( i{+}1)  {+} I_{\overline{E} ,a}^{db}( i{+}5)\\
   &    \quad  {+}I_{\underline{E} ,b}^{db}( i{+}3){+}I_{\underline{E} ,a}^{db}( i{+}7)  {-}I_{F,b,a}^{db}( i{+}2){-}I_{F,a,b}^{db}( i{+}6){+} \frac{1}{2} I_{G,b,a}^{db}( i{+}2){+} \frac{1}{2} I_{G,a,b}^{db}( i{+}6){+} ({-})^i  b I_{\overline{A}}^{bt}( i)\\
   & \quad   {+} ({-})^i  a I_{\overline{A}}^{bt}( i{+}4){+}({-})^ia I_{\underline{A}}^{bt}( i){+} ({-})^i b I_{\underline{A}}^{bt}( i{+}4) {+} ({-})^i a I_{B}^{bt}(i{+}2) {+} ({-})^i b I_{B}^{bt}(i{+}6)\Big)\Bigg)\, .
    \end{split}
\end{equation}

%\begin{equation}
%    \begin{split}
%        &A^{1(2){-}loop}_8=\left(\sum_{a,b=\pm,\pm}\text{LS}_{8,a,b}[1]+\text{LS}_{8,a,b}[3]\right)\times I_{8,1(2)}(0)\\
%        &\quad + \sum_{i=1}^4 \Bigg( \text{LS}_{8,+,+}[i]  {+}  \text{LS}_{8,-,+}[i] \Pi_{8,1}   {+} % \text{LS}_{8,-,+}[i] \Pi_{8,1} {+}  \text{LS}_{8,-,-}[i]\Pi_{8,1} \Pi_{8,2} \Bigg) I_{8,1(2)}(i).
%    \end{split}
%\end{equation}

\noindent\textbf{Prescriptive unitarity}

The full eight-point amplitude provides a non-trivial example of how triangulating through $\mathbb{T}_n$ gives a representation that satisfies prescriptive unitarity. More precisely, the resulting form consists of integrands that evaluate to 1 on one of the maximal cut solutions, and zero on the others. Let's take the two-loop maximal cut as an example, which involves 4 solutions.  Consider the cut 
\begin{equation}
\langle AB 12 \rangle =\langle AB 34 \rangle =\langle AB 56 \rangle =\langle CD 56 \rangle =\langle CD 78 \rangle =\langle CD 12 \rangle =0.
\end{equation}
The contributing integrand consist of $I_{G,\pm,\pm}^{db}( 4)$ which appears in the loop-form associated with chambers $[1\cap2]$ and $[3\cap2]$. The cut solutions are
\begin{equation}
    \begin{split}
        &Z_A=Z_1+\alpha Z_2, \quad Z_B=\beta Z_3 +Z_4,\\
        &Z_C=Z_5+\gamma Z_6, \quad Z_D=\delta Z_7 +Z_8,\\
    \end{split}
\end{equation}
with 
\begin{equation}
    \begin{split}
        &\alpha_{\pm}=-\frac{\langle1356\rangle}{\langle2356\rangle}\pm \frac{\sqrt{-\langle1234\rangle\langle3456\rangle\langle5612\rangle}}{\langle\!\langle 24\rangle\!\rangle \langle2356\rangle},\quad \beta=-\frac{\langle1456\rangle+\alpha \langle2456\rangle}{\langle1356\rangle+\alpha \langle2356\rangle},\\
        &\gamma_{\pm}=-\frac{\langle1257\rangle}{\langle1267\rangle}\pm\frac{\sqrt{-\langle1256\rangle\langle5678\rangle\langle7812\rangle}}{\langle\!\langle68\rangle\!\rangle\langle1267\rangle}, \quad \delta=-\frac{\langle1258\rangle+\gamma \langle1268\rangle}{\langle1257\rangle+\gamma \langle 1267\rangle}.
    \end{split}
\end{equation}
We denote the cut solution by $\text{cut}_{\pm,\pm}$ according to the sign of $\alpha_\pm$, $\gamma_\pm$. One can straightforwardly check that 
\begin{equation}
\left. I_{G,\pm,\pm}^{db}( 4)\right|_{\text{cut}_{a,b}}=\delta_{\pm,a}\delta_{\pm,b}\,,\quad\quad
\end{equation}
as required.

Equipped with the explicit cut solution, one can further verify that the leading singularity from the cut, matches associated tree-form $\mathcal{B}$. Using the map in eq.~\eqref{eq: BtoLS}, we see that the tree form associated with $\mathcal{B}_{1\cap 2}{+}\mathcal{B}_{3\cap 2}$ is $\text{LS}_{8,+,+}[2]$. The action of $\Pi_{8,1}$, $\Pi_{8,2}$ and $\Pi_{8,3}$ then leads to $\text{LS}_{8,-,+}$, $\text{LS}_{8,+,-}$ and $\text{LS}_{8,-,-}$  respectively. Now let's consider the gluing of the five tree amplitudes in the cut:
$$
\vcenter{\hbox{\scalebox{0.8}{


}}}$$
Substituting the four cut-solutions one indeed finds:
\begin{equation}
    \begin{split}
        &\frac{1}{J}\left.\int\prod_{k_i} d\eta_{k_i} \frac{\delta^{(6)}(Q_1)}{\langle23\rangle\langle3 k_2\rangle} \frac{\delta^{(6)}(Q_2)}{\langle45\rangle\langle5 k_3\rangle} \frac{\delta^{(6)}(Q_3)}{\langle67\rangle\langle7 k_5\rangle}\frac{\delta^{(6)}(Q_4)}{\langle81\rangle\langle1 k_6\rangle} \frac{\delta^{(6)}(Q_5)}{\langle k_3 k_4\rangle\langle k_4 k_6\rangle}\right|_{\text{cut}_{\pm,\pm}}\\
        =& \ \text{LS}_{8,\pm,\pm}[2]\, .
    \end{split}
\end{equation}
with Jacobian factor $J=\sqrt{-p_{2,3}^2 p_{4,5}^2 p_{6,7,8,1}^2}\sqrt{-p_{6,7}^2p_{8,1}^2 p_{2,3,4,5}^2}$. Therefore, the tree form exactly matches the leading singularity.

\section{Conclusion and outlook}

In this paper, we have proposed certain new positive geometries relevant for the $n$-point all-loop planar integrand of ABJM theory. The canonical form of the all-loop ABJM amplituhedron can be obtained in three steps:
\begin{itemize}
\item With an overall sign flip, we obtain the tree positive region $\mathbb{T}_n$ by taking the dimensional reduction to $D=3$ of that of $n$-point sYM amplituhedron, which turns out to be non-empty only for middle sector $k=\frac n 2-2$; restricted on a suitable $3k$-dim subspace, we conjecture that it gives the tree ABJM amplituhedron, related via T-duality to the ABJM momentum amplituhedron of~\cite{He:2021llb, Huang:2021jlh}.
\item By reducing the loop part to $D=3$ in exactly the same way, we conjecture that the geometry is the all-loop ABJM amplituhedron; concretely we use intersections of all possible BCFW cells (chambers) to triangulate the $3L$-dim loop geometry and obtain a unique loop form for any points in a given chamber. 
\item By wedging the form of each tree chamber with the $L$-loop form, which turns out to be given by local integrals, and summing over all chambers, we obtain the integrand of the positive sector; we further include all the $2^k$ sectors ($k=1,2$ for $n=6,8$) related to the positive one by parity operations, and that gives the complete results of ABJM $L$-loop integrands at $n=6,8$. 
\end{itemize}
We have provided strong evidence for these claims, {\it e.g.} by showing how geometries guarantee all-loop vanishing cuts, soft cuts, and most importantly unitarity cuts.  We have triangulated loop geometry based on tree chambers and computed canonical forms for $L=1$, $n=6,8,10$ (which involves summing over $1, 4, 25$ chambers, respectively), and  for $L=2$, $n=6,8$; in all cases, we obtain correct loop integrands, and the method is rather efficient. We have also shown that the same bipartite structures of multi-loop negative geometries seen earlier for $n=4$~\cite{He:2022cup} extend nicely to all multiplicities and plays an important role in our two-loop computations. Perhaps the most immediate next step is to go to higher multiplicities and loops: we expect to triangulate $L=2$ geometries for $n\geq 10$ once tree chambers are systematically understood; going to $L=3$ for at least $n=6$ poses no more difficulties with bipartite structures the key in simplifying the computations. 

There are many open questions to be clarified:
\begin{itemize}
    
    \item \textbf{Chambers}: Currently, we have used BCFW cells and their overlap to triangulate  $\mathbb{T}_n$ in ABJM. Such triangulation is in a sense ``hand-picked" by the loop geometry, where each region yields a unique loop-form at given loop order. An obvious question is then whether such BCFW triangulation is sufficient for all loops. The question can be rephrased as whether all $n{-}3$ dimensional cells in the momentum orthogonal Grassmannian corresponds to BCFW cells. Note that we have a negative answer for the  corresponding question for $\mathcal{N}=4$ sYM: indeed the four-mass box leading singularity at N${^2}$MHV~\cite{Arkani-Hamed:2012zlh}
    
    $$\includegraphics[scale=0.4]{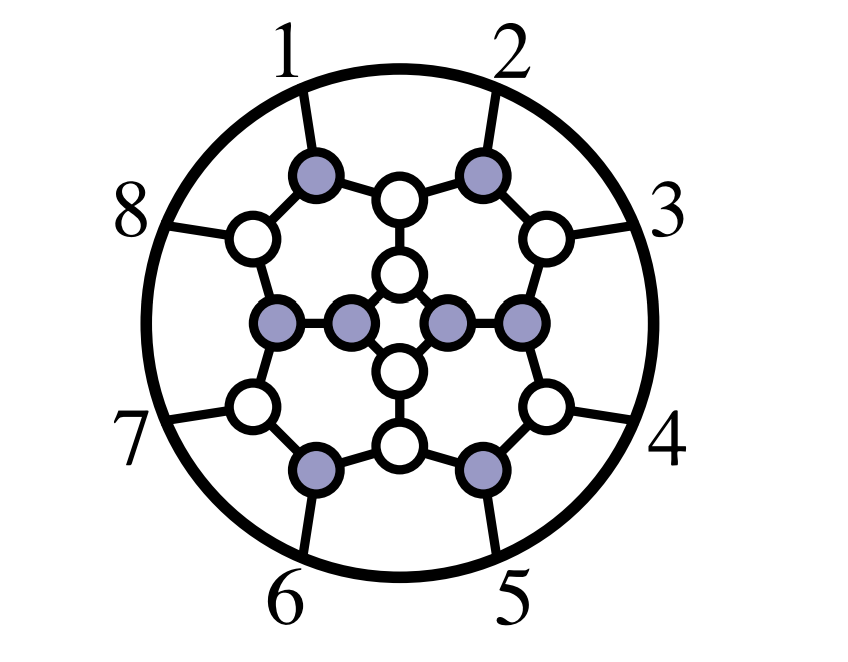}\,,$$
    is not a BCFW cell. 
    \item \textbf{Prescriptive unitarity for $\mathcal{N}=4$ sYM}: Note that triangulations of the loop amplituhedron using chambers of $\mathbb{T}_{n,k}$ can be straightforwardly applied to ${\cal N}=4$ sYM. We will use the one-loop 6-point NMHV amplitude as an example. It is known that all $2n{-}4$ cells for this configuration are BCFW cells, and hence we expect that the chambers of NMHV $\mathbb{T}_{6,1}$ are comprised of the overlap of the BCFW cells. Denoting the BCFW cells as $[\,i\,]$, the tree amplitude is given by 
\begin{equation}
    [1]\cup [3]\cup [5]=[2]\cup [4]\cup [6].
\end{equation}
Therefore, NMHV $\mathbb{T}_{6,1}$ can be partitioned into nine chambers:
$$
\vcenter{\hbox{\scalebox{1.2}{


}}}$$

\noindent
The canonical form of the overall geometry is equal to the sum of the canonical forms of these nine chambers, each with its own loop geometry. The canonical form for the chamber $1\cap 2$ can be presented in the following local form,  
\begin{equation}
    [p\cap q]\times \left( \sum_{\scalebox{0.75}{$\begin{array}{c} i=\{p,q\} \\ j=\{1,2\}\end{array}$}}I^{\text{1mb}}_j(i)+I^{\text{2mh}}_j(i)+\sum_{i=1}^6 I^{\text{tri}}(i)\right)\quad .
\end{equation}
Here $I^{\text{1mb}}_j(i)$ and $I^{\text{2mh}}_j(i)$ refer to the one-mass box and two-mass-hard integrals, which we list in appendix~\ref{app:local_integral} for completeness. Using that $[1]=[1\cap 2]+[1\cap 4]+[1\cap 6]$, and $[2]=[1\cap 2]+[3\cap 2]+[5\cap 2]$, summing with the other eight chambers, we arrive at the following NMHV$_6$ one-loop  integrands
\begin{equation}
 A^{1{-}loop}_6=\sum_{i=1}^6 [i] \times \left(\sum_{j=1,2}I^{\text{1mb}}_j(i)+I^{\text{2mh}}_j(i) {+}\sum_{k=1}^6 \frac{1}{2}I^{\text{tri}}(k)\right).
\end{equation}
Note that all leading singularities are present, reflecting the fact such triangulations leads to representation satisfying prescriptive unitarity. 

What is more interesting is the eight-point N${^2}$MHV case, where the four-mass-box leading singularity has $\Gamma[C]=2$. Since we only expect a single positive branch for $\mathcal{N}=4$, the two solutions must participate in the dissection of $\mathbb{T}_{8,2}$. It will be interesting to see how prescriptive unitarity is realized through proper chamber analysis for $\mathbb{T}_{8,2}$~\cite{sYMchamber}.    

\item \textbf{Soft-collinear singularities}: The geometric definition of the loop integrand renders many physical properties manifest, including the recursive nature of soft-cuts, the correct factorization of maximal cuts as well as vanishing cuts. However, there is a glaring exception, the absence of non-factorizable soft-collinear singularities. In fact, it is this property, combined with the previously mentioned cuts, that fixes the integrand at two-loops up to eight point~\cite{Caron-Huot:2012sos, He:2022lfz}. It will be interesting to understand this further. 

\item \textbf{Parity operation for general $n$:} As we have discussed for $k=\frac{n}{2}{-}2$, we need $k$ parity operators. We have found the correct operators for $k=1,2$. However, when generalizing to the higher multiplicities, some situations where changing the sign on $Z_i$s does not interchange the solutions. For example, at $n=12$ the following maximal cut 
$$
\vcenter{\hbox{\scalebox{1}{


}}}$$
its square root is simply 
\begin{equation}
    \sqrt{-p_{1,2,3,4}^2\,p_{5,6,7,8}^2\,p_{9,10,11,12}^2}
\end{equation}
which could not be written as products of spinor brackets. It is important to figure out how to modify and  identify the parity operators in general and how their inclusion can be understood as a necessity from a geometric point of view. 

%\item \textbf{T duality map and its inverse.} To avoid issues related to the dimension crossing of the two geometries as $k$ is varied, it may be more appropriate to focus on the positive tree region when exploring the relationship between the momentum amplituhedron and the amplituhedron. The behavior of collinear poles in the amplituhedron under T-duality maps is a fascinating subject of study. In particular, it is intriguing to observe how these poles disappear under the map. Moreover, comparing the geometries of sYM and ABJM under such maps is of interest, since the latter does not possess collinear poles even in the momentum amplituhedron. 
    
\item \textbf{The ABJM tree amplituhedron:} Finally, the existence of loop amplituhedron for ABJM suggests that there is a well-defined subspace for $\mathbb{T}_n$ whose canonical form gives the tree amplitude. It would be nice to have this construction. 

This would also help address our observation that the loop canonical form constructed from a given block in $\mathbb{T}_n$ only exhibits local poles associated with on-shell diagrams that defined the block. For $\mathcal{N}=4$ sYM, such phenomenon can be understood from the fact that the amplituhedron subspace is defined as:
    \begin{equation}
    \mathcal{Y}=\mathcal{C}\cdot \mathcal{Z}, \quad \mathcal{C}=
\left(\begin{array}{c} C \\ D_{\ell_1} \\ D_{\ell_2}\\ \vdots\end{array}\right)
    \end{equation}
    where $(C, D_{\ell_i})$ are the tree and loop parameterization respectively. Since all ordered  $(k{+}2\ell)\times (k{+}2\ell)$ minors of $\mathcal{C}$ must be positive, it is natural that the positive cells of $C$ dictates the allowed boundaries for the associated form in $D$, as the former must appear as the boundary of the form in $D$. 
    
\item \textbf{BDS geometry:} As observed for two-loops $n=4$~\cite{Chen:2011vv}, $n=6$~\cite{Caron-Huot:2012sos}, and $n=8$~\cite{He:2022lfz}, the infrared divergence of the two-loop amplitude is fully captured by the BDS ansatz of $\mathcal{N}=4$ sYM~\cite{Bern:2005iz}. Furthermore, the BDS piece is fully captured by a set of local integrands that are characterized by the presence of IR divergence and free of all unphysical cuts. Given that our triangulation exposes the possibility of local triangulation of the amplituhedron, it would be interesting to see if one can characterize a subregion of $\mathbb{L}_n$ that gives loop integrands for the BDS piece.   
\end{itemize}

\section*{Acknowledgement}
It is our great pleasure to thank Nima Arkani-Hamed, Jacob Bourjaily, Tomasz Lukowski, Matteo Parisi, Justinas Rumbutis, Jonah Stalknecht, Jaroslav Trnka,  Congkao Wen, Akshay Yelleshpur, Shun-Qing Zhang, and Yaoqi Zhang for stimulating discussions. The authors would also like to thank DIAS for hosting the conference ``Amplituhedron at 10"  where the results of this paper were discussed. SH thanks IAS (Princeton) for hospitality during the completion of the work; his research is supported in part by National Natural Science Foundation of China under Grant No. 11935013, 12047502, 12047503, 12247103, 12225510. Y-t H and C-k Kuo are funded by the National Science and Technology Council of Taiwan, Grant No. 111-2811-M-002 -125.

\appendix

\section{Momentum-twistor Grassmannian formula for ABJM}\label{sec:Tdual}
This appendix provides a comprehensive review of the Grassmannian formula for ABJM in momentum twistor space~\cite{Elvang:2014fja}. As discussed in section~\ref{sec:loop_geo} and the outlook,  the complete tree geometry has not yet been established, and this formula may be crucial for such a definition. It is possible that uplifting the formula to the product map with $Z$ could yield a subspace of the tree region, akin to the original amplituhedron in ${\cal N}=4$ sYM. Below we have gathered some essential properties of this formula%,which could be useful in defining the complete tree amplituhedron in future studies
.

The momentum-twistor space Grassmannian formula is subject to the delta functions:
\begin{equation}\label{eq:twisor_delta_function_3d}
        \sum_{j=1}^n \;\mathcal{D}_{\bar{\alpha},j}Z_j=0, \quad \sum_{i,j=1}^n\tilde{g}_{i,j} \mathcal{D}_{\bar{\alpha},i} \mathcal{D}_{\bar{\beta},j}=0, \quad \text{for}\  \bar{\alpha},\bar{\beta}=1,\cdots,k\,.
\end{equation}
The second constraint is the orthogonal condition defining $\widetilde{\text{OG}}(k,2k{+}4)$. Here we use $\widetilde{\quad}$ to indicate the orthogonal condition for the Grassmannian is defined with respect to a kinematic dependent $n\times n$ metric, i.e. $\tilde{g}=[\tilde{g}]_{n\times n}$ is defined by $\tilde{g}_{i,j}\equiv g_{i,j}/({(-)^k}\det{[g]_{n\times n}})^{1/(k+1)}$ with $g_{i,j}=\text{sgn}[j-i]\langle\!\langle i j \rangle\!\rangle$, $\det{[g]_{n\times n}}= 2^{2k}\prod_{i=1}^n\langle\!\langle i\, i{+}2 \rangle\!\rangle$.

The $\mathcal{R}$-invariant-like ABJM amplitude (in which a prefactor has been extracted from the amplitude)  can be defined as the contour integral over the Grassmannian $G(k,n)$:
\begin{equation}\label{eq:3d_twistor_integral}
    \mathcal{R}_n=\int_\gamma \frac{d^{k\times n }\mathcal{D}_{\bar{\alpha},j}}{\text{Vol}(\text{GL}(k))} \frac{1}{m_1 m_2 \cdots m_k} \delta^{k (k+1)/2}\left (\mathcal{D}\cdot \tilde{g}\cdot \mathcal{D}^T\right)  \delta^{4k}(\mathcal{D}\cdot Z) \delta^{3k}(\mathcal{D}\cdot \chi),
\end{equation}
 where $m_l$ represent the $l$-th consecutive minors:
 \begin{equation}
     m_l\equiv\epsilon_{i_1,i_2,\cdots, i_{k}}\mathcal{D}_{l+1}^{i_1} \mathcal{D}_{l+2}^{i_2}\cdots \mathcal{D}_{l+k}^{i_{k}}.
 \end{equation}
To localize the integral \eqref{eq:3d_twistor_integral}, the matrix $\mathcal{D}$ needs to be restricted to $4k$ cells (after solving the orthogonal condition).

Compared to the original formula in the momentum space, the matrix $\mathcal{D}$ is projected out $\lambda$-plane. In momentum space, We fix the first two rows of $C$ matrix to be $(-1)^{i-1} \lambda$
\begin{equation}
    C=\begin{pmatrix}
    \lambda_1^1 & -\lambda_2^1 & \cdots &\lambda_{n-1}^1 &-\lambda_n^1\\
    \lambda_1^2 & -\lambda_2^2 & \cdots &\lambda_{n-1}^2 &-\lambda_n^2\\
    c_{1,1} & c_{1,2} & \cdots &c_{1,n-1} &c_{1,n}\\
    \vdots & \vdots & \ddots & \vdots & \vdots\\
    c_{k,1} & c_{k,2} & \cdots &c_{k,n-1} &c_{k,n}
    \end{pmatrix},
\end{equation}
and  it is subject to the conditions:
\begin{equation}\label{eq:spinor_delta_function_3d}
    \sum_{j=1}^n C_{\alpha,j}\lambda_j=0,\quad \sum_{j=1}^n (-1)^{j-1} C_{\alpha,j} C_{\beta,j}=0, \quad \text{for}\  \alpha,\beta=1,\cdots,k+2\,.
\end{equation}
For $\alpha$ and $\beta$ equal to 1 or 2, the delta function conditions are automatically satisfied as a consequence of momentum conservation, $\sum_i (-1)^{i-1} \lambda_i^a \lambda_i^b = 0$.  The integral representation of amplitude in $\lambda$-space is 
\begin{equation}
    \begin{split}
        \mathcal{L}_n=\int_\gamma \frac{d^{(k+2)\times n}C_{\alpha,i}}{\text{Vol}(\text{GL}(k{+}2))} \frac{1}{M_1 M_2 \cdots M_k} &\delta^{(k+2)(k+3)/2}(C\cdot \left[{(-1)}^a\delta_{a,b}\right]_{n\times n}  \cdot C^T)\\
        &\quad \delta^{2(k{+}2)}(C\cdot \lambda) \delta^{3(k{+}2)}(C\cdot \eta).
    \end{split}
\end{equation}
where $M_l$ represent the $l$-th consecutive minors:
 \begin{equation}
     M_l\equiv\epsilon_{i_1,i_2,\cdots, i_{k+2}}C_{l}^{i_1} C_{l+1}^{i_2}\cdots C_{l+k+1}^{i_{k{+}2}}.
 \end{equation}

The $C$ matrix and $\mathcal{D}$ matrix are related by the projection matrix $Q_{a,b}$, given by
\begin{equation}
    Q_{a,b}=\frac{\langle b{-}1\, b{+}1\rangle\delta_{a,b}+\langle b\, b{-}1\rangle \delta_{a,b+1}+\langle b{+}1\, b\rangle \delta_{a,b-1}}{\langle b{+}1\, b\rangle \langle b\, b{-}1\rangle}
\end{equation}
through
\begin{equation}
    \mathcal{D}_{\bar{\alpha},j} = \sum_{i=1,\cdots,n} (-1)^{i+j-1} C_{\bar{\alpha},i}  Q_{i,j}.
\end{equation}
It's worth noting that the projection through $Q_{a,b}$ preserves the cell structure and transforms the positive matrix $C$ into a positive matrix $\mathcal{D}$.
Additionally, there is a relationship between the minors of $C$ and $\mathcal{D}$, and  we define the ratio of them to be
\begin{equation}
J_l\equiv\frac{M_l}{m_l}=\langle l, l{+}1\rangle \langle l{+}1, l{+}2\rangle\cdots \langle l{+}k, l{+}k{+}1\rangle.
\end{equation}

The two integral formulas, $\mathcal{R}_n$ and $\mathcal{L}_n$, are related up to a Jacobian factor and delta functions that enforce momentum conservation and super momentum conservation, 
\begin{equation}
    \mathcal{L}_n=\delta^3(p)\delta^{(6)}(Q)\frac{1}{\prod_{i=1,\cdots,k+2}J_i}\mathcal{R}_n.
\end{equation}
Specifically, for $n=6$
\begin{equation}
    \mathcal{L}_6=\frac{\delta^3(p)\delta^{(6)}(Q)}{\langle 12\rangle\langle 23\rangle^2\langle 34\rangle^2\langle 45\rangle} \int \frac{d^{6} \mathcal{D}_i}{\text{Vol}(\text{GL}(1))}\frac{1}{\mathcal{D}_2\mathcal{D}_3\mathcal{D}_4} \delta\left (\mathcal{D}\cdot \tilde{g}\cdot \mathcal{D}^T\right)  \delta^{4}(D\cdot Z) \delta^{3}(D\cdot \chi),
\end{equation}
In momentum space, since the minors and their conjugates are equal, the minors of the positive matrix $\mathcal{D}$ and their conjugates are closely related. This allows us to choose alternative sets of minors in the denominator of the Grassmannian formula to simplify the Jacobian factor. Specifically, we can write
\begin{equation}
    \mathcal{L}_6= \frac{\delta^3(p)\delta^{(6)}(Q)}{\prod_i\langle i\, i{+}1\rangle}\int \frac{d^{6} \mathcal{D}_i}{\text{Vol}(\text{GL}(1))}\frac{1}{\mathcal{D}_2\mathcal{D}_4\mathcal{D}_6} \delta\left (\mathcal{D}\cdot \tilde{g}\cdot \mathcal{D}^T\right)  \delta^{4}(D\cdot Z) \delta^{3}(D\cdot \chi).
\end{equation}

To facilitate the comparison between the two formulas,   we consider the component amplitudes. For $n=6$, we have
\begin{equation}
   \frac{1}{2} \left.\frac{M_1^3}{M_1 M_2 M_3}\right|_{C=C^*}=\frac{1}{2} \left.\frac{J_1^3}{\prod_{i=1,2,3} J_i}\frac{1}{m_1 m_2 m_3}\right|_{D=D^*};
\end{equation}
and for $n=8$ in the cell $M_1=0$ ($m_1=0$) for $C$ ($\mathcal{D}$),
\begin{equation}
    \frac{1}{4\sqrt{-p_{1,2,3,4}^2}}\left.\frac{M_3^3}{M_2 M_3 M_4}\right|_{C=C^*_{M_1=0}}=\frac{J_1}{4\sqrt{-p_{1,2,3,4}^2}} \left.\frac{1}{\prod_{i=1,2,3,4} J_i}\frac{m_3^3}{m_2 m_3 m_4}\right|_{D=D^*_{m_1=0}}.
\end{equation}
In the equation above, the terms $\left(4\sqrt{-p_{1,2,3,4}^2}\right)^{-1}$ and $J_1\left(4\sqrt{-p_{1,2,3,4}^2}\right)^{-1}$ represent the Jacobian factors resulting from localizing the delta function in eqs.~\eqref{eq:spinor_delta_function_3d} and \eqref{eq:twisor_delta_function_3d}, respectively.

We can see that the selection of minors from the $\mathcal{D}$ matrix has a significant impact on the overall pulling-out factor. Opting for a more symmetric approach, such as choosing each minor to be half the size, can yield favorable results:
\begin{equation}
    \begin{split}
        \mathcal{R}_n=\int_\gamma \frac{d^{k\times n }\mathcal{D}_{\bar{\alpha},j}}{\text{Vol}(\text{GL}(k))} \frac{1}{\sqrt{m_1 m_2 \cdots m_n}} \delta^{k (k+1)/2}\left (\mathcal{D}\cdot \tilde{g}\cdot \mathcal{D}^T\right)  \delta^{4k}(D\cdot Z) \delta^{3k}(D\cdot \chi),
    \end{split}
\end{equation}
with prefactor $1/(\langle12\rangle\langle23\rangle\cdots \langle n1\rangle)^{(k+1)/2}$ such that
\begin{equation}
    \mathcal{L}_n=\delta^3(p)\delta^{(6)}(Q)\left(\frac{1}{\langle12\rangle\langle23\rangle\cdots \langle n1\rangle}\right)^{(k+1)/2}\mathcal{R}_n.
\end{equation}

\section{Sectors of $\mathbb{T}_n$ for even $n$}\label{sec:AppendixTn}
In Section \ref{sec:tree_geo} we have shown that  $\mathbb{T}_n$ is an empty set when $n$ is odd; here we would like to demonstrate that for even $n$ $\mathbb{T}_n$ only exists for the middle sector with $k=\frac{n}{2}-2$.

Recall that in $D=3$ we can factor out a common factor $\langle\!\langle a{-}1\, a{+}1\rangle\!\rangle$ for each sign-flip sequence $\big\{\langle a{-}1\, a\, a{+}1\, j\rangle{=}\langle\!\langle a{-}1\, a{+}1\rangle\!\rangle\langle\!\langle a\, j\rangle\!\rangle\big\}$, without affecting the sign-flip number. Therefore, we can translate the sign-flipping condition on the sequences of double two-brackets by using twisted cyclic symmetry with $ Z_{n+i}\equiv(-1)^{k-1}Z_i \equiv \hat{Z}_{i}$:
\begin{equation}
    \big\{\langle\!\langle a j\rangle\!\rangle\big\}_{j= a+2}^{a+n-2}\quad \text{for $a=1,\cdots, n$.}
\end{equation}

It is useful to arrange the sign-flip sequence into a zigzag shape, as shown in eq.~\eqref{eq:sequence_6pt}.  The definite sign of the boundaries ensures that the minor of every $2\times 2$ adjacent block is strictly negative:
\begin{equation}
\cdots\ 
    \begin{array}{ccccccccc}
     \langle\!\langle a\,i  \rangle\!\rangle &\quad  & \langle\!\langle a\,i{+}1  \rangle\!\rangle      \\
    \langle\!\langle a{+}1\,i  \rangle\!\rangle & & \langle\!\langle a{+}1\,i{+}1  \rangle\!\rangle   
     \end{array}
     \ \cdots \quad  \Rightarrow\ 
     \begin{array}{c}
         \langle\!\langle a\, i\rangle\!\rangle \langle\!\langle a{+}1\, i{+}1\rangle\!\rangle{-} \langle\!\langle a\, i{+}1\rangle\!\rangle \langle\!\langle a{+}1\, i\rangle\!\rangle\\
         = \langle a\, a{+}1\, i\, i{+}1\rangle<0.
     \end{array}
\end{equation}
This also implies that $2\times 2$ adjacent blocks with the following signs are forbidden:
\begin{equation}
    \begin{pmatrix}
    + & &- \\
    + & &+ 
    \end{pmatrix},\ 
    \begin{pmatrix}
    + & &+ \\
    - & &+ 
    \end{pmatrix},\ 
    \begin{pmatrix}
    - & &- \\
    + & &- 
    \end{pmatrix},\ 
    \begin{pmatrix}
    - & &+ \\
    - & &- 
    \end{pmatrix}.
\end{equation}

We will see that for even $n$, only the middle $k$ is a nonempty set in $\mathbb{T}_n$, which be easily shown in the case of $n=6$: 
\begin{equation}\label{eq:sequence_6pt}
    \begin{array}{ccccccccc}
     \langle\!\langle 13  \rangle\!\rangle &\quad \quad & \langle\!\langle 14  \rangle\!\rangle  &\quad\quad   &\textcolor{red}{\langle\!\langle 15 \rangle\!\rangle} \\
    + &  & &  &\ \  ({-})^{k}\\
     & & \textcolor{blue}{ \langle\!\langle 24  \rangle\!\rangle} &  &\textcolor{violet}{\langle\!\langle 25  \rangle\!\rangle} &\quad \quad &\textcolor{blue}{\langle\!\langle 26  \rangle\!\rangle}\\
      & &- & &? & &\ \  ({-})^{k+1}\\
      & & & &\textcolor{red}{\langle\!\langle 35  \rangle\!\rangle} & &\langle\!\langle 36  \rangle\!\rangle &\quad \quad &\langle\!\langle 3\hat{1}  \rangle\!\rangle\\
      & & & &+ & & & &\ \  ({-})^{k}
     \end{array}
\end{equation}
 The sequences  $\{\langle\!\langle24\rangle\!\rangle,\langle\!\langle25\rangle\!\rangle,\langle\!\langle26\rangle\!\rangle\}$ and $\{\langle\!\langle51\rangle\!\rangle,\langle\!\langle52\rangle\!\rangle,\langle\!\langle53\rangle\!\rangle\}$ always have a conjugate sign flip ($k_1+k_2=n-4$), regardless of the sign of $\langle\!\langle25\rangle\!\rangle$. Therefore, the middle sector $k=n/2-2=1$ is the only allowed one.\\

Moving to the case $n=8$, we can always show that if a sign-flip sequence is given for the rows, a column sequence with the conjugate sign-flip number exists in this zigzag arrangement, for any non-middle $k$. Specifically, when $k=0$: 
\begin{equation}
\vcenter{\hbox{\scalebox{1}{

}}}
\end{equation}
By requiring that the sequences $\{\langle\!\langle 1i\rangle\!\rangle\}$, $\{\langle\!\langle 2i\rangle\!\rangle\}$, and $\{\langle\!\langle 3i\rangle\!\rangle\}$  have zero sign-flips, each sequence becomes monotonic in sign. This, in turn, forces the sequence $\{\langle\!\langle 6i\rangle\!\rangle\}$  to be alternating, with the maximum number of sign-flips.

When $k=4$:
\begin{equation}
\vcenter{\hbox{\scalebox{1}{
%
}}}
\end{equation}
Similar to its conjugate case $k=0$, imposing that the three sequences $\{\langle\!\langle 1i\rangle\!\rangle\}$, $\{\langle\!\langle 2i\rangle\!\rangle\}$, and $\{\langle\!\langle 3i\rangle\!\rangle\}$  have the maximum number of sign-flips results in each sequence becoming alternating in sign. As a consequence, the sign-flip number of the sequence $\{\langle\!\langle 6i\rangle\!\rangle\}$  is guaranteed to be zero.

When $k=1$,  it is more convenient to divide it into two cases according to the sign of $\langle\!\langle26\rangle\!\rangle$:
\paragraph{Case 1: } $\langle\!\langle 26\rangle\!\rangle>0$
\begin{equation}
\vcenter{\hbox{\scalebox{1}{

}}}
\end{equation}
Requiring the sequence $\{\langle\!\langle 2 i\rangle\!\rangle\}$  has one sign-flip, $\langle\!\langle27\rangle\!\rangle$ is determined to be negative. Using the constraint that any $2\times 2$ block's minor is negative,  the sign of $\langle\!\langle 16\rangle\!\rangle$ is then determined to be positive. Consequently, the sequence $\{\langle\!\langle 6 i\rangle\!\rangle\}$ has $3$ sign-flips.

\paragraph{Case 2: } $\langle\!\langle 26\rangle\!\rangle<0$

\begin{equation}
\vcenter{\hbox{\scalebox{1}{
%
}}}
\end{equation}
This case $\langle\!\langle25\rangle\!\rangle$ is determined to be negative and $\langle\!\langle36\rangle\!\rangle$ to be positive, which is deduced from the sequence $\{\langle\!\langle 2i\rangle\!\rangle\}$ has one sign flips and any minor of $2\times2$ block is negative. We conclude that the sign-flip number of sequence $\{\langle\!\langle 6i\rangle\!\rangle\}$ is $3$.

When $k=3$, we can perform a similar analysis as for $k=1$, and conclude that the column sequence $\{\langle\!\langle 6i \rangle\!\rangle\}$ has only one sign flip. The corresponding zigzag diagrams and their sign are displayed below:
\paragraph{Case 1: } $\langle\!\langle 26\rangle\!\rangle>0$
\begin{equation}
\vcenter{\hbox{\scalebox{1}{

}}}
\end{equation}
\paragraph{Case 2: } $\langle\!\langle 26\rangle\!\rangle<0$
\begin{equation}
\vcenter{\hbox{\scalebox{1}{
%
}}}
\end{equation}

We have also performed similar analysis for the sequences $\{\langle\!\langle ai\rangle\!\rangle\}$ at $n=10$ and $n=12$, and found that given a sequence has $k$ sign-flips it is impossible to require all $2\times 2$ block minors to be positive for any $k\neq n/2-2$. This result strongly suggests that only the middle sector with $k=n/2-2$ has a non-empty positive tree region $\mathbb{T}_n$ for any even $n$.

\paragraph{Claim} The positive tree region $\mathbb{T}_n$ exists only for even  $n$ and is located at the middle sector $k=\frac{n}{2}-2$.\\

\paragraph{Positive kinematics for 3d sYM} 
As emphasized, the negative sign of $\langle i i{+}1 j j{+}1\rangle <0$ plays a crucial role in excluding non-physical kinematic configurations in ABJM. If we were to stay with positive $\langle i i{+}1 j j{+}1\rangle >0$, one could find solutions for all $n$ and $k$. This is not surprising, since one would be just considering the geometry in a three-dimensional subspace, {\it i.e.} 3d sYM.

It is easy to see that solutions exist for all $n$ and $k$. If we revert back to $\langle i i{+}1 j j{+}1\rangle >0$, then the conditions~\eqref{eq:consecutive twistors} will be modified to   consecutive even and odd brackets have a uniform sign:
\begin{equation}
    \begin{split}
        &\langle\!\langle 13 \rangle\!\rangle, \ \langle\!\langle 35 \rangle\!\rangle, \ \cdots,\  (-)^{k-1}\langle\!\langle n{-}1 1\rangle\!\rangle\gtrless0,\\
        &\langle\!\langle 24 \rangle\!\rangle, \ \langle\!\langle 46 \rangle\!\rangle, \ \cdots,\  (-)^{k-1}\langle\!\langle n 2\rangle\!\rangle\gtrless0.
    \end{split}
\end{equation}
Thus we no longer have those contradictions that lead to the absence of kinematic configurations for odd $n$ or $k\neq \frac{n}{2}-2$. 
%\textbf{T-duality map for 3d Yang-Mills...}

Flipping the sign of $\langle i i{+}1 j j{+}1\rangle$ results in a consistent way of satisfying symplectic conditions for all $n$ and $k$. Nonetheless, it seems that such a modification makes it difficult to extend the kinematic region to loops. One possible reason for this is that 3d sYM is no longer dual conformal. Also, it remains an open question whether a subspace exists within this positive 3d kinematic region that is directly related to the dimensional reduction sYM amplitude. This interesting problem is left for future investigation.

\section{Details on two-particle cuts}\label{sec:TwoParticle}
In this appendix, we will give details about two-particle cuts at the form level based on the discussion on regions, which will allow us to verify unitarity explicitly for generic $n$. The form  will factorize as region and  should satisfy 
\begin{equation}\label{eq:2cut_formula}
    \begin{split}
         &\mathcal{B}_a \Omega_a^{(L)}=\frac{dz}{ g(z)}  \prod_{l=1}^{L-1}\langle(AB)_l d^2 A_l\rangle\langle(AB)_l d^2 B_l\rangle \delta(\Omega_{IJ}A_l^I B_l^J)\\
        &\sum_{\text{allowed sets}} \mathcal{B}_b\Omega_{b}^{(L_{1})}(Z_1,{\cdots}, Z_i, A,B, Z_{j-1},{\cdots},Z_n) \times  \mathcal{B}_c \Omega_{c}^{(L_{2})}(A,Z_{i+1},\cdots,Z_j, B).
    \end{split}
\end{equation}

The factorization of the loop form follows trivially from eq.~\eqref{eq:loop_2_cut} since the uncut loop attaching to  the left and right regions are disjoint ($\{L_1\}\cap\{L_2\}=\emptyset$). So, the inequalities of the uncut loop on the left and right sets are unaffected by each other, and their form  just trivially product together \textit{i.e.} $\Omega_b \times \Omega_c$. However, the external left and right sets share the same twistor variables $\{Z_i, A, B\}$, thus their tree forms are not trivially factorized. Besides, the lack of a subspace prevents us from directly obtaining the tree form, which makes identifying the tree form of the two-cut as the product of amplitudes more involved. 

Here we will finish the computation of the forms for $n=6,8$ cases discussed in section~\ref{sec:loop_geo} and focus on the form of tree region and loop $(AB)$. We compare the result with gluing super-amplitudes associated with the positive solution in the momentum space.

%\begin{equation}
%    \begin{split}
%        \mathcal{M}_n^L=\frac{dx dz}{g(x,z)} \prod_{a=1}^{L-1}\langle(AB)_a d^2 A_a\rangle\langle(AB)_a d^2 B_a\rangle
%        &\sum_{\text{allowed sets}} \mathcal{M}_{\mathcal{L}}^{L_{1}}(Z_1,{\cdots}, Z_i, A,B, Z_{j-1},{\cdots},Z_n)\\
%        &\quad \quad  \times \mathcal{M}_{\mathcal{R}}^{L_{2}}(A,Z_{i+1},\cdots,Z_j, B).
%    \end{split}
%\end{equation}

\paragraph{Unitarity at $n=6$.}
The inequalities involving the twistors $\{Z_i, A,B\}$ in the double-cut geometry $\langle AB 23\rangle{=}\langle AB45\rangle{=}0$ can be replaced by the external left and right sets  $\mathcal{L}=\{1,2,A,B,5,6\}$, $\mathcal{R}=\{A,3,4,B\}$,  living in the region of amplituhedron: 
\begin{equation}
    \begin{gathered}
        \text{Left:}\ \langle i\, i{+}1\, j\, j{+}1 \rangle_{\mathcal{L}}<0,\ \{\langle 123i \rangle_{\mathcal{L}}\}\ \text{has $1$ sign flip}, \ \langle\!\langle i i{+}1\rangle\!\rangle_{\mathcal{L}}=0.\\
        \text{Right:}\ \langle 1234\rangle_{\mathcal{R}}<0,\ \langle\!\langle i i{+}1\rangle\!\rangle_{\mathcal{R}}=0.
    \end{gathered}
\end{equation}
We will denote the resulting one-form for this 6-4 channel as $\omega_6$.

From the generalized unitarity, the double cut of loop integrand equals the left and right amplitude integrate out the cut momentum and Grassmann variables
\begin{equation}
        \begin{split}
            \text{LS}_{6,+} \omega_6= \int d^3\eta_{k_1} d^3\eta_{k_2} &\frac{\delta^{(6)}(Q_L)\big(\sum_{\bar{p},\bar{q},\bar{r}=,2,4,6}\langle\bar{p}\bar{q}\rangle_L\eta_{\bar{r},L}{+}\sum_{p,q,r=,1,3,5}\langle pq\rangle_L\eta_{r,L}\big)}{c_{2,5}^Lc_{4,1}^Lc_{6,3}^L}\\
             &\qquad \qquad \cdot \frac{\delta^{(6)}(Q_R)}{\langle 12\rangle_R \langle23\rangle_R}\;\delta(k^2_1)\delta(k^2_2)\,d^3\ell
        \end{split}
\end{equation}
where $c_{\bar{r}s}=\langle \bar{r}{-}2\, \bar{r}{+}2\rangle-\sum_{i=1,3,,5}\langle\bar{r}i\rangle\langle s i\rangle$, and the labelling $L$, $R$ imply that the spinor variables is taking from $\lambda_L=\{\lambda_1,\lambda_2,k_1,k_2,\lambda_5,\lambda_6\},\ \lambda_R=\{-k_1,\lambda_3,\lambda_4,-k_2\}.$ On the RHS the Jacobian factor from solving the cut condition can be expressed as 
\begin{equation}
\delta(k^2_1)\delta(k^2_2)\,d^3\ell=\frac{1}{\langle\!\langle35\rangle\!\rangle z} \frac{\langle23\rangle\langle45\rangle}{\langle AB I\rangle}dz
\end{equation}
where we've used the cut solution in eq.~\eqref{eq: cutsol}.

To convert the generalized unitarity result into  factorization form in twistor space, we pull out the prefactor of LS$_{6,+}$ and two tree amplitudes in the RHS, which is defined in appendix~\ref{sec:Tdual}:
\begin{equation}
    \begin{split}
        &\text{LS}_{6,+}=\left(\frac{\delta^3(p)\delta^{(6)}(Q)}{\prod_{i=1}^6 \langle i\, i{+}1\rangle}\right) \left. \frac{\delta(\mathcal{D}\cdot \chi)}{\sqrt{\mathcal{D}_{1} \mathcal{D}_2 \mathcal{D}_3 \mathcal{D}_4 \mathcal{D}_5 \mathcal{D}_6}}\right|_{[\mathcal{D}]_{1\times 6}Z=0,\, \mathcal{D}{\cdot} \tilde{g} {\cdot} \mathcal{D}^T=0},\\
    \end{split}
\end{equation}
\begin{equation}
    \begin{split}
        & \frac{\delta^{(6)}(Q_L)\big(\sum_{\bar{p},\bar{q},\bar{r}=,2,4,6}\langle\bar{p}\bar{q}\rangle_L\eta_{\bar{r},L}{+}\sum_{p,q,r=,1,3,5}\langle pq\rangle_L\eta_{r,L}\big)}{c_{2,5}^Lc_{4,1}^Lc_{6,3}^L}\\
        &\qquad\ =\left(\frac{\delta^3(p_{\mathcal{L}})\delta^{(6)}(Q_{\mathcal{L}})}{\prod_{i=1}^6 \langle i\, i{+}1\rangle}\right) \left. \frac{\delta(L\cdot \chi_{\mathcal{L}})}{\sqrt{L_{1} L_2 L_3 L_4 L_5 L_6}}\right|_{[L]_{1\times 6}Z=0, \, L{\cdot} \tilde{g}_{\mathcal{L}} {\cdot} L^T=0},
    \end{split}
\end{equation}
and the four-point amplitude becomes unity after stripping off the prefactor.

Intermixing fermionic variables $\eta$ and $\chi$ in the equation requires carefully dealing transformation factor of them when integrating $\eta_{k_1},\, \eta_{k_2}$. To simplify the calculation, we choose the component amplitude $A_6(\bar{\phi},\phi,\bar{\psi},\psi,\bar{\psi},\phi)$ by integrating out $\eta_1$, $\eta_2$, and $\eta_6$. The contribution of this component can be easily related to each other in terms of the minors of Grassmannian formula in momentum and that in momentum-twistor space:
\begin{equation}
    \begin{split}
        (612)=\langle61\rangle\langle23\rangle\mathcal{D}_1,\ (612)_{L}=\langle61\rangle_{\mathcal{L}}\langle23\rangle_{\mathcal{L}}L_1.
    \end{split}
\end{equation}

By collecting a bunch of prefactors and Jocabian factors into the measure function $g(z)$, the remaining part becomes 
\begin{equation}
    \begin{split}
        \left. \frac{\mathcal{D}_1^3}{\sqrt{\mathcal{D}_{1} \mathcal{D}_2 \mathcal{D}_3 \mathcal{D}_4 \mathcal{D}_5 \mathcal{D}_6}}\right|_{\scalebox{0.75}{$\begin{gathered}  [\mathcal{D}]_{1\times 6}Z=0\,\\ \mathcal{D}{\cdot} \tilde{g} {\cdot} \mathcal{D}^T=0 \end{gathered}$}} \omega_6=&\frac{dz}{ g(z)}  \cdot \left. \frac{L_1^3}{\sqrt{L_{1} L_2 L_3 L_4 L_5 L_6}}\right|_{\scalebox{0.75}{$\begin{gathered}  [L]_{1\times 6}Z_{\mathcal{L}}=0\,\\ L{\cdot} \tilde{g} {\cdot} L^T=0 \end{gathered}$}}
    \end{split}
\end{equation}
here the measure function $g(z)=\frac{1}{ x(z) z^{3/2}} \sqrt{\frac{\langle23\rangle}{\langle45\rangle}}$ is coming from
\begin{equation}
    \begin{gathered}
    \left(\frac{\langle61\rangle_\mathcal{L}\langle12\rangle_\mathcal{L}\langle23\rangle_\mathcal{L}}{\langle61\rangle\langle12\rangle\langle23\rangle}\right)^3\left(\prod_{i=1}^6\frac{1}{\langle i\, i{+}1\rangle}\right)^{-1} \left(\prod_{i=1}^6\frac{1}{\langle i\, i{+}1\rangle_{\mathcal{L}}}\right) \frac{\langle12\rangle_{\mathcal{R}}^3}{\langle12\rangle_{\mathcal{R}}\langle23\rangle_{\mathcal{R}}}\\
    \times\frac{1}{\langle\!\langle35\rangle\!\rangle z} \frac{\langle23\rangle\langle45\rangle}{\langle AB I\rangle} \left(\frac{\langle A B I\rangle}{\langle  k_1 k_2\rangle}\right)^{-1/2} .
    \end{gathered}
\end{equation}
We note that the last term in the contribution of $g(z)$ is the scaling factor of $A$, $B$ to the cut momentum $k_1$, $k_2$ such that the first component of $A$, $B$ can be identified as spinors.

\paragraph{Unitarity at $n=8$.} We can use a similar strategy for $n=8$. The difference is that the eight-point has four chambers. Each chamber will give a partial contribution in eq.~\eqref{eq:2cut_formula} of the full amplitude. For example, the $8$-4 channel from the double cut $\langle AB34\rangle=\langle AB 56\rangle=0$ requires to sit the kinematic on  $(1\cap 2)$, $ (3\cap 2)$, $(1\cap 4)$,  and $(3\cap 4)$ and the inequalities are replaced by
\begin{equation}
    \begin{gathered}
        \text{Left:}\ \langle i\, i{+}1\, j\, j{+}1 \rangle_{\mathcal{L}}<0,\ \{\langle 123i \rangle_{\mathcal{L}}\}\ \text{has $2$ sign flip}, \ \langle\!\langle i i{+}1\rangle\!\rangle_{\mathcal{L}}=0.\\
        \text{Right:}\ \langle 1234\rangle_{\mathcal{R}}<0,\ \langle\!\langle i i{+}1\rangle\!\rangle_{\mathcal{R}}=0\,.
    \end{gathered}
\end{equation}
with external left and right sets  $\mathcal{L}=\{1,2,3,A,B,6,7,8\}$, $\mathcal{R}=\{A,4,5,B\}$. The first two chambers will give the same loop form, denoted by $\omega_{8}(2)$, and the last two chambers will give the loop form, denoted by $\omega_{8}(4)$.

In momentum space, the sum of  the form  $(1\cap 2)$ and $ (3\cap 2)$ follows the unitarity and gives
\begin{equation}
        \begin{split}
            \text{LS}_{8,+,+}[2]\, \omega_8(2)= \int d^3\eta_{k_1} d^3\eta_{k_2} &\left.\frac{1}{\sqrt{\big(p_{2,3,4,5}^2\big)_L}}\frac{\delta^{(12)}(C_{L} \eta_L) }{(1234)_L(3456)_L(4567)_L}\right|_{\scalebox{0.75}{$\begin{gathered}  [C_L]^{+,+}_{4\times 8}\lambda_{\mathcal{L}}{=}0,\, (2345)_L{=}0 \\ C_L^T{\cdot} C_L{=}0.\end{gathered}$}}\\
             &\  \cdot \frac{\delta^{(6)}(Q_R)}{\langle 12\rangle_R \langle23\rangle_R}   \;\delta(k^2_1)\delta(k^2_2)\,d^3\ell
        \end{split}
\end{equation}
Similarly, the sum of form $(1\cap 4)$ and $ (3\cap 4)$ is
\begin{equation}
        \begin{split}
            \text{LS}_{8,+,+}[4]\, \omega_8(4)= \int d^3\eta_{k_1} d^3\eta_{k_2} &\left.\frac{1}{\sqrt{\big(p_{4,5,6,7}^2\big)_L}}\frac{\delta^{(12)}(C_{L} \eta_L) }{(1234)_L(2345)_L(3456)_L}\right|_{\scalebox{0.75}{$\begin{gathered}  [C_L]^{+,+}_{4\times 8}\lambda_{\mathcal{L}}{=}0,\, (2345)_L{=}0 \\ C_L^T{\cdot} C_L{=}0.\end{gathered}$}}\\
             &\  \cdot \frac{\delta^{(6)}(Q_R)}{\langle 12\rangle_R \langle23\rangle_R}   \;\delta(k^2_1)\delta(k^2_2)\,d^3\ell\,.
        \end{split}
\end{equation}
Here,  the spinor variables $\lambda_L=\{\lambda_1,\lambda_2,\lambda_3,k_1,k_2,\lambda_6,\lambda_7,\lambda_8\}$, $\lambda_R=\{-k_1,\lambda_3,\lambda_4,k_2\}.$ The Jocobian factor from solving the cut condition is
\begin{equation}
    \delta(k^2_1)\delta(k^2_2)\,d^3\ell=\frac{1}{\langle\!\langle46\rangle\!\rangle z} \frac{\langle34\rangle\langle56\rangle}{\langle AB I\rangle} dz.
\end{equation}

Taking the component by integrating out $\eta_6$, $\eta_7$, $\eta_8$, $\eta_1$, stripping off the prefactor and collecting them into $g(z)$ give
\begin{equation}
    \begin{split}
        &\left.\frac{J_{2,\mathcal{D}}\,m_{7}^3}{\sqrt{m_{1}m_{3}m_{4}m_{5} m_{7}m_{8}}}\right|_{\scalebox{0.75}{$\begin{gathered}   [\mathcal{D}]^{+,+}_{2\times 8}Z=0, \\ \mathcal{D}{\cdot} \tilde{g} {\cdot} \mathcal{D}^T{=}0, m_{2}{=}0.\end{gathered}$}} \omega_8(2)= \frac{dz}{g(z)}  {\cdot} \left.\frac{J_{2,L}\,\big(m_{7}^{L}\big)^3}{\sqrt{m_{1}^Lm_{3}^L m_{4}^L m_{5}^L m_{7}^Lm_{8}^L}}\right|_{\scalebox{0.75}{$\begin{gathered}  [L]^{+,+}_{2\times 8}Z_{\mathcal{L}}=0, \\ L{\cdot} \tilde{g}_{\mathcal{L}} {\cdot} L^T{=}0, m_{2}^L{=}0.\end{gathered}$}}\\
        &\left.\frac{J_{2,\mathcal{D}}\,m_{7}^3}{\sqrt{m_{1}m_{2}m_{3}m_{5} m_{6}m_{7}}}\right|_{\scalebox{0.75}{$\begin{gathered}   [\mathcal{D}]^{+,+}_{2\times 8}Z=0, \\ \mathcal{D}{\cdot} \tilde{g} {\cdot} \mathcal{D}^T{=}0, m_{4}{=}0.\end{gathered}$}} \omega_8(4)= \frac{dz}{g(z)}  {\cdot} \left.\frac{J_{2,L}\,\big(m_{7}^{L}\big)^3}{\sqrt{m_{1}^L m_{2}^L m_{3}^L  m_{5}^Lm_{6}^L m_{7}^L}}\right|_{\scalebox{0.75}{$\begin{gathered}  [L]^{+,+}_{2\times 8}Z_{\mathcal{L}}=0, \\ L{\cdot} \tilde{g}_{\mathcal{L}} {\cdot} L^T{=}0, m_{4}^L{=}0.\end{gathered}$}}
    \end{split}
\end{equation}
where $\mathcal{D}$ and $L$ are chosen to be localized on the positive solutions.  The  Jocabian factors are 
\begin{equation}
    \begin{split}
        &J_{2,\mathcal{D}}=\left(\frac{\langle1256\rangle}{\langle12\rangle\langle56\rangle} \prod_{i=2,6} \frac{1}{\langle i\, i{+}1\rangle\langle i{+}1 \,i{+}2\rangle\langle i{+}2\, i{+}3\rangle}\right)^{-1/2}\, ,\\
        &J_{2,L}=\left(\frac{\langle1256\rangle_{\mathcal{L}}}{\langle12\rangle_{\mathcal{L}}\langle56\rangle_{\mathcal{L}}} \prod_{i=2,6} \frac{1}{\langle i\, i{+}1\rangle_{\mathcal{L}}\langle i{+}1 \,i{+}2\rangle_{\mathcal{L}}\langle i{+}2\, i{+}3\rangle_{\mathcal{L}}}\right)^{-1/2} \Bigg(\frac{\langle ABI\rangle}{\langle k_1 k_2\rangle}\Bigg)^{1/2}\, ,\\
        &J_{4,\mathcal{D}}=\left(\frac{\langle3478\rangle}{\langle34\rangle\langle78\rangle} \prod_{i=4,8} \frac{1}{\langle i\, i{+}1\rangle\langle i{+}1 \,i{+}2\rangle\langle i{+}2\, i{+}3\rangle}\right)^{-1/2}\, ,\\
        &J_{4,L}=\left(\frac{\langle3478\rangle_{\mathcal{L}}}{\langle34\rangle_{\mathcal{L}}\langle78\rangle_{\mathcal{L}}} \prod_{i=4,8} \frac{1}{\langle i\, i{+}1\rangle_{\mathcal{L}}\langle i{+}1 \,i{+}2\rangle_{\mathcal{L}}\langle i{+}2\, i{+}3\rangle_{\mathcal{L}}}\right)^{-1/2} \Bigg(\frac{\langle ABI\rangle}{\langle k_1 k_2\rangle}\Bigg)^{1/2}\, ,
    \end{split}
\end{equation}
and the measure function $g(z)=\frac{1}{x^{3/2}z^{3/2}} \sqrt{\frac{\langle34\rangle}{\langle56\rangle}}$ is coming from
\begin{equation}\label{eq:g_ftn_8}
    \begin{gathered}
        \left(\frac{\langle78\rangle_\mathcal{L}\langle81\rangle_\mathcal{L}\langle12\rangle_\mathcal{L}}{\langle78\rangle\langle81\rangle\langle12\rangle}\right)^3 \left(\prod_{i=1}^8\frac{\langle i\, i{+}1\rangle\langle i{+}1\, i{+}2\rangle\langle i{+2}\, i{+}3\rangle}{\langle i\, i{+}1\rangle_{\mathcal{L}}\langle i{+}1\, i{+}2\rangle_{\mathcal{L}}\langle i{+2}\, i{+}3\rangle_{\mathcal{L}}}\right)^{1/2} \frac{\langle14\rangle_{\mathcal{R}}^3}{\langle12\rangle_{\mathcal{R}}\langle23\rangle_{\mathcal{R}}}\\
        \times\frac{1}{\langle\!\langle46\rangle\!\rangle z} \frac{\langle34\rangle\langle56\rangle}{\langle AB I\rangle} \left(\frac{\langle A B I\rangle}{\langle  k_1 k_2\rangle}\right)^{-1} .
    \end{gathered}
\end{equation}

Now, we go beyond the $n$-4 factorization channel and consider 6-6 channel at $n{=}8$, choosing the double cut to be $\langle A B 23\rangle=\langle A B 67\rangle=0$. 
The double-cut geometry can be replaced by the inequalities involving external left and right sets  $\mathcal{L}=\{1,2,A,-B,-7,-8\}$, $\mathcal{R}=\{A,3,4,5,6,B\}$,  living in the region of amplituhedron: 
\begin{equation}
    \begin{gathered}
        \text{Left:}\ \langle i\, i{+}1\, j\, j{+}1 \rangle_{\mathcal{L}}<0,\ \{\langle 123i \rangle_{\mathcal{L}}\}\ \text{has $1$ sign flip}, \ \langle\!\langle i i{+}1\rangle\!\rangle_{\mathcal{L}}=0.\\
        \text{Right:}\ \langle i\, i{+}1\, j\, j{+}1 \rangle_{\mathcal{R}}<0,\ \{\langle 123i \rangle_{\mathcal{R}}\}\ \text{has $1$ sign flip}, \ \langle\!\langle i i{+}1\rangle\!\rangle_{\mathcal{R}}=0.
    \end{gathered}
\end{equation}
The form on this channel  only has support on $(3\cap 2)$ and $ (3\cap 4)$. These two chambers have the same loop form, denoted by $\omega_8(3)$.

We sum the form of double-cut geometry of the chambers $(3\cap 2)$ and $ (3\cap 4)$, and it satisfies
\begin{equation}
        \begin{gathered}
            \text{LS}_{8,+,+}[3]\, \omega_8(3)= \int d^3\eta_{k_1} d^3\eta_{k_2} \frac{\delta^{(6)}(Q_L)\big(\sum_{\bar{p},\bar{q},\bar{r}=,2,4,6}\langle\bar{p}\bar{q}\rangle_L\eta_{\bar{r},L}{+}\sum_{p,q,r=,1,3,5}\langle pq\rangle_L\eta_{r,L}\big)}{c_{2,5}^Lc_{4,1}^Lc_{6,3}^L}\\
               \cdot \frac{\delta^{(6)}(Q_R)\big(\sum_{\bar{p},\bar{q},\bar{r}=,2,4,6}\langle\bar{p}\bar{q}\rangle_R\eta_{\bar{r},R}{+}\sum_{p,q,r=,1,3,5}\langle pq\rangle_R\eta_{r,R}\big)}{c_{2,5}^Rc_{4,1}^Rc_{6,3}^R}  \;\delta(k^2_1)\delta(k^2_2)\,d^3\ell\,.
        \end{gathered}
\end{equation}
Here,  the spinor variables $\lambda_L=\{\lambda_1,\lambda_2,k_1,-k_2,-\lambda_7,-\lambda_8\}$, $\lambda_R=\{-k_1,\lambda_3,\lambda_4,\lambda_5,\lambda_6,-k_2\}$, and the Jacobian factor resulting from solving the cut condition can be written as 
\begin{equation}
    \delta(k^2_1)\delta(k^2_2)\,d^3\ell=\frac{1}{\langle\!\langle37\rangle\!\rangle z{+}\langle\!\langle36\rangle\!\rangle} \frac{\langle23\rangle\langle67\rangle}{\langle AB I\rangle} dz\,.
\end{equation}
Notably, the term $1/(\langle\!\langle37\rangle\!\rangle z{+}\langle\!\langle36\rangle\!\rangle )$ is also the same as the Jocabian factor of the symplectic condition $\delta(\Omega_{IJ}A^I B^J)$.

Here we integrate out $\eta_1,\eta_2,\eta_3, \eta_8$, which corresponds to integrating out $\eta_{1,\mathcal{L}(\mathcal{R})}$, $\eta_{2,\mathcal{L}(\mathcal{R})}$, and $\eta_{6,\mathcal{L}(\mathcal{R})}$ in the left and right sets,. Pulling out the prefactor from the amplitude and collecting them together, we arrive
\begin{equation}
    \begin{split}
        &\left.\frac{J_{3,\mathcal{D}}\,m_{8}^3}{\sqrt{m_{1}m_{2}m_{4}m_{5} m_{6} m_{8}}}\right|_{\scalebox{0.75}{$\begin{gathered}   [\mathcal{D}]^{+,+}_{2\times 8}Z=0,\\ \mathcal{D}{\cdot} \tilde{g} {\cdot} \mathcal{D}^T{=}0,  m_{3}{=}0.\end{gathered}$}} \omega_8(3)\\
        =& \frac{dz}{ g(z) } \cdot \left.\frac{L_1^3}{\sqrt{L_1 L_2 L_3 L_4 L_5 L_6}}\right|_{\scalebox{0.75}{$ \begin{gathered} [L]^+_{1\times 6}Z_{\mathcal{L}}=0\\
        L{\cdot} \tilde{g}_{\mathcal{L}} {\cdot} L^T{=}0
        \end{gathered}$}} \times \left.\frac{R_1^3}{\sqrt{R_1 R_2 R_3 R_4 R_5 R_6}}\right|_{\scalebox{0.75}{$ \begin{gathered}
            [R]^+_{1\times 6}Z_{\mathcal{R}}=0\\
            R{\cdot} \tilde{g}_{\mathcal{R}} {\cdot} R^T{=}0
        \end{gathered} $}}.
    \end{split}
\end{equation}
The  Jocabian factors are 
\begin{equation}
    \begin{split}
        &J_{3,\mathcal{D}}=\left(\frac{\langle2367\rangle}{\langle23\rangle\langle67\rangle} \prod_{i=3,7} \frac{1}{\langle i\, i{+}1\rangle\langle i{+}1 \,i{+}2\rangle\langle i{+}2\, i{+}3\rangle}\right)^{-1/2}\, ,
    \end{split}
\end{equation}
and $g(z)=\frac{1 }{x(z)z (\langle\!\langle37\rangle\!\rangle z{+}\langle\!\langle36\rangle\!\rangle )}  \sqrt{\frac{\langle23\rangle\langle8123\rangle\langle2345\rangle\langle5678\rangle}{\langle67\rangle\langle1234\rangle}}$ is coming from
\begin{equation}
    \begin{gathered}
\left(\frac{\langle61\rangle_\mathcal{L}\langle12\rangle_\mathcal{L}\langle61\rangle_\mathcal{R}\langle12\rangle_\mathcal{R}}{\langle81\rangle\langle12\rangle\langle23\rangle}\right)^3 \left(\frac{\prod_{i=1}^8\langle i\, i{+}1\rangle\langle i{+}1\, i{+}2\rangle\langle i{+2}\, i{+}3\rangle}{\prod_{i=1}^6 \langle i\, i{+}1\rangle_{\mathcal{L}}\langle i{+}1\, i{+}2\rangle_{\mathcal{L}} \langle i\, i{+}1\rangle_{\mathcal{R}}\langle i{+}1\, i{+}2\rangle_{\mathcal{R}}}\right)^{1/2} \\
        \times  \frac{1}{\langle\!\langle37\rangle\!\rangle z{+}\langle\!\langle36\rangle\!\rangle } \frac{\langle23\rangle\langle67\rangle}{\langle AB I\rangle} \left(\frac{\langle A Z_{\text{ref}} I\rangle}{\langle  k_1 \lambda_{\text{ref}}\rangle}\right)^{2} \left(\frac{\langle  B Z_{\text{ref}} I\rangle}{\langle  k_2 \lambda_{\text{ref}}\rangle}\right)^{-1}.
    \end{gathered}
\end{equation}
Here, $Z_{\text{ref}}$ is the reference twistor with the first two components $\lambda_{\text{ref}}$.

\section{Local integrand basis}\label{app:local_integral}
This appendix defines the two-loop local integrals used for $n=6,8$ ABJM theory and one-loop integrals for $n=6$  $\mathcal{N}=4$ sYM. For simplicity, we only explicitly present the numerators of these integrals, and the denominators can be easily inferred from the corresponding diagram. For instance,  the integral $I^{\text{critter}}(i)$ at $n=6$ (shown below) is
\begin{equation}
    I^{\text{critter}}(i):=\int_{\ell_1,\ell_2} \frac{n^{\text{critter}}(i)}{(\ell_1\cdot i{-}1)(\ell_1\cdot i)(\ell_1\cdot i{+}1)(\ell_1\cdot \ell_2)(\ell_2\cdot i{+}2)(\ell_2\cdot i{+}3)(\ell_2\cdot i{+}4)}+(\ell_1\leftrightarrow \ell_2)
\end{equation}
The integrals are written in their dual form, and readers can refer to the notation introduced in~\cite{Caron-Huot:2012sos,He:2022lfz}.

\paragraph{Six points:} The local integrals contributing to the two-loop six-point are

\begin{eqnarray*}
    \vcenter{\hbox{\scalebox{0.7}{


}
}}
\end{eqnarray*}
\begin{equation}
    \begin{split}
        n^{\text{critter}}(i):=&\big(\epsilon(\ell_1,i{-}1,i,i{+}1,\ ^\mu)\epsilon(\ell_2,i{+}2,i{+}3,i{+}4,\ _\mu){+}(\ell_1\cdot i)(\ell_2\cdot i{+}3)\\
        &\quad  \quad (i{-}1\cdot i{+}1)(i{+}2\cdot i{+}4)\big)/2 \, ,\\
        n^{\text{crab}}(i):=&\big(\epsilon(\ell_1,i{-}1,i,i{+}1,\ ^\mu)\epsilon(\ell_2,i{+}3,i{+}4,i{-}1,\ _\mu){+}(\ell_1\cdot i)(\ell_2\cdot i{+}4)\\
        &\quad  \quad (i{-}1\cdot i{+}1)(i{+}3\cdot i{-}1)\big)/2 \, ,\\
        n^{\text{2mh}}_{\pm}(i):=&\Big(\epsilon(\ell_1,i{-}1,i,i{+}1,\ ^\mu)\epsilon(\ell_2,i{+}1,i{+}3,i{-}1,\ _\mu){-}(\ell_1\cdot i)(i{-}1\cdot i{+}1) \\
        &\ \big((\ell_2\cdot i{-}1) (i{+}1\cdot i{+3}){+}(\ell_2\cdot i{+}1)(i{-}1\cdot i{+}3){-}(\ell_2\cdot i{+}3)(i{-}1\cdot i{+}1)\big)\Big)/2 \\
        &\ \pm (-1)^{i}\epsilon(\ell_1,i{-}1,i,i{+}1,\ell_2)\sqrt{(i{+}1\cdot i{+}3\cdot i{-}1)}/\sqrt{2}\, ,\\
        n^{bt}_{1}(i):=&\epsilon(\ell_1,i{-}1,i,i{+}1,i{+}2)\sqrt{(i\cdot i{+}2\cdot i{+}4)}/\sqrt{2}(i\cdot i{+}2)\, ,\\
        n^{bt}_{2}(i):=&\epsilon(\ell_1,i{-}1,i,i{+}1,i{+}4)\sqrt{(i\cdot i{+}2\cdot i{+}4)}/\sqrt{2}(i\cdot i{+}4)\,.
    \end{split}
\end{equation}
with the shorthand notation $\sqrt{(i\cdot j \cdot k)}:=\sqrt{(i\cdot j)(j\cdot k)(k\cdot i)}$.

\paragraph{Eight points:} For two-loop eight points, the local integrals are

\begin{eqnarray*}
    \vcenter{\hbox{\scalebox{0.7}{


}
}}
\end{eqnarray*}
\begin{equation}
    \begin{split}
        n^{db}_A(i):=& \big(\epsilon(\ell_1,i{-}1,i,i{+}1,\ ^\mu)\epsilon(\ell_2,i{+}2,i{+}3,i{+}4,\ _\mu){+}(\ell_1\cdot i)(\ell_2\cdot i{+}3)\\
        &\quad \quad (i{-}1\cdot i{+}1)(i{+}2\cdot i{+}4)\big)/2 \, ,\\
        n^{db}_B(i):=&\big(\epsilon(\ell_1,i{-}1,i,i{+}1,\ ^\mu)\epsilon(\ell_2,i{+}1,i{+}2,i{+}3,\ _\mu){+}(\ell_1\cdot i)(\ell_2\cdot i{+}2)\\
        &\quad \quad (i{-}1\cdot i{+}1)(i{+}1\cdot i{+}3)\big)/2\, ,\\
        n^{db}_C(i):=&\epsilon(\ell_1,i{-}1,i,i{+}1,\ ^\mu)\epsilon(\ell_2,i{+}3,i{+}4,i{+}5,\ _\mu)/2 \, .
    \end{split}
\end{equation}

\begin{eqnarray*}
    \vcenter{\hbox{\scalebox{0.7}{


}
}}
\end{eqnarray*}
\begin{equation}
    \begin{split}
        n^{db}_{\overline{D},\pm}(i){:=}&\Big(\epsilon(\ell_1,i{-}1,i,i{+}1,\ ^\mu)\epsilon(\ell_2,i{+}1,i{+}3,i{-}1,\ _\mu) {+}(\ell_1\cdot i)(\ell_2\cdot i{+}3)(i{-}1\cdot i{+}1)^2\\
        &\, {-}(\ell_1\cdot i)(\ell_2\cdot i{+}1)(i{-}1\cdot i{+}1)(i {-}1\cdot i{+}3){-}(\ell_1\cdot i)(\ell_2\cdot i{-}1)(i{-}1\cdot i{+}1)\\
        &\, (i {+}1\cdot i{+}3)\Big)/2 {\pm} ({-})^i \epsilon(\ell_1,i{-}1,i,i{+}1,\ell_2) \sqrt{(i{+}1,i{+}3,i{-}1)}/\sqrt{2} \, ,\\
        n^{db}_{\underline{D},\pm}(i){:=}& \Big(\epsilon(\ell_1,i{-}1,i,i{+}1,\ ^\mu)\epsilon(\ell_2,i{+}1,i{+}5,i{-}1,\ _\mu){+}(\ell_1\cdot i)(\ell_2\cdot i{+}5)(i{-}1\cdot i{+}1)^2\\
        &\, {-}(\ell_1\cdot i)(\ell_2\cdot i{+}1)(i{-}1\cdot i{+}1)(i {-}1\cdot i{+}5) {-}(\ell_1\cdot i)(\ell_2\cdot i{-}1)(i{-}1\cdot i{+}1)\\
    &\,  (i {+}1\cdot i{+}5)\Big)/2{\pm}({-})^i\epsilon(\ell_1,i{-}1,i,i{+}1,\ell_2)\sqrt{(i{+}1,i{+}5,i{-}1)}/\sqrt{2}\, .
    \end{split}
\end{equation}

\begin{equation}
    \begin{split}
        n^{db}_{\overline{E},\pm}(i):=&\Big(\epsilon(\ell_1,i{-}1,i,i{+}1,\ ^\mu)\epsilon(\ell_2,i{+}3,i{+}5,i{-}1,\ _\mu)+(\ell_1\cdot i)(\ell_2\cdot i{+}5)\\
        &\ (i{-}1\cdot i{+}1)(i {-}1\cdot i{+}3){-}(\ell_1\cdot i)(\ell_2\cdot i{+}3)(i{-}1\cdot i{+}1)(i {-}1\cdot i{+}5) \Big)/2\\
        &\pm({-})^i\epsilon(\ell_1,i{-}1,i,i{+}1,\ell_2)\sqrt{(i{+}3,i{+}5,i{-}1)}/\sqrt{2}\, ,\\
        n^{db}_{\underline{E},\pm}(i):=&\Big(\epsilon(\ell_1,i{-}1,i,i{+}1,\ ^\mu)\epsilon(\ell_2,i{+}1,i{+}3,i{+}5,\ _\mu)-(\ell_1\cdot i)(\ell_2\cdot i{+}5)\\
    &\ (i{-}1\cdot i{+}1)(i {+}1\cdot i{+}3){+}(i{-}1\cdot i{+}1)(i {+}1\cdot i{+}5)(\ell_1\cdot i)(\ell_2\cdot i{+}3) \Big)/2\\
    &\pm ({-})^i \epsilon(\ell_1,i{-}1,i,i{+}1,\ell_2) \sqrt{(i{+}1,i{+}3,i{+}5)}/\sqrt{2}\, .
    \end{split}
\end{equation}

\begin{eqnarray*}
    \vcenter{\hbox{\scalebox{0.7}{


}
}}
\end{eqnarray*}
\begin{equation}
    \begin{split}
        &n^{db}_{F,\pm,\pm^\prime}(i){:=} \Big(\epsilon(\ell_1,i{-}1,i,i{+}1,\, ^\mu)\epsilon(\ell_2,i{+}2,i{+}4,i{+}6,\, _\mu){-}(\ell_1\cdot i)(i{-}1{\cdot} i{+}1)\big( (\ell_2\cdot i{+}2) (i{+}4{\cdot} i{+}6) \\
        &\,  {-}(\ell_2\cdot i{+}4)(i{+}2\cdot i{+}6){+}(\ell_2\cdot i{+}6)(i{+}2\cdot i{+}4)\Big)/2{-}{\pm}{\pm^\prime}(\ell_1\cdot i)(\ell_2\cdot y_{i{+}1,i{+}6})\sqrt{(i {\cdot} i{+}2{\cdot}  i{+}4 {\cdot} i{+}6)}\\
        &\,  {\pm}({-})^{i-1}\epsilon(\ell_1,i{-}1,i,i{+}1,\ell_2) \sqrt{(i{+}2 {\cdot} i{+}4 {\cdot} i{+}6)}/\sqrt{2}{\pm}^\prime({-})^{i} \frac{(\ell_1\cdot i)(i{-}1\cdot i{+}1)\sqrt{(i{\cdot} i{+}2{\cdot} i{+}6)}}{\sqrt{2}(i\cdot i{+}2)(i \cdot i{+}6)(i{+}4 \cdot y_{i{+}1,i{+}6})}\\
        &\,  \Big(\epsilon(\ell_2,i{-}1,i{+}1,i{+}2,i{+}4)(i{+}4\cdot i{+}6) {-}\epsilon(\ell_2,i{+}6,i{+}1,i{+}2,i{+}4)(i{+}4\cdot i{-}1)\Big) \, ,
    \end{split}
\end{equation}
where we define $\sqrt{(i\cdot j\cdot k\cdot l)}:=\sqrt{(i\cdot j)(j\cdot k)(k\cdot l)(l\cdot i)}$. The double labeling $\pm$ in the subscript of $n^{db}_{F,\pm,\pm^\prime}(i)$ corresponds to the sign of the first (resp. second) term in the definition. The same convention applies to the double subscripts in $n^{db}_{G,\pm,\pm^\prime}(i)$ below.

\begin{equation}
    \begin{split}
    &n^{db}_{G,\pm,\pm^\prime}(i){:=} \Big(\epsilon(\ell_1,i{-}2,i,i{+}2,\, ^\mu)\epsilon(\ell_2,i{+}2,i{+}4,i{-}2,\, _\mu){-}(\ell_1\cdot i)(\ell_2\cdot i{-}2)(i{-}2{\cdot} i{+}2)(i{+}2{\cdot} i{+}4)\\
    &\quad \quad \quad \, {+}(\ell_1\cdot i{-}2)(i\cdot i{+}2)\big((\ell_2\cdot i{-}2)(i{+}2\cdot i{+}4){-}(\ell_2\cdot i{+}4)(i{-}2\cdot i{+}2)\big){+}\big( (\ell_1\cdot i)(i{-}2\cdot i{+}2)\\
    &\quad \quad \quad \, {-}(\ell_1\cdot i{+}2)(i{-}2\cdot i)\big)\big( (\ell_2\cdot i{+}4)(i{-}2\cdot i{+}2){-}(\ell_2\cdot i{+}2)(i{-}2\cdot i{+}4)\big)\Big)/2+{\pm}{\pm}^\prime(\ell_1\cdot \ell_2)\\
    &\quad \quad \quad \, \sqrt{(i{-}2\cdot i\cdot i{+}2)}\sqrt{(i{+}2\cdot i{+}4\cdot i{-}2)}\pm({-})^i\epsilon(\ell_1,i{-}2,i,i{+}2,\ell_2)\sqrt{(i{+}2\cdot i{+}4\cdot i{-}2)}/\sqrt{2}\\
    &\quad \quad \quad \, \pm^\prime({-})^{i-1}\epsilon(\ell_1,i{+}2,i{+}4,i{-}2,\ell_2)\sqrt{(i{-}2\cdot i\cdot i{+}2)}/\sqrt{2}\, .
    \end{split}
\end{equation}

\begin{eqnarray*}
    \vcenter{\hbox{\scalebox{0.7}{


}
}}
\end{eqnarray*}
\begin{equation}
    \begin{split}
        &n^{bt}_{\overline{A}}(i):=\epsilon(\ell_1,i{-}1,i,i{+1},i{+}6)\sqrt{(i\cdot i{+}4\cdot i{+}6)}/\sqrt{2}(i\cdot i{+}6)\,,\\
        &n^{bt}_{\underline{A}}(i):=\epsilon(\ell_1,i{-}1,i,i{+1},i{+}2)\sqrt{(i\cdot i{+}2\cdot i{+}4)}/\sqrt{2}(i\cdot i{+}2)\, ,\\
        &n^{bt}_{B}(i):=\epsilon(\ell_1,i{-}1,i,i{+1},i{+}2)\sqrt{(i\cdot i{+}2\cdot i{+}6)}/\sqrt{2}(i\cdot i{+}2)\, .
    \end{split}
\end{equation}

\noindent Finally, we list the local integrands used for the NMHV $\mathcal{N}=4$ sYM, 
\begin{equation}
    \begin{split}
        &I^{\text{1mb}}_1(i):=\int_{A,B} \frac{\langle AB (i{-}2\, i{-}1\, i )\cap(i\, i{+}1\, i{+}2)\rangle\langle X\, i{+}1\, i{-}1\rangle}{\langle AB\, i{-}2\, i{-}1\rangle \langle AB\, i{-}1\, i\rangle \langle AB\, i\, i{+}1\rangle \langle AB\, i{+}1\, i{+}2\rangle \langle AB X\rangle}\\
        &I^{\text{1mb}}_2(i):=\int_{A,B} \frac{\langle AB\, i{+}4\, i{+}2\rangle\langle X\, (i{+}1\, i{+}2\, i{+}3 )\cap( i{+}3\, i{+}4\, i{+}5)\rangle}{\langle AB\, i{+}1\, i{+}2\rangle \langle AB\, i{+}2\, i{+}3\rangle \langle AB\, i{+}3\, i{+}4\rangle \langle AB\, i{+}4\, i{+}5\rangle \langle AB X\rangle}\\
        &I^{\text{2mh}}_1(i):=\int_{A,B} \frac{\langle AB \, i{+}2\, (i\, i{+}1)\cap\big( i{+}2\, i{+}3 \,(i{+}4\,i{+}5\,i{+}1)\cap X -(i{+}2,i{+}3)\leftrightarrow (i{+}4,i{+}5) \big) \rangle}{\langle AB\, i\, i{+}1\rangle \langle AB\, i{+}1\, i{+}2\rangle \langle AB\, i{+2}\, i{+}3\rangle \langle AB\, i{+}4\, i{+}5\rangle \langle AB X\rangle}\\
        &I^{\text{2mh}}_2(i):=\int_{A,B} \frac{\langle AB \, i{-}2\, (i{-}1\, i)\cap\big( i{-}3\, i{-}2 \,(i{+}1\,i{+}2\,i{-}1)\cap X -(i{-}3,i{-}2)\leftrightarrow (i{+}1,i{+}2) \big) \rangle}{\langle AB\, i{-}3\, i{-}2\rangle \langle AB\, i{-}2\, i{-}1\rangle \langle AB\, i{-}1\, i\rangle \langle AB\, i{+}1\, i{+}2\rangle \langle AB X\rangle}.
    \end{split}
\end{equation}

$$
\vcenter{\hbox{\scalebox{1}{


}}}$$

\noindent
and the triangle integral $I^{\text{tri}}(i)$ is defined by
\begin{equation}
    I^{\text{tri}}(i):=\int_{A,B}\frac{\langle i{-}1\,i\,i{+}1\,i{+}2\rangle \langle X\, i{+}1\,i\rangle}{\langle AB\, i{-}1\, i\rangle \langle AB\, i\, i{+}1\rangle\langle AB\, i{+}1\, i{+}2\rangle \langle AB X\rangle}.
\end{equation}

\bibliographystyle{utphys}
\bibliography{bib}

\end{document}